\documentclass[letterpaper,english]{IEEEtran}
\usepackage[T1]{fontenc}
\usepackage[latin9]{inputenc}
\usepackage{amsmath}
\usepackage{mathtools}
\usepackage{amsthm}
\usepackage{amssymb}
\usepackage{color}
\usepackage{hyperref}
\mathtoolsset{showonlyrefs}
\usepackage{bm}
     
\usepackage{enumerate}

\makeatletter
\theoremstyle{plain}
\newtheorem{thm}{\protect\theoremname}
\newtheorem*{thm*}{\protect\theoremname}
\theoremstyle{definition}
\newtheorem{defn}[thm]{\protect\definitionname}
\newtheorem*{defn*}{\protect\definitionname}
\theoremstyle{plain}

\theoremstyle{definition}
\newtheorem{example}[thm]{\protect\examplename}
\newtheorem{assumption}{Assumption}
\theoremstyle{plain}
\newtheorem{lem}[thm]{\protect\lemmaname}
\ifx\proof\undefined
\newenvironment{proof}[1][\protect\proofname]{\par
\normalfont\topsep6\p@\@plus6\p@\relax
\trivlist
\itemindent\parindent
\item[\hskip\labelsep\scshape #1]\ignorespaces
}{%
\endtrivlist\@endpefalse
}
\providecommand{\proofname}{Proof}
\fi
\theoremstyle{remark}
\newtheorem{rem}[thm]{\protect\remarkname}

\newcommand{\ce}{c}

\newcommand{\E}{\mathbb{E}}
\newcommand{\eps}{\epsilon}
\newcommand{\gexit}{G}

\newcommand{\expt}{\ensuremath{\mathbb{E}}}
\newcommand{\ex}[1]{\ensuremath{\mathbb{E} \left[ #1\right]}}
\newcommand{\pr}[1]{\ensuremath{\mathbb{P}\left[ #1\right]}}
\DeclareMathOperator{\BER}{\mathsf{BER}}
\DeclareMathOperator{\var}{\mathsf{Var}}
\DeclareMathOperator{\snr}{\mathsf{snr}}

\renewcommand{\mid}{\,|\,}

\newcommand{\vU}{\bm{U}}
\newcommand{\vX}{\bm{X}}
\newcommand{\vY}{\bm{Y}}
\newcommand{\vZ}{\bm{Z}}
\newcommand{\vnot}{\bm{0}}
\newcommand{\vb}{\bm{b}}
\newcommand{\vc}{\bm{c}}
\newcommand{\vi}{\bm{i}}
\newcommand{\vu}{\bm{u}}
\newcommand{\vv}{\bm{v}}
\newcommand{\vx}{\bm{x}}
\newcommand{\vy}{\bm{y}}

\newcommand{\cX}{\mathcal{X}}
\newcommand{\cY}{\mathcal{Y}}
\newcommand{\cZ}{\mathcal{Z}}

\newcommand{\cM}{\mathcal{M}}

\newcommand{\cC}{\mathcal{C}}
\newcommand{\cG}{\mathcal{G}}
\newcommand{\cH}{\mathcal{H}}

\newcommand{\normal}{\mathcal{N}}
\DeclareMathOperator{\mmse}{\mathsf{mmse}}

\usepackage{pgfplots}
\pgfplotsset{compat=1.12}
\usetikzlibrary{positioning}

\newif\ifextfig
\extfigtrue

\makeatother

\usepackage{babel}
\providecommand{\examplename}{Example}
\providecommand{\lemmaname}{Lemma}
\providecommand{\propositionname}{Proposition}
\providecommand{\definitionname}{Definition}
\providecommand{\remarkname}{Remark}
\providecommand{\theoremname}{Theorem}

\allowdisplaybreaks

\renewcommand{\var}{\mathrm{Var}}

\newif\ifresponse
\responsefalse

\ifresponse
\colorlet{high}{blue}
\else
\colorlet{high}{black}
\fi

\newif\ifbib
\bibfalse

\newif\ifbio
\biofalse

\begin{document}

\title{Reed--Muller Codes on BMS Channels \\ Achieve Vanishing Bit-Error Probability \\ for All Rates Below Capacity}
\author{Galen Reeves and Henry D. Pfister\thanks{This is an updated version of~\cite{Reeves-arxiv21} where the title has been changed to match the journal version~\cite{Reeves-it23}. The work of G.~Reeves and H.~D.~Pfister was supported in part by the National Science Foundation (NSF) under Grant Numbers 1718494, 1750362, and 1910571.
Any opinions, findings, recommendations, and conclusions expressed in this material are those of the authors and do not necessarily reflect the views of these sponsors.
G.~Reeves is a member of the Department of Electrical and Computer Engineering and the Department of Statistical Science, Duke University (email: galen.reeves@duke.edu).
H.~D.~Pfister is a member of the Department of Electrical and Computer Engineering and the Department of Mathematics, Duke University (email: henry.pfister@duke.edu).
}
}
\maketitle

\begin{abstract}
This paper considers the performance of Reed--Muller (RM) codes transmitted over binary memoryless symmetric (BMS) channels under bitwise maximum-a-posteriori (bit-MAP) decoding.
Its main result is that, for a fixed BMS channel, the family of binary RM codes can achieve a vanishing bit-error probability at rates approaching the channel capacity.
This partially resolves a long-standing open problem that connects information theory and error-correcting codes.
In contrast with the earlier result for the binary erasure channel, the new proof does not rely on hypercontractivity.
Instead, it combines a nesting property of RM codes with new information inequalities relating
the generalized extrinsic information transfer function and the extrinsic minimum mean-squared error.
\end{abstract}

\begin{IEEEkeywords}
binary memoryless channels,
capacity-achieving codes,
GEXIT functions,
MAP decoding,
MMSE,
Reed--Muller codes.
\end{IEEEkeywords}

\section{Introduction}

Reed--Muller (RM) codes have been the subject of considerable research since their introduction by Muller in~\cite{Muller-ire54} and their majority-logic decoding by Reed in~\cite{Reed-ire54}.
Almost 70 years after their discovery, RM codes remain an active area of research in theoretical computer science and coding theory.
In 2007, Costello and Forney described ``the road to channel capacity''~\cite{Costello-proc07} and wrote that:
\begin{quote}
[I]n recent years it has been recognized that ``RM codes are not so bad''.
RM codes are particularly good in terms of performance versus complexity with trellis-based decoding and other soft-decision decoding algorithms [...]

Indeed, with optimum decoding, RM codes may be ``good enough'' to reach the Shannon limit on the AWGN channel.
[...]
It seems likely that the real coding gains of the self-dual RM codes with optimum decoding approach the Shannon limit [...], but to our knowledge this has never been proved.
\end{quote}
We note that their observations preceded the introduction of polar codes~\cite{Arikan-it09} by roughly one year and, after polar codes, there was a significant renaissance in research on RM codes~\cite{Arikan-comlett08,Arikan-itw10}.
This paper considers the performance of long RM codes transmitted over binary memoryless symmetric (BMS) channels under bitwise maximum-a-posteriori (bit-MAP) decoding and proves that their intuition was indeed correct.

For a BMS channel, the output sequence is generated by passing all symbols in the input sequence through independent identically-distributed channels whose noise processes do not depend on the input symbol.
Some examples are the binary erasure channel (BEC), the binary symmetric channel (BSC), and the binary-input additive white Gaussian noise (BIAWGN) channel. 
The primary technical result can be summarized by the following theorem, which follows easily from Theorem~\ref{lem:BER_bound}.
\begin{thm} \label{thm:main_simple}
Consider any BMS channel with capacity $C \in (0,1)$.
For every sequence of RM codes with strictly increasing blocklength and rate converging to $R \in [0,C)$, the bit-error rate (BER) under bit-MAP decoding converges to zero. 
\end{thm}

This essentially settles a rather old question in coding theory and shows that binary Reed--Muller codes can achieve capacity on any BMS channel under bit-MAP decoding!
We note that this conclusion was certainly more expected than its alternative because~\cite{Kudekar-it17} already established this result for the special case of the BEC.
We note that our result falls short of the stronger condition that the block-error probability vanishes.

For a detailed discussion of relevant prior work until 2017,  see~\cite{Kudekar-it17}. 
Since then, there have been a few papers that address this question directly or indirectly~\cite{Sberlo-soda20,Abbe-it20,hkazla-stoc21}.
In~\cite{Sberlo-soda20}, new bounds on RM weight enumerators are introduced and used to prove that low rate and high rate RM codes can correct a large number of erasures and errors.
A very different approach is pursued in~\cite{Abbe-it20} by treating RM codes from a polar coding perspective and showing that almost all of the effective channels polarize.
This approach shows that RM codes must be close to a ``twin code'' that achieves capacity on BMS channels.
Finally, the results of~\cite{hkazla-stoc21} very cleverly combine a number of earlier results (including the BEC result from~\cite{Kudekar-it17}) to establish that the bit (and block) error probabilities of RM codes will vanish on general BMS channels but only for rates bounded away from capacity.
For a good tutorial that covers RM codes and relevant prior work until 2020, we suggest~\cite{Abbe-it20a}.

The proof for the BEC case in~\cite{Kudekar-it17} requires only linearity and doubly-transitive symmetry for the code.
To achieve this, it relies on the sharp threshold property for symmetric boolean functions and the Extrinsic Information Transfer (EXIT) Area Theorem~\cite{Kudekar-it17}.
One new element in this work is that our proof also relies on the RM nesting property which says that longer RM codes can be punctured down to shorter RM codes of the same order.
But, this does not follow directly from the doubly-transitive symmetry of the code. 
Another difference between the new proof and~\cite{Kudekar-it17} is highlighted by the fact that the new proof holds for the BEC but does not make use of hypercontractivity (which seems to be crucial for the boolean function result).
Lastly, the new proof does not extend to all sequences of doubly-transitive codes nor does it imply that the block-error rate converges to 0.
However, we are optimistic that an extension to block-error rate is possible, perhaps using techniques from~\cite{Kudekar-isit16,Hassani-isit18,hkazla-stoc21}.

There has also been significant recent interest in finding low-complexity decoders for RM codes with near-optimal performance~\cite{Santi-isit18,Hashemi-istc18,Ivanov-isit19,Lian-isit20,Thangaraj-isit20,Ye-it20,Kamenev-itw20,Geiselhart-com21,Huang-arxiv21}.
We do not delve into the details but mention only that many of these approaches also exploit the symmetry and nesting properties of RM codes.

\subsection{Primary Contributions and Overview}

The main contribution of this paper is to establish that RM codes can achieve capacity for any BMS channel. %
Our proof uses many ideas developed previously in the context of generalized EXIT (GEXIT) analysis. For example, we focus on a family of BMS channels and use the GEXIT area theorem.
However, there are a number of steps in our proof that appear to be new. Since these steps may be of interest in their own right we summarize them briefly here. 

A major theme in this work is our focus on the impact of extra observations on the estimate of a single codeword bit (e.g., see Lemma~\ref{lem:mmse_to_var}). The precise form of the ``extra observation'' varies from place to place. In some cases it corresponds to a second look at a single position in the codeword and in other cases it corresponds to a second look at a collection of symbol positions. But the underlying idea is the same --- an additional observation cannot make a meaningful difference in the ability to estimate the bit of interest if either of the following conditions is met:
\begin{enumerate}[(i)]
    \item the expected information from the first observation is very small so that a second independent observation is unlikely to tell us much more; or
    \item the expected information from the first observation is nearly maximal and a second independent observation cannot contribute much more.
\end{enumerate}

To fully utilize this observation, a crucial step is harnessing the nesting property of RM codes to provide a strong upper bound on the variance of the conditional mean of a codeword bit given the observation~(e.g., see Section~\ref{sec:two_look_bound}).
In particular, we embed the RM code of interest in a longer RM code with a slightly lower rate and show that the two codes must behave very similarly for almost all channel noise parameters as the block length grows.

To make these arguments precise, one needs to compare the associated GEXIT functions with and without extra observations. While there are numerous functional properties associated with mutual information and entropy in the context of an additional observation, the challenge faced in our setting is that the GEXIT function corresponds to a \emph{difference} in mutual information, and in this setting many of the usual properties no longer hold. 

The technical tools that allow us to overcome this difficulty form a collection of generalized I-MMSE relations, which are introduced in Section~\ref{sec:I-MMSE}.
They allow us to bound the GEXIT function in terms of a quantity, called the extrinsic MMSE, which is easier to analyze. In particular, the extrinsic MMSE satisfies a data processing inequality and has a sub-additivity property, which follows as a natural consequence of the Efron-Stein inequality.

Here is a list of key elements in the proof along with brief descriptions: 
\begin{itemize}
\item Lemma~\ref{lem:rm_overlap} describes the RM nesting property as used in the proof.

\item Lemma~\ref{lem:twolook} derives the two-look formula, which is the foundation for our GEXIT analysis.

\item Lemma~\ref{lem:gexit_expansion} defines a series expansion for the GEXIT formula and Lemma~\ref{lem:gexit_diff_expansion} uses the two-look formula to relate the GEXIT function to the extrinsic MMSE.

\item Lemma~\ref{lem:integral_decay} derives an integral constraint on the extrinsic MMSE function of RM codes that shows it must transition quickly from 0 to 1 as the blocklength increases.

\item Lemma~\ref{lem:Mt_area_thm} uses the GEXIT area theorem to compare the transition point of the extrinsic MMSE to the capacity of the BMS channel. 

\item Theorem~\ref{lem:BER_bound} proves the main result by deriving a non-asymptotic upper bound on BER of an RM code on any BMS channel. 
\end{itemize}

For readers who are familiar with~\cite{Kudekar-it17} and boolean functions, 
Appendix~\ref{sec:bec_comparison} discusses the method introduced in this paper, for the special case of the BEC, and compares it with the approach in~\cite{Kudekar-it17}.

\subsection{Other Consequences}

Our work also has some additional consequences when combined with other recent results:
\begin{itemize}
\item Since our main result shows that RM codes achieve capacity on the BSC, it follows that Quantum Reed--Muller (QRM)  codes~\cite{Kumar-isit16} achieve the hashing bound on a modified depolarizing channel where $X$ and $Z$ errors occur independently with the same probability~\cite[p.~568]{Wilde-2013}.

\item Combined with the duality result of Renes for classical-quantum (CQ) channels~\cite{Renes-it18}, our result can also be applied to an RM code on a pure-state CQ channel.
Due to the sequential nature of the implied quantum detection model, however, the established decay rate for the bit-error probability is not immediately sufficient to guarantee a vanishing bit-error rate for all code symbols.
But, for linear codes on a pure-state CQ channel, one can show that the optimal measurements for each bit actually commute~\cite{Brandsen-draft22,Piveteau-arxiv21}.
Thus, the bit-error probability is the same for all bits and does not depend on the decoding order!
Hence, RM codes achieve a vanishing bit-error probability on the pure-state CQ channel for all rates below capacity.

\item  If one can prove the stronger result that the block error probability vanishes up to capacity, then one can apply Renes's quantum duality~\cite[Theorem~3]{Renes-it18} to show that RM codes achieve strong secrecy up to capacity for the pure-state wire-tap channel.

\end{itemize}

\subsection{Open Problems}

This results of this paper naturally suggest the following interesting open problems:
\paragraph{Block Error Rate}
Can the main result be strengthened to show that the block error rate vanishes under the same conditions?

For RM codes on BMS channels, the question of when a vanishing BER implies a vanishing block error rate was addressed some years back in~\cite{Kudekar-isit16}.
While simple arguments work if the BER decays faster than $d_{\min}/N$, their result shows that a much slower rate is actually sufficient.
Unfortunately, the decay rate we achieve in this paper is still too slow.

\emph{Note:} After this paper was accepted for publication, an arXiv preprint~\cite{Abbe-arxiv23} was posted with an argument that RM codes achieve vanishing block error probability for all rates below capacity.
This new work builds on our observation (see Lemmas~\ref{lem:RM_rate} and~\ref{lem:rm_overlap}) that long RM codes can be punctured, in many different ways, down to much shorter RM codes of \emph{roughly the same rate}.
Instead of using two looks to bound a variance (as we do), their proposed decoder first constructs, for each bit, a large set of weakly correlated looks and outputs their majority vote.
Then, as a second stage, the decoder finds the nearest codeword to the first output sequence.
The key technical achievement in~\cite{Abbe-arxiv23} is showing that the BER of the first output sequence decays fast enough so that the second output equals the transmitted codeword with high probability.

\paragraph{Other Codes and Channels}
Can this approach be extended to work for other codes and channels?

For example, there are known codes (e.g., multidimensional product codes and Berman codes~\cite{Natarajan-ncc22}) which have a nesting property that is compatible with the approach in this paper but issues arise because they are not doubly transitive.
Also, there may be families of affine invariant codes with a compatible nesting property (e.g., see~\cite{Ivanov-arxiv21}).

The extension to non-binary RM codes over symmetric non-binary channels is also interesting.
Many of the key properties should generalize but it could be challenging to extend the theory of GEXIT functions.
The class of symmetric binary-input classical-quantum channels also seems feasible but the extension of GEXIT functions would face similar challenges.

\emph{Note:} After this paper was accepted for publication, a paper extending this result to non-binary RM codes over symmetric non-binary channels was accepted for publication~\cite{Reeves-isit23,Reeves-arxiv23}.
The new paper shows that non-binary RM codes achieve capacity on sufficiently
symmetric non-binary channels with respect to symbol error rate.
The new proof also simplifies the approach in this paper in a variety of ways that may be of independent interest.

Finally, it is interesting to consider whether the new techniques in this paper can be applied to study sharp threshold behaviors such as the ``all-or-nothing phenomenon'' arising in other high-dimensional structured inference problems  \cite{reeves:2019b,reeves:2019e,barbier:2020, niles-weed:2020}.

\subsection{Notation}

The real numbers and extended real numbers are denoted by $\mathbb{R}$ and $\overline{\mathbb{R}} \coloneqq \mathbb{R} \cup \{\pm \infty\}$.
The natural numbers are denoted by $\mathbb{N} \coloneqq \{1,2,\ldots\}$ and $\mathbb{N}_0 \coloneqq  \mathbb{N} \cup \{0\}$.
For $N \in \mathbb{N}_0$, a range of natural numbers is denoted by $[N] \coloneqq \{0,1,\ldots,N-1\}$.
Also, $\mathbb{F}_2$ is used to denote the Galois field with 2 elements (i.e., the integers $\{0,1\}$ with addition and multiplication modulo 2).
For a set $\cX$, the $N$-element vector $\vx \in \mathcal{X}^N$ is denoted by boldface and is indexed from 0 so that $\vx=(x_0,,\ldots,x_{N-1})$.
For an $M$-element index set $A=\{a_0,a_1,\ldots,a_{M-1}\}\subseteq [N]$ with $a_0<a_1<\cdots<a_{M-1}$, we define the subvector $x_A = (x_{a_0},x_{a_1},\ldots,x_{a_{M-1}}) \in \cX^M$ without using boldface. A single random variable is denoted by a capital letter (e.g., $X,Y,Z$). Vectors of random variables are denoted by boldface capital letters (e.g., $\vX,\vY,\vZ$). For a bounded random variable $X$, the $L^p$-norm ($p \ge 1$) is denoted by  $\|X\|_p \coloneqq ( \ex{|X|^p})^{1/p}$.

\section{Reed--Muller Codes}

\subsection{Background}

\label{subsec:rm_codes}
A length-$N$ binary code is a set $\mathcal{C}\subseteq \{0,1\}^N$
of length-$N$ binary vectors called codewords.
Such a code allows the transmission of $|\mathcal{C}|$ different messages each of which is associated with a codeword $\vc \in \mathcal{C}$.
A codeword $\vc = (c_0,\ldots,c_{N-1})$ is transmitted using a sequence of channel uses where the $i$-th code symbol,  $c_i$, determines the input for the $i$-th channel use.
The rate of the code $\cC$ is defined to be $\frac{1}{N}\log_2 | \cC|$.

For a length-$N$ binary linear code $\mathcal{C}$ with dimension $K$, the code rate equals $K/N$ and a generator matrix $G \in \mathbb{F}_2^{K\times N}$ defines an encoder $E \colon \mathbb{F}_2^K \to \mathcal{C}$ that maps an information vector $\vu\in \mathbb{F}_2^K$ to a codeword via $\vu \mapsto \vu G$.
The Reed--Muller code RM$(r,m)$ is a binary linear code of length $N=2^{m}$ and rate
\begin{equation} \label{eq:rm_rate}
R(r,m) \coloneqq \frac{1}{2^{m}}\sum_{i=0}^{r}\binom{m}{i}.
\end{equation}
Below, we introduce facts about RM codes as they are needed.
For a thorough discussion,
see~\cite{Macwilliams-1977,Lin-2004}.
\begin{example}
\label{exa:RM13_a} Let $G_{r,m}$ be the generator matrix of RM$(r,m)$. The generator matrix (in the standard RM order) of RM$(1,3)$ is given by
\[
G_{1,3}=\begin{bmatrix}1 & 1 & 1 & 1 & 1 & 1 & 1 & 1\\
0 & 1 & 0 & 1 & 0 & 1 & 0 & 1\\
0 & 0 & 1 & 1 & 0 & 0 & 1 & 1\\
0 & 0 & 0 & 0 & 1 & 1 & 1 & 1
\end{bmatrix}=\begin{bmatrix}G_{1,2} & G_{1,2}\\
0 & G_{0,2}
\end{bmatrix},
\]
where 
\[
G_{1,2}=\begin{bmatrix}1 & 1 & 1 & 1\\
0 & 1 & 0 & 1\\
0 & 0 & 1 & 1
\end{bmatrix} \text{ and} \;\; G_{0,2}=\begin{bmatrix}1 & 1 & 1 & 1
\end{bmatrix}.
\]
\end{example}

RM codes can be described in many different ways.
One way is via the one-to-one correspondence between the set of codewords in RM$(r,m)$ and the set of $\mathbb{F}_2$-multilinear polynomials in $m$ indeterminates whose total degree is at most $r$.
For this correspondence, the mapping from a polynomial $p$ to a codeword $c$ is given by evaluating $p$ at all points $\vv\in \mathbb{F}_2^m$.
In particular, the $i$-th code symbol is given by $c_i = p(\vv)$ where $\vv=(v_0,\ldots,v_{m-1})\in \mathbb{F}_2^m$ is the binary expansion of $i = \tau(\vv)$ for
\[ \tau(\vv) \coloneqq \sum_{j\in[m]}v_j 2^j. \]

For $0\leq r \leq m$, let $\mathcal{P}_{r,m}$ be the vector space  (over $\mathbb{F}_2$) of multilinear polynomials in $m$ indeterminates with degree at most $r$. 
This vector space is spanned by the subset of multilinear monomials
\[ \mathcal{M}_{r,m} \coloneqq \left\{ \prod_{j\in S} v_j \, \middle| S \in \mathcal{S}_{r,m} \right\}, \]
where $\mathcal{S}_{r,m} \coloneqq \{S\subseteq [m]\,|\,|S| \leq r\}$.
Each polynomial in $\mathcal{P}_{r,m}$ is defined by a set of coefficients  $\{\alpha_S \in \mathbb{F}_2 \}_{S\in \mathcal{S}_{r,m}}$ with respect to the monomial basis and its evaluation is given by
\begin{equation}\label{eq:msg_poly_eval}
\vv \mapsto \sum_{S\in \mathcal{S}_{r,m}} \alpha_S \prod_{j\in S} v_j.
\end{equation}
This viewpoint can be unified with the generator matrix perspective by noting that, for $S\in \mathcal{S}_{r,m}$, the coefficient $\alpha_S$ can be seen as an information bit that modulates the row in the generator matrix associated with the monomial $S$.
In particular, that row can be computed by evaluating the monomial $S$ at all points in $\mathbb{F}_2^m$.  
To generate the codebook, one first enumerates all information vectors $\vu\in \mathbb{F}_2^K$ (or equivalently all polynomials in $\mathcal{P}_{r,m}$)
and then multiplies each by $G$
(or equivalently each polynomial is evaluated at the $2^{m}$ points in $\mathbb{F}_2^m$).

\begin{example} \label{exa:RM13_b}
Continuing Example~\ref{exa:RM13_a}, we observe that the only degree-0 monomial is $1$.
Thus, the first and only row of $G_{0,2}$ can be computed by evaluating $p(\vv)=1$ for all $\vv\in \{0,1\}^2$.
This also explains the first  row of $G_{1,2}$.
The second and third rows of $G_{1,2}$ are associated with evaluating $p(\vv)=v_0$  and $p(\vv)=v_1$ for all $\vv\in \{0,1\}^2$ in the order given by $\tau(\vv)$.
Likewise, for $G_{1,3}$, the rows are associated with evaluating the monomials $\{1,v_0,v_1,v_2\}$, respectively, for all $\vv\in \{0,1\}^3$ with the order given by $\tau(\vv)$.
\end{example}

RM codes have many algebraic and combinatorial properties.
One of these is a nesting property that will play a particularly important role in this work.
To describe this property, we will consider a few different ways that $\mathcal{C}=\mathrm{RM}(r,m)$ can be punctured down to the code RM$(r,m-k)$.

\begin{defn}[Punctured Code]
For a length-$N$ binary code $\mathcal{C}$ and a subset $I\subseteq[N]$, we denote by $\cC_I$ the punctured code formed by only keeping symbol indices with positions in $I$.
Formally, we write
\[ \mathcal{C}_{I} \coloneqq \{\vc' \in \{0,1\}^{|I|} \,:\, \exists \vc\in \mathcal{C}, c_I = \vc'\}. \]
\end{defn}

\begin{rem}
One can imagine a puncturing operation that also includes the reordering of code bits.
But, this is not needed for our results.
So, we restrict our attention to the case where the bits whose indices are not in $I$ are punctured and the remaining bits are renumbered but kept in the same order.
\end{rem}

The code $\mathcal{C}=\mathrm{RM}(r,m)$ can be punctured down to RM$(r,m-k)$ in multiple different ways.
For example, there exist $I,I'\subset[2^{m}]$ such that $\mathcal{C}_{I}$ and $\mathcal{C}_{I'}$ are both equal to RM$(r,m-k)$.
We emphasize that we mean equality here (rather than equivalence) and this does depend on the ordering of the code bits.
Fortunately, in the construction below, the correct bit order is given by enumerating $I$ (and $I'$) in increasing order and this agrees with our definition of $\cC_I$.
We will see below that this statement follows naturally from two well-known properties of RM codes.
The first property is encapsulated in the following lemma.

\begin{lem}[RM Puncturing] \label{lem:rm_punct}
If one punctures the code $\cC = \mathrm{RM}(r,m)$ by only keeping symbol positions with indices in the set $I = \tau(V)$, where $V=\left\{ 0,1\right\} ^{m-k}\times\left\{ 0\right\} ^{k}$, then $\cC_I = \mathrm{RM}(r,m-k)$.
Moreover, puncturing a uniform random codeword from $\cC$ results in a uniform random codeword from $\cC_I$.
\end{lem}
\begin{proof}
To see this, we can split the monomials into two groups.
Let the first set of monomials be the subset of $\cM_{r,m}$ that only contains the variables $v_0,\ldots,v_{m-k-1}$ and observe that this equals $\cM_{r,m-k}$.
That means the second set, which  contains all the rest, is given by $\cM' = \cM_{r,m} \setminus \cM_{r,m-k}$.
The key observation is that the monomials in $\cM'$ all evaluate to 0 on the set $V$ because all points in $V$ have $v_{m-k}=\cdots=v_{m-1}=0$.
Thus, for $\vc\in \cC$ and $i\in I$, only the monomials in $\cM_{r,m-k}$ contribute to the value of $c_i$.
This implies that the codewords in $\cC_I$ are formed by evaluating the set of $\mathbb{F}_2$-multilinear polynomials in $m-k$ indeterminates whose total degree is at most $r$ at the points in $V$.
Moreover, this notation orders the vector $c_I$ so that the evaluation at $\vv\in V$ appears before the evaluation at $\vv' \in V$ iff $\tau(\vv) < \tau(\vv')$. 
This is the natural binary ordering on $(v_0,\ldots,v_{m-k-1})$ and, hence, $\cC_I$ is precisely equal to $\mathrm{RM}(r,m-k)$.
Another important point is that exactly $2^{|\cM'|}=|\cC|/|\cC_I|$ codewords in $\cC$ are mapped to each codeword in $\cC_I$.
This holds because, if the information bits associated with $\cM_{r,m-k}$ are fixed, then the punctured codeword $c_I$ is fixed.
But, by choosing the information bits associated with the monomials in $\cM'$, one can generate $2^{|\cM'|}$ different codewords in $\cC$ that have the same $c_I$.
\end{proof}

The second property is that, for an invertible binary matrix $Q\in\mathbb{F}_{2}^{m\times m}$ and a vector $\vb\in\mathbb{F}_{2}^{m}$, the degree of a polynomial is preserved by the affine change of variables $\vv \mapsto \pi_{Q,\vb}(\vv)$ where  $\pi_{Q,\vb}\colon\left\{ 0,1\right\} ^{m}\to\left\{ 0,1\right\} ^{m}$ is defined by
\[
[\pi_{Q,\vb} (\vv)]_{i}
= \sum_{j=1}^{m}Q_{i,j}v_{j}+b_{i}.
\]
Thus, the set of all multilinear polynomials with degree at most $r$ is mapped to itself by this change of variables and the permutation $\tau(\pi_{Q,\vb}(\tau^{-1}(i)))$ defines an automorphism of the RM code in terms of symbol indices~\cite[p.~398]{Macwilliams-1977}.

Combining these two properties, one finds that, if an evaluation subset $V\subseteq\left\{ 0,1\right\} ^{m}$ is an $\mathbb{F}_{2}$-subspace with dimension $m-k$, then there is an invertible binary matrix $Q$ (and hence a linear automorphism $\pi_{Q,\vnot}$) that maps $V$ to $\left\{ 0,1\right\} ^{m-k}\times\left\{ 0\right\} ^{k}$.Thus, each $m-k$ dimensional subspace $V \subseteq \left\{ 0,1\right\} ^{m}$ defines an ordered subset of indices that reveals an RM$(r,m-k)$ code inside an RM($r,m$) code. Moreover, each smaller code contains the code symbol $c_0$ from the larger code because all of these automorphisms map $\vnot$ to $\vnot$.
This operation can be seen as a puncturing, according to our definition, if the ordered subset of indices is in increasing order.

\begin{figure}
\centering
\ifextfig
\scalebox{0.9}{\includegraphics{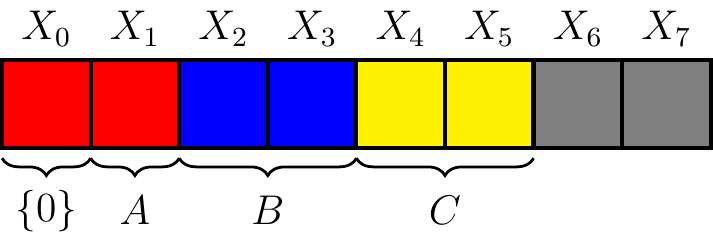}}
\else
\scalebox{0.9}{%
\begin{tikzpicture}[xscale=0.9,yscale=-0.9]

\draw[very thick,black,fill=red] (0,0) rectangle ++(1,1);
\draw[very thick,black,fill=red]  (1,0) rectangle ++(1,1);

\foreach \x in {6,7} {
    \draw[very thick,black,fill=gray]  (\x,0) rectangle ++(1,1);
}

\foreach \x in {2,3} {
    \draw[very thick,black,fill=blue]  (\x,0) rectangle ++(1,1);
}

\foreach \x in {4,5} {
    \draw[very thick,black,fill=yellow]  (\x,0) rectangle ++(1,1);
} 

\foreach \x in {0,1,...,7} {
    \node at (\x+0.5,-0.35) {\large $X_{\x}$};
}

\draw [
    thick,
    decoration={
        brace,
        mirror,
		amplitude=5pt,
        raise=1cm
    },
    decorate
] (0,0) -- (1,0)
node [pos=0.5,anchor=north,yshift=-1.2cm] {\large $\{0\}$}; 

\draw [
    thick,
    decoration={
        brace,
        mirror,
		amplitude=5pt,
        raise=1cm
    },
    decorate
] (1,0) -- (2,0)
node [pos=0.5,anchor=north,yshift=-1.25cm] {\large $A$}; 

\draw [
    thick,
    decoration={
        brace,
        mirror,
		amplitude=5pt,
        raise=1cm
    },
    decorate
] (2,0) -- (4,0)
node [pos=0.5,anchor=north,yshift=-1.25cm] {\large $B$}; 

\draw [
    thick,
    decoration={
        brace,
        mirror,
		amplitude=5pt,
        raise=1cm
    },
    decorate
] (4,0) -- (6,0)
node [pos=0.5,anchor=north,yshift=-1.25cm] {\large $C$}; 
\end{tikzpicture}}
\fi

\caption{Diagram showing two copies of RM$(1,2)$ inside RM$(1,3)$ based on Example~\ref{exa:RM13_c}. The first copy of RM$(1,2)$ is supported on the red and blue indices $I=\{0\} \cup  A\cup B =\{0,1,2,3\}$ and the second copy is supported on the red and yellow indices $I'= \{0\}  \cup A\cup C = \{0,1,4,5\}$. \label{fig:rm12_in_rm13}
}
\end{figure}
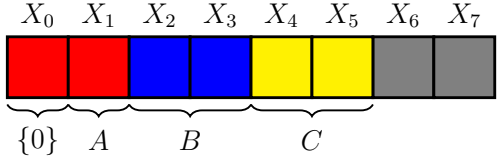

\begin{example} \label{exa:RM13_c}
Continuing the example, we observe that the generator matrix decomposition in Example~\ref{exa:RM13_a} implies that, for all $\vc\in \text{RM}(1,3)$, we have $(c_0,c_1,c_2,c_3) \in \text{RM}(1,2)$. This also follows from evaluating all degree at most $1$ polynomials in $3$ indeterminates on the set $V=\mathrm{span}\left\{ (1,0,0),(0,1,0)\right\}$.
This gives the first 4 symbols of all codewords in RM$(1,3)$ because $I=\tau(V)=\{0,1,2,3\}$.
We can also extract columns $0,1,4,5$ from $G_{1,3}$ to get the submatrix
\[
G'=\begin{bmatrix}1 & 1 & 1 & 1\\
0 & 1 & 0 & 1\\
0 & 0 & 0 & 0\\
0 & 0 & 1 & 1
\end{bmatrix}.
\]
Since the row space of $G'$ equals the row space of $G_{1,2}$, we observe that any codeword of RM$(1,3)$ also contains a codeword of RM$(1,2)$ in these bit positions.
Alternatively, we could evaluate the above set of polynomials on the set  $V'=\mathrm{span}\left\{ (1,0,0),(0,0,1)\right\}$.
This calculation gives the code symbols indexed by $0,1,4,5$ for all codewords of RM$(1,3)$ because $I'=\tau(V')=\{0,1,4,5\}$.
Due to symmetry, however, we will obtain the same set of codewords as we obtained by evaluating on the set $V$.
Thus, we see again that for all $\vc\in\text{RM}(1,3)$ we have $(c_0,c_1,c_4,c_5)\in \text{RM}(1,2)$.
Figure~\ref{fig:rm12_in_rm13} illustrates this example using the notation defined in Lemma~\ref{lem:rm_overlap} (which outlines a general version of this construction).
\end{example}

For an RM$(r,m)$ code, the number of information symbols is equal to
\[K = |\mathcal{S}_{r,m}| = \sum_{i=0}^r \binom{m}{i} \] because~\eqref{eq:msg_poly_eval} implies that we can assign one information bit to each $\alpha_S$ for $S\in \mathcal{S}_{r,m}$.
This information symbol determines whether or not the monomial defined by $S$ is present in the associated polynomial.
This justifies the rate formula in~\eqref{eq:rm_rate}.
Notice that the rate formula equals the cumulative distribution function (cdf) of a binomial random variable, with $m$ equiprobable trials, evaluated at $r$. 
For large $m$, the central limit theorem implies that $R(r,m)$ transitions from roughly 0.025 to roughly 0.975 as $r$ ranges from $\lfloor m/2-\sqrt{m} \rfloor$ to $\lceil m/2+\sqrt{m}\, \rceil$ because the standard deviation of the binomial is $\sqrt{m}/2$ and, for a Gaussian, roughly 95\% percent of the probability lies within 2 standard deviations of the mean. It can also be useful to consider sequences of RM codes where the $n$-th code is RM$(r_{n},m_{n})$ with $m_{n} \to \infty$ and $r_n = m_n/2 +  \alpha \sqrt{m_n} /2 + o(\sqrt{m_n}) $.
In particular, for such sequences, the central limit theorem implies that $R(r_n,m_n) \to \Phi(\alpha)$, where $\Phi(\alpha) = (2\pi)^{-1/2} \int_{-\infty}^\alpha \exp( - z^2/2) \, dz$ is the cdf of a standard Gaussian random variable. 

The above rate calculation appeared earlier in~\cite[Remark~24]{Kudekar-it17} and we mention it here for completeness.
There, it is observed that, for any code rate $R\in(0,1)$, the rate calculation implies one can construct a sequence of RM codes with increasing $m$ whose code rate converges to $R$.

\subsection{New Observations}
\label{sec:new_obs}

The following lemma characterizes the change in code rate due to perturbations of the $m$ parameter.

\begin{lem}[RM Rate Change] \label{lem:RM_rate}
For the codes $\mathrm{RM}(r,m)$ and $\mathrm{RM}(r,m+k)$ with $k \geq 1$, we have
\begin{align}
R(r,m) - R(r,m+k) &\leq \frac{3k+4}{5\sqrt{m}}.
\end{align}
\end{lem}
\begin{proof}
Recall that $R(r,m)$ is equal to the cdf, evaluated at $r$, of the sum of $m$ independent symmetric Bernoulli random variables. Sharp bounds on the normal approximation for the symmetric binomial distribution show that $|R(r,m) - \Phi(\alpha(r,m))| \le 1/\sqrt{2 \pi  m}$ 
where $\alpha(r,m) \coloneqq  (2 r-m)/\sqrt{m}$ and $\Phi(z) \coloneqq (2\pi)^{-1/2} \int_{-\infty}^z \exp( - u^2/2) \, du$ is the cdf of the standard Gaussian distribution~\cite[Corollary~1.2]{Hipp-siamtpa08}.  Using two applications of this bound, we obtain
\begin{align}
R(r,m) &- R(r,m+k) \\
&  \le  \Phi(\alpha(r,m)) - \Phi(\alpha(r,m+k)) \\ & +  \frac{1}{\sqrt{ 2 \pi m}}   +   \frac{1}{\sqrt{ 2 \pi (m+k)}}.
\end{align}
To bound the difference between the Gaussian cdfs, we can write
\begin{align}
 &\Phi(\alpha(r,m)) - \Phi(\alpha(r,m+k)) \\ & \quad =  
\int_{m+k}^{m} \frac{ d \Phi(\alpha(r,x))}{d x} \, dx
\\ & \quad =    \int_{m}^{m+k} \frac{x+2r}{2x^{3/2}} \Phi'( \alpha(r,x)) \, dx \\
& \quad \le \frac{1}{\sqrt{2 \pi}}  \int_{m}^{m+k} \frac{x+2r}{2x^{3/2}}  \, dx
\le  \frac{1}{\sqrt{2 \pi}} \cdot  \frac{ k ( m  +2r)}{2  m^{3/2}},
\end{align}
where we use $\Phi'(z) \le 1/\sqrt{2 \pi}$ in the first inequality  and the fact that the integrand is non-increasing in the second inequality.  Noting that $r \le m$, we can simplify to get the bound
\begin{align}
 R(r,m) - R(r,m+k)
 &\le  \frac{3k}{2\sqrt{2\pi m}} +\frac{2}{\sqrt{2\pi m}}  \\ &= \frac{3 k + 4}{\sqrt{8 \pi m }} < \frac{3k+4}{5\sqrt{m}},
\end{align}
where the last step follows from $\sqrt{8\pi} > 5$.
\end{proof}

The above observations have a surprising consequence that, as far as we know, has not been exploited previously.
Notice that, if the code sequence $\mathrm{RM}(r_{n},m_{n})$ satisfies $R(r_n,m_n) \to R$ for $R \in (0,1)$,
then the rate of the code sequence $\mathrm{RM}(r_{n},m_{n}+k_{n})$ also converges to $R$ for $k_n = o(\sqrt{m_n})$.
But, $\mathrm{RM}(r_n,m_n)$ can be formed from $\mathrm{RM}(r_{n},m_{n}+k_{n})$ by puncturing all but the first $2^{m_n}$ symbols.
Thus, we have a code sequence whose rate converges to $R$ where throwing away a fraction $1-2^{-k_n}$ of the symbols gives another code sequence whose rate converges to $R$.
This is quite surprising because one might expect that puncturing a significant fraction of the bits should increase the code rate by a significant amount.%

\begin{figure*}
\centering
\ifextfig
\scalebox{0.9}{\includegraphics{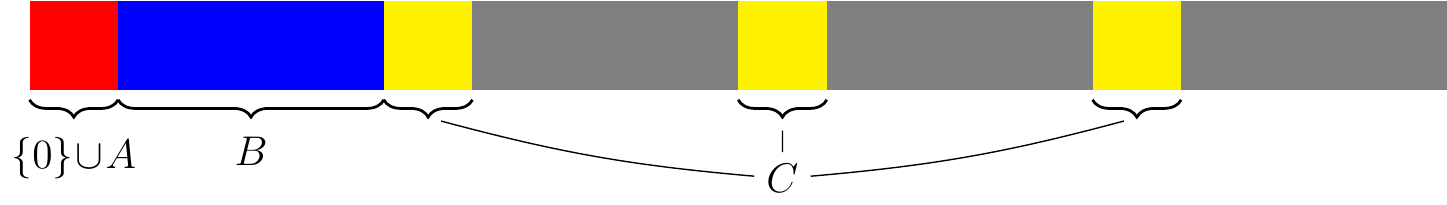}}
\else
\scalebox{0.9}{%
\begin{tikzpicture}[xscale=0.9,yscale=-0.9]

\fill[red] (0,0) rectangle ++(1,1);

\foreach \x in {1,2,...,3} {
  \foreach \j in {1,2,...,3} {
    \fill[gray] (4*\x+\j,0) rectangle ++(1,1);
  }
}

\foreach \x in {1,2,...,3} {
    \fill[yellow] (4*\x,0) rectangle ++(1,1);
\draw [
    thick,
    decoration={
        brace,
		mirror,
		amplitude=5pt,
        raise=1cm
    },
    decorate
] (4*\x,0) -- (4*\x+1,0)
node  [pos=0.5,anchor=north,yshift=-1.05cm] (c\x) {}; 
}

\node (c) at (8.5,2) {\large $C$};
\draw (c1) edge [bend left=5] (c);
\draw (c2) -- (c);
\draw (c3) edge [bend right=5] (c);

\foreach \x in {1,2,...,3} {
    \fill[blue] (\x,0) rectangle ++(1,1);
}

\draw [
    thick,
    decoration={
        brace,
        mirror,
		amplitude=5pt,
        raise=1cm
    },
    decorate
] (1,0) -- (4,0)
node [pos=0.5,anchor=north,yshift=-1.25cm] {\large $B$}; 

\draw [
    thick,
    decoration={
        brace,
		mirror,
		amplitude=5pt,
        raise=1cm
    },
    decorate
] (0,0) -- (1,0)
node [pos=0.5,anchor=north,yshift=-1.25cm] {\large $\{0\} \!\cup\! A$}; 
\end{tikzpicture}}
\fi

\caption{Diagram highlighting two copies of RM$(r,m)$ inside RM$(r,m+k)$ for $k=2$ as outlined in Lemma~\ref{lem:rm_overlap}. The first copy is supported on the red and blue indices $I=\{0\} \cup A\cup B$ and the second copy is supported on the red and yellow indices $I'=\{0\} \cup A\cup C$.
The condition $k=2$ is indicated by the fact that $|B|=(2^k - 1)|\{0\} \cup A|$.
There is also a copy of RM$(r,m-k)$ supported on $I \cap I' = \{0\} \cup A  $. \label{fig:rm_in_rm_k2}
}
\end{figure*}
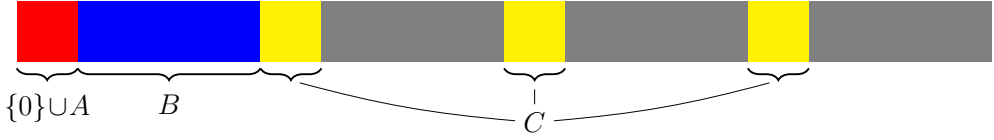

\begin{lem}
\label{lem:rm_overlap} For an $\text{\ensuremath{\mathrm{RM}}}(r,m+k)$ code $\mathcal{C}$ with $r\leq m$ and $1 \leq k\leq m$, there are multiple distinct puncturing patterns that result in an RM$(r,m)$ code.
In particular, there are subsets $A,B,C \subset [2^{m+k}]$
that define puncturing patterns $I = \{0\} \cup A \cup B = [2^m]$ and $I'= \{0\} \cup A \cup C$ such that $\cC_I = \cC_{I'} = \mathrm{RM}(r,m)$.
In addition, $|I \cap I'| = 2^{m-k}$, $\cC_{I \cap I'} = \mathrm{RM}(r,m-k)$, and a uniform distribution on $\mathrm{RM}(r,m+k)$ induces a uniform distribution on the punctured codes $\cC_I$ and $\cC_{I'}$.
This construction is shown in Figure~\ref{fig:rm_in_rm_k2}.
\end{lem}

\begin{proof}
Let $V=\mathbb{F}_{2}^{m}\times \{0\}^k$ (i.e., all length-$(m+k)$ binary vectors where the last $k$ entries are zero) be the subspace of $\mathbb{F}_{2}^{m+k}$ spanned by the first $m$ canonical basis vectors and let $I = \tau(V) = [2^m]$ be the associated set of codeword indices.
This implies that $\mathcal{C}_{I}$ is given by evaluating the set of degree at most $r$ polynomials in $m+k$ variables on the subset $V$.
Using Lemma~\ref{lem:rm_punct}, we see that $\mathcal{C}_{I}$ is equal to RM$(r,m)$.

Similarly, let $V'=\mathbb{F}_{2}^{m-k} \times \{0\}^{k}  \times \mathbb{F}_2^{k}$ (i.e., all length-$(m+k)$ binary vectors satisfying $v_{m-k} = v_{m-k+1} = \cdots = v_{m-1} = 0$) be the subspace of $\mathbb{F}_{2}^{m+k}$ spanned by the first $m-k$ and last $k$ canonical basis vectors.
For the associated set of codeword indices, $I' = \tau(V')$, this implies that $\mathcal{C}_{I'}$ is given by evaluating the set of degree at most $r$ polynomials in $m+k$ variables on the subset $V'$.
Since $V'$ is a subspace of $\mathbb{F}_2^{m+k}$, the automorphism argument earlier in the previous section implies that $\mathcal{C}_{I'}$ is equivalent to RM$(r,m)$ (i.e., equal up to bit order).
The code $\mathcal{C}_{I'}$ will equal an RM$(r,m)$ code if the implied order of the evaluation points equals the natural binary ordering on $(v_0,\ldots,v_{m-k-1},v_m,\ldots,v_{m+k-1}) \in \mathbb{F}_2^m$.
Fortunately, the natural binary ordering for $V'$ is mapped to an increasing sequence of integer indices in $I'=\tau(V')$ and this is the ordering used by our definition of the punctured code.  Thus, $\mathcal{C}_{I'}$ equals RM$(r,m)$. 
We note that the condition $k\geq 1$ is necessary to avoid the $k=0$ case where $V = V'$.

Since $V \cap V' = \mathbb{F}_{2}^{m-k} \times \{0\}^{2k}$ is a subspace with $2^{m-k}$ points, it follows that $T \coloneqq I\cap I' = \tau(V\cap V') = [2^{m-k}]$.
Similar to the argument above, this shows that $\mathcal{C}_{T}$ is equal to RM$(r,m-k)$.
Then, we can define the set $A=T\setminus\{0\}$ to be the non-zero overlap, the set $B=I \setminus T$ to be the indices needed to complete $I$, and the set $C=I' \setminus T$ to be the indices needed to complete $I'$.
Finally, as noted in Lemma~\ref{lem:rm_punct}, a uniform distribution on $\cC$ generates a uniform distribution on $\cC_I$ (and hence $\cC_{I'})$. 
\end{proof}

The following example hints at how this property can be exploited to analyze RM codes.
\begin{example}
\label{exa:RM13_two_est} Continuing Example~\ref{exa:RM13_c}, consider the case where a random codeword $X=(X_0,\ldots,X_7) \in \mathrm{RM}(1,3)$ is transmitted over a BEC and received as $Y=(Y_0,\ldots,Y_7)\in \{0,1,?\}^7$. Then, one can estimate $X_{0}$ from $Y_{1},Y_{2},Y_{3}$ using only the fact that $(X_0,X_1,X_2,X_3)\in \text{RM}(1,2)$. One can also estimate $X_{0}$ from $Y_{1},Y_{4},Y_{5}$ using only the fact that $(X_0,X_1,X_4,X_5)\in \text{RM}(1,2)$.
In this case, $X_{0}$ will be recovered if either of these estimates is not an erasure. Moreover, the performance given by combining the two estimates is strictly better than that given by either single estimate unless the two estimates are perfectly correlated.
See Figure~\ref{fig:rm12_in_rm13} for a graphical representation of this construction.
\end{example}

Since there are multiple copies of RM$(r,m)$ embedded inside of RM$(r,m+k)$ with the same bit 0,
one can utilize two of them separately to compute estimates of bit 0 based on the RM$(r,m)$ code structure. 
These \emph{two looks} can be combined to get a better estimate of bit 0.
Unless they are perfectly correlated, they will actually provide a strict improvement over one estimate.
The interplay between the rate difference, $R(r,m)-R(r,m+k)$, and the two-look phenomenon plays a key role in this work.

\paragraph{Discussion of the code rate after puncturing.}

The above results show that it is possible to remove (i.e., puncture) half of the code symbols from an RM$(r,m+1)$ code to get an RM$(r,m)$ and this puncturing increases the code rate by less than $2/\sqrt{m}$.
The original code has $2^{2^{m+1} R(r,m+1)}$ codewords and the punctured code has $2^{2^m R(r,m)}$ codewords.
This implies that, on average, roughly
\begin{align*}
    \frac{2^{2^{m+1} R(r,m+1)}}{2^{2^{m} R(r,m)}} &= 2^{2^m (2 R(r,m+1)-R(r,m))} \\ &\geq 2^{2^m (R(r,m+1)-2/\sqrt{m})}
\end{align*}
codewords in the original code must collapse onto a single codeword in the punctured code.

For a length-$N$ binary code $\mathcal{C}$, an information set is a subset $I\subseteq [N]$ such that $|\mathcal{C}_I|=|\mathcal{C}|$ and $\mathcal{C}_I = \{0,1\}^{|I|}$~\cite{Huffman-2003}.
It follows that the number of codewords is unchanged if one punctures a set of symbols whose indices are disjoint from a fixed information set.
Thus, in our example, if the code rate is less than $1/2$ and we puncture a set of $N/2$ code symbols that is disjoint from a fixed information set, then the code rate must increase by a factor of 2.
This is because the number of distinct codewords is unchanged but the code length is reduced by a factor of 2.

One can also ask whether random linear codes have some of the same properties.
For now, we will ignore the fact that such codes have a very small probability of being transitive and focus only on the rate after puncturing.
For the random generator matrix model, applying a fixed puncturing pattern to a random generator matrix, with design rate $R$ and length $N$, simply gives a random generator matrix from the same ensemble with length $N'$ and design rate $R' = R N / N'$.
Thus, the existence of a puncturing pattern that nearly preserves the rate can be related to the concentration of the code rate around its expected value.
For this model with design rate $R'$ and any $\delta > 0$, one can show that the probability that the true rate, for a fixed puncturing pattern, is less than $R' - \delta$ is upper bounded by $2^{h_b(\delta) N -\delta R' N^2}$.
Since there are $\binom{N}{N'} \leq 2^N$ puncturing patterns, the union bound implies that, with high probability as $N$ increases,
there is no puncturing pattern that
results in rate smaller than $R'-\delta$.
In particular, the probability that one exists is upper bounded by $2^{(h_b(\delta)+1) N -\delta R' N^2}$ which goes to zero as $N$ increases.

Still, RM codes are not alone with this property.
One can show that certain sequences of multidimensional product codes also have this property.
There is also a generalization of RM codes, known as Berman codes, that have this property~\cite{Natarajan-ncc22}.
It would be interesting to study whether or not there are other algebraic code constructions with this property.

It is worth noting that a property related to the above discussion was recently discussed in the context of successive-cancellation list decoding for polar-like codes~\cite{Ivanov-arxiv21}.
In that work, it is shown that an affine-invariant code with rate $R$ and length-$2^{m+1}$ can be transformed into a code of length-$2^m$ whose rate $R'$ approaches $R$ for large $m$.
When applied to RM codes, the transformation process they consider is a mapping from RM$(r,m+1)$ to RM$(r-1,m)$.
Thus, some of the ideas in~\cite{Ivanov-arxiv21} might provide an avenue for extending some of our results to affine-invariant codes.

\section{Background}
\subsection{Binary Memoryless Symmetric Channels}

An information channel $W$ is defined by an input alphabet $\cX$, an output alphabet $\cY$,
and a transition probability 
that maps elements of the input alphabet to probability measures on the output alphabet. We follow the convention of representing the transition probability using a density function $w(y|x)$ with respect to a base measure on the output alphabet (e.g., counting measure if output distribution is discrete or Lebesgue measure if $\cY=\mathbb{R}^d$ and the output distribution is continuous).
In this paper, we restrict our attention to the channels satisfying the following.

\begin{defn}[BMS Channel~{\cite[p.~178]{RU-2008}}] \label{def:bms} A channel with binary input alphabet $\cX = \{\pm 1\}$ and output alphabet $\cY = \overline{\mathbb{R}}$ is said to be symmetric if the transition probability satisfies $w(y \mid +1) = w(-y \mid -1)$ for all $y \in \cY$. 
A binary memoryless symmetric (BMS) channel consists of a sequence of channels uses such that:
\begin{itemize}
    \item Each channel use has binary input alphabet $\{\pm 1\}$ and is symmetric. 
    \item Conditional on the input to the $i$-th channel, the output of the $i$-th channel is independent of all of the other channel uses. 
\end{itemize}
\end{defn}

Every channel satisfying Definition~\ref{def:bms} can be expressed as a multiplicative noise channel (though this representation is not unique). Specifically, if the input is a random vector $\vX \in \{\pm 1 \}^N$ then the output $\vY \in \cY^N$ is given by 
\begin{align}
\vY = \vX \odot \vZ  \label{eq:multiplicative_channel}
\end{align}
where $\odot$ denotes the Hadamard (entrywise) product and $\vZ \in \cZ^N$ is an independent random vector whose entries are independent with $Z_i$ drawn according to the distribution of the output in the $i$-th channel given the input is $+1$~\cite[p.~182]{RU-2008}. Minus a few exceptions, we will assume that the $\{Z_i\}$ are identically distributed, and thus each channel is described by the same transition probability. 
 
Examples of BMS channels include the following: \begin{itemize}
    \item Binary erasure channel (BEC) where $Z_i \in \{0,1\}$ transmits the input faithfully if $Z_i=1$ and outputs an erasure if $Z_i = 0$. 
    \item Binary symmetric  channel (BSC) where $Z_i \in \{\pm 1 \}$ transmits the input faithfully if $Z_i = 1$ and flips the input if $Z_i = -1$.
    \item Additive white Gaussian noise (AWGN) channel where $Z_i \sim \normal( 1, \sigma^2_i)$ for some noise power $\sigma_i^2$. 
\end{itemize}
While the definition of BMS channels can be extended to output alphabets beyond $\overline{\mathbb{R}}$ (e.g., which do not satisfy the multiplicative noise decomposition above), it turns out that every BMS channel defined in the more general sense has a real sufficient statistic that satisfies the definition given above.
So there is no loss of generality in restricting our attention to channels satisfying \eqref{eq:multiplicative_channel}. See  Appendix~\ref{sec:bms_general} for details.  

\subsection{MMSE and Bit Error Rate}

For a binary random variable $X\in\{\pm 1\}$ and an observation $Y$ defined on the same probability space, the minimum-mean squared error (MMSE) in estimating $X$ from $Y$ is associated with
\begin{align} 
\mmse (X \mid Y) \coloneqq&  \| X - \E[X \mid Y] \|_2^2 \\ =& 1 - \| \E[ X \mid Y] \|_2^2, \label{eq:mmse_X_given_Y}
\end{align}
where the $L^p$ norm of a random variable is denoted by $\|\cdot\|_p = \ex{|\cdot|^p}^{1/p}$. The second expression is obtained from the first by expanding the square,  using nested conditioning, and noting that $\|X\|^2_2=1$. More generally, for a Markov chain $X - Y - Z$, we have the useful identities
\begin{subequations}
\label{eq:L2diff}
\begin{align}
    &\| \ex{ X \mid Y} - \ex{ X \mid Z}\|_2^2 \\
    &\quad=   \| \ex{ X \mid Y}\|^2 -\| \ex{ X \mid Z}\|_2^2 \label{eq:L2diffa}\\
    &\quad=   \mmse(X \mid Z) - \mmse(X \mid Y), \label{eq:L2diffb}
\end{align}
\end{subequations}
where \eqref{eq:L2diffa} follows from expanding the square and then using the nested expectation $\ex{ \ex{X \mid Y} \mid Z} = \ex{ X \mid Z} $, which implies that $\ex{ \ex{X \mid Z} (\ex{X \mid Y} - \ex{X \mid Z} )} =0$, and \eqref{eq:L2diffb} follows from \eqref{eq:mmse_X_given_Y}\noeqref{eq:L2diffa}\noeqref{eq:L2diffb}

Another performance metric of interest is the bit-error probability of the MAP decision rule $\phi(y) \coloneqq \arg\max_{x \in \{\pm 1\}} \pr{ X = x \mid Y = y}$, which is given by
\begin{align*} 
\BER(X \mid Y) \coloneqq& \pr{ X \ne \phi(Y) } \\
=& \frac{1}{2} \left( 1  -  \|\E[ X \mid Y] \|_1 \right).
\end{align*}
Here, the second expression is a consequence of the relationship between the conditional probability and the conditional mean defined by
\[
\pr{ X = 1 \mid Y = y}  = \frac{1}{2} (1+ \ex{ X \mid Y = y}).
\]
See the proof of Lemma \ref{lem:mmse_ber} for details.

For digital communication systems with error-correcting codes, the bit-error probability (of codeword bits) after decoding is an important performance metric.
The system may be considered reliable if this probability can be made arbitrarily small.
A more stringent requirement is that the block-error probability (i.e., the probability that any bit in the codeword is not correct) can be made arbitrarily small.
In this work, we focus on the bit-error rate.

\begin{lem}
\label{lem:mmse_ber}
For a binary random variable $X\in\{\pm 1\}$ observed as $Y$, the quantities $\mmse(X\mid Y) \in [0,1]$ and $\BER(X\mid Y) \in [0,\frac{1}{2}]$ satisfy
\begin{align}
 \frac{ 1 -  \sqrt{ 1-  \mmse(X \mid Y)} }{2} \le        \BER(X \mid Y) \le \frac{1}{2} \mmse(X \mid Y).\!
\end{align}
Thus, for a sequence of observations of $X$, the BER converges to 0 (respectively $\frac{1}{2}$) if and only if the MMSE converges to 0 (respectively $1$).
\end{lem}
\begin{proof}
See \hyperlink{prf:mmse_ber}{proof} in Section~\ref{sec:proof_background}.  \renewcommand{\qedsymbol}{}
\end{proof}

\subsection{Generalized Extrinsic Information Transfer Functions}
\label{sec:gexit_intro}

The generalized extrinsic information transfer (GEXIT) function \cite{RU-2008,Measson-it09} provides a powerful tool for the analysis of communication problems. This section briefly reviews some of the main ideas used in the analyses of GEXIT functions as well as some related concepts involving I-MMSE relations.  

Rather than focusing on a specific information channel $W$,  the main object of interest is a family of channels $ \{W(t)\}$ indexed by a real-valued parameter $t \in [0,1]$, where each $W(t)$ represents a channel from a common input alphabet $\cX$  (not necessarily binary) to a common output alphabet $\cY$. For concreteness, it is assumed throughout that $t = 0$ is a perfect channel (i.e., the input is determined uniquely by the output) and $t= 1$ is an uninformative channel (i.e., the output is independent of the input). For a given number of channel uses $N$, the problem is described as follows: %
  \begin{itemize}
      \item The input  $\vX  \in \cX^N$ is a random vector with distribution $p_X$. For communication systems, this is typically the uniform distribution over a subset of the input space defined by a code. 
      \item The output $\vY \in \cY^N$ is an observation of $\vX$ through a memoryless channel where each $Y_i$ is an observation of $X_i$ through the channel $W(t_i)$ for some $t_i \in [0,1]$.
  \end{itemize}
In some cases, we use the notation $\vY(t_{0}, \dots, t_{N-1})$ to make the dependence on the channel parameters explicit. Instead, if all channel parameters take the common value $t$, then we use the notation $\vY(t)$. 

Once the input distribution and the family of channels have been specified, the high-level idea is to study how certain quantities, such as the entropy and the bit error rate, depend on the underlying channel parameters. Under the assumptions on the channel family outlined above, the conditional entropy of the input given the output satisfies the boundary conditions $H(\vX \mid \vY(0))  = 0$  (a perfect channel) and $H(\vX \mid \vY(1)) = H(\vX)$ (an uninformative channel). If we also assume that the family of channels depends smoothly on the parameter $t$ (e.g., that the mapping $(t_0, \dots, t_{N-1}) \mapsto  H(\vX \mid \vY(t_0, \dots, t_{N-1}))$ is differentiable on $[0,1]^N$) then we can use the fundamental theorem of calculus and the law of the total derivative to obtain the following decomposition: 
\begin{align}
    &H(\vX)  = \int_0^1 \left\{ \frac{d}{d s}   H(\vX \mid \vY(s)) \right\}_{s = t}  \,  dt  \\
    &\;=  \sum_{i=0}^{N-1} \int_0^1   \underbrace{ \left\{    \frac{\partial}{ \partial s_i} H(\vX \mid \vY(s_0, \dots, s_{N-1}  )) \right\}_{s_0 = \dots = s_{N-1} = t}}_\text{GEXIT function of entry $i$} \,  \!\!\!\!\!\!\!\!\! dt, \label{eq:Hdecomp2}
\end{align}
where in the second line, the partial derivative is taken with respect to parameter in the $i$-th channel.

There are two special cases where the partial derivatives in \eqref{eq:Hdecomp2} can be recognized as measures of uncertainty associated with the $i$-th entry of the input: 
\begin{itemize}
\item \textbf{Erasure:} Consider the family of erasure channels where $\cY = \cX \cup \{?\}$ and the probability of erasure is equal to $t$. In this case,  it is straightforward to show that the $i$-th partial derivative in \eqref{eq:Hdecomp2} is equal to  $H(X_i \mid Y_{\sim i}(t))$ where the subscript $\sim i$ denotes the subvector with $i$-th element omitted. The mapping described by $t \mapsto H(X_i \mid Y_{\sim i}(t))$ is called the extrinsic information transfer (EXIT) function and it has found many uses in the literature~\cite{Ashikhmin-it04,Measson-it08,Kudekar-it11,Kudekar-it17}.

\item \textbf{AWGN:} Consider the family of AWGN channels defined by $\cX = \cY = \mathbb{R}$ where each $X_i$ is observed as $Y_i (t_i) = \sqrt{\snr(t_i)} X_i + V_i$ with $V_i$ equal to i.i.d.\ standard Gaussian noise.
We assume that $\snr(t)$ is a non-increasing function of $t$ with $\snr(t) \to \infty$ as $t \to 0$ and $\snr(1) =0$. In this case, it follows from the  I-MMSE relation \cite{Guo-it05} that the $i$-th  partial derivative in \eqref{eq:Hdecomp2} is equal to $-\frac{1}{2} \snr'(t) \mmse(X_i \mid \vY(t))$ where $\snr'(t)$ is the derivative of $\snr(t)$ and $\mmse(X_i \mid \vY(t))$ %
is the minimum mean-squared error of the $i$-th input. This relationship has played a key role in the analysis of coding problems as well as high-dimensional inference problems involving Gaussian noise~\cite{Bustin-eurasip09,Wu-it11,Deshpande-isit14,Reeves-it19,reeves:2020c}. %
\end{itemize}

Going beyond the BEC and AWGN, the partial derivatives in \eqref{eq:Hdecomp2} associated with a general channel no longer have such a simple interpretation. Nevertheless, many of the ideas developed in the context of the BEC and AWGN cases are still applicable. 
Historically, the idea of GEXIT functions is introduced and developed by M\'{e}asson, Montanari, Richardson, and Urbanke in~\cite{Measson-it09}. %
 
\begin{defn}[GEXIT function] \label{def:gexit}
Let $\vX \in \cX^N$ be  a random vector with distribution $p_X$ and let $\vY(t_0, \dots, t_{N-1}) \in \cY^N$ be an observation of $\vX$ through a memoryless channel where each $Y_i$ is an observation of $X_i$ through the channel $W(t_i)$. 
The GEXIT function $G_i \colon [0,1] \to \mathbb{R}$ for entry $i\in [N]$ is defined to be the partial derivative w.r.t.\ the channel parameter for the $i$-th output:
\begin{align}
    \gexit_i(t) \coloneqq \frac{\partial}{ \partial s_i} H(\vX \mid \vY(s_0, \dots, s_{N-1}  )) \Big \vert_{s_0 = \dots = s_{N-1} = t}.  %
\end{align}
\end{defn}

The full power of GEXIT analysis is realized when the input distribution and the channel satisfy certain symmetry properties that imply the GEXIT functions are all identical, i.e., $\gexit_0 = \dots = \gexit_{N-1}$. In this case, \eqref{eq:Hdecomp2}~implies that, for all $i\in[N]$, we have
\begin{equation} %
\int_0^1 \gexit_i (t) \, dt = \frac{1}{N}H(\vX).
\end{equation}
For the BEC and AWGN channel, this result connects a well-known reliability measure associated with the single element $X_i$ to the entropy of the entire vector $\vX$. A sufficient condition under which the GEXIT functions are identical %
is that the distribution of the input vector has transitive symmetry. %

\begin{defn}[Symmetry and Transitivity] 
Let $S_N$ be the set of permutations (i.e, bijective functions) mapping $[N]$ to itself. 
The symmetry group of a random vector $\vX = (X_0, \dots, X_{N-1})$ is defined to be
\[ \cG \coloneqq \{ \pi \in S_N \,:\,  (X_{\pi(0)},\ldots,X_{\pi(N-1)}) \stackrel{d}{=} \vX \}, \]
where $\stackrel{d}{=}$ indicates equality in distribution. 
We say that $\vX$ has transitive symmetry if $\cG$ is transitive (i.e., for all $i,j \in [N]$, there is a $\pi \in \cG$ such that $\pi (i) = j$).
We say that $\vX$ has a doubly-transitive symmetry if $\cG$ is doubly transitive (i.e., for distinct $i,j,k \in [N]$, there is a $\pi \in \cG$ such that $\pi(i) = i$, and $\pi(j)=k$).
\end{defn}

\section{Preliminary Results} 

\subsection{BMS Families Ordered by Degradation} \label{sec:BMS_families}

Our approach builds upon the GEXIT analysis outlined in Section~\ref{sec:gexit_intro}. Rather than focusing on a particular BMS channel we study a family of BMS indexed by a parameter $t \in [0,1]$ where $t =0$ is a perfect channel and $t = 1$ is an uninformative channel.  We also require that the family to be ordered with respect to degradation in the sense that $W(t)$ is degraded with respect to $W(s)$ for all $0 \le s \le t \le 1$. Equivalently, for any distribution on the input $X\in \{\pm 1\}$ there exists a joint distribution on $(X, Y(s), Y(t))$ such that
\begin{itemize}
\item $Y(s)$ is an observation of $X$ through channel $W(s)$
\item $Y(t)$ is an observation of $X$ through channel $W(t)$
\item $X - Y(s) - Y(t)$ forms a Markov chain.
\end{itemize}
We remark that this degradation assumption is also standard in the literature on GEXIT analysis~\cite{RU-2008}. See Appendix~\ref{sec:degradation} for a precise definition of channel degradation and some of its consequences.

There are a few well-known examples of channel families that are ordered by degradation. Some examples are the family of BECs where the erasure probability transitions from $0$ to $1$, the family of BSCs where the crossover probability transitions from $0$ to $1/2$, and the family of AWGN channels where the noise power transitions from 0 to $+\infty$. 

The Shannon capacity of a BMS channel is equal to the mutual information between the input and output when the input is uniformly distributed~\cite[p.~193]{RU-2008}. For a family of BMS channels, two important metrics are provided by the entropy and the MMSE for a uniform input distribution:

\begin{defn} \label{defn:BS_metrics} Let $\{W(t)\}_{t \in [0,1]}$ be a family of BMS channels that is ordered w.r.t.\ degradation  where $W(0)$ is the perfect channel and $W(1)$ is the uninformative channel. 
Let $X_\mathrm{u}$ be uniformly distributed on $\{ \pm 1\} $ and let  $Y_\mathrm{u}(t)$ is an observation of $X_\mathrm{u}$ through the channel $W(t)$.  Note that, under these assumptions,   $\ex{X_\mathrm{u} \mid Y_\mathrm{u} (0)} = X_\mathrm{u}$ and $\ex{X_\mathrm{u} \mid Y_\mathrm{u} (1)} = 0$. %
The entropy function $\cH \colon [0,1] \to [0,1]$ and MMSE function $\cM \colon [0,1] \to [0,1]$ are defined according to
\begin{align}
  \cH(t) & \coloneqq H(X_\mathrm{u} \mid Y_\mathrm{u}(t) ) \label{eq:cH}\\
  \cM(t) &\coloneqq \mmse(X_\mathrm{u} \mid Y_\mathrm{u}(t)).
\label{eq:cM}  %
\end{align}
The family is said to be absolutely continuous if the entropy function is absolutely continuous,  i.e., if there exists a function $\cH' \colon [0, 1] \to \mathbb{R}$ such that $\cH(b)  - \cH(a) = \int_a^b \cH'(t) \, dt$ for all $0 \leq a \leq b \leq 1$. 
\end{defn}

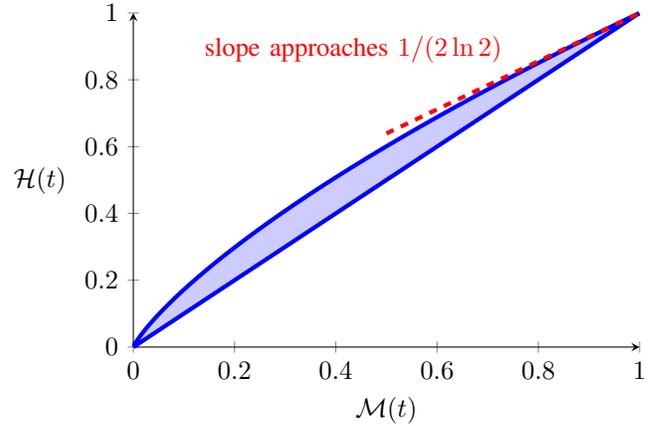
\begin{figure}
\centering
\ifextfig
\includegraphics{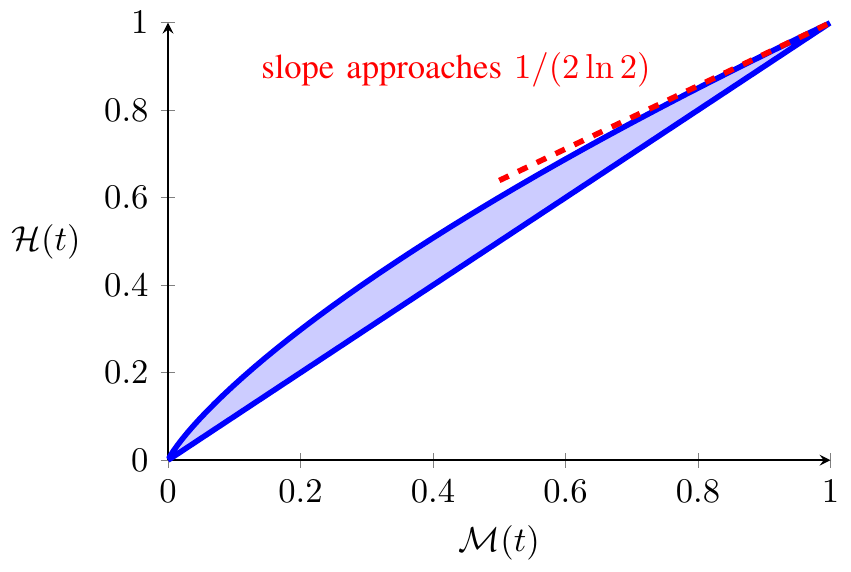}
\else
\begin{tikzpicture}
\begin{axis}[%
width=2.65in,
height=1.75in,
scale only axis,
xmin=0,
xmax=1,
xlabel={$\mathcal{M}(t)$},
ymin=0,
ylabel={$\mathcal{H}(t)$},
ylabel style={rotate = -90},
 axis lines =left,
 clip  = false,
]

\addplot [ultra thick, blue,fill=blue, fill opacity=0.2]
  table[row sep=crcr]{%
0.001	0.00335282294525208\\
0.00399699699699696	0.0114102462746929\\
0.00799299299299294	0.0208343012828235\\
0.013986986986987	0.0336689431996783\\
0.02097997997998	0.047487343411468\\
0.028971971971972	0.0622809851598793\\
0.037962962962963	0.0780129440185297\\
0.047952952952953	0.0946390941694945\\
0.0589419419419419	0.112114708771593\\
0.07092992992993	0.130396781111385\\
0.0839169169169169	0.149444822428295\\
0.0979029029029029	0.169221057515273\\
0.112887887887888	0.189690371584232\\
0.128871871871872	0.210820156654773\\
0.145854854854855	0.232580123095155\\
0.163836836836837	0.254942106038215\\
0.182817817817818	0.27787987993606\\
0.202797797797798	0.301368986710645\\
0.223776776776777	0.325386579198803\\
0.245754754754755	0.349911279796663\\
0.268731731731732	0.374923053392297\\
0.292707707707708	0.400403093350947\\
0.317682682682683	0.426333719239042\\
0.343656656656657	0.452698285014623\\
0.37062962962963	0.479481096507188\\
0.398601601601602	0.506667337125087\\
0.427572572572573	0.534243000846049\\
0.458541541541542	0.563117012540876\\
0.49050950950951	0.592318249154237\\
0.523476476476476	0.621837146925655\\
0.557442442442442	0.651664522093413\\
0.592407407407407	0.681791557186881\\
0.628371371371371	0.712209786601362\\
0.665334334334334	0.742911081886185\\
0.703296296296296	0.773887637047559\\
0.742257257257257	0.805131954074854\\
0.782217217217217	0.836636828832191\\
0.824175175175175	0.869163453242135\\
0.867132132132132	0.901912103180215\\
0.911088088088088	0.934877691879384\\
0.956043043043043	0.968055275086457\\
0.999	0.999278532206853\\
0 0 \\
};
\addplot [ ultra thick, dashed, red] coordinates { (.5,0.6393) (1,1) } node[midway,above left] {slope approaches $1/(2 \ln 2)$};
\end{axis}
\end{tikzpicture}
\fi

    \caption{ \label{fig:joint_range}
    Joint range of entropy and MMSE functions associated with a family of BMS channels as described by \eqref{eq:cH_to_cM}. The upper boundary is attained by the BSC and the lower boundary is attained by the BEC~\cite{Jiang-it08a}. Also, the derivative ratio $\cH'(t)/\cM'(t)$ is bounded from below by $1/(2 \ln 2)$. }
\end{figure}

The entropy and MMSE functions have a number of important functional properties.  Under the assumed degradation ordering, both functions are non-decreasing with $\cH(0) = \cM(0) = 0$ and $\cH(1) = \cM(1) = 1$. For each $t \in[0,1]$, the Shannon capacity of the channel $W(t)$ is equal to $C(t) = 1- \cH(t)$. It is known that the extremal relationships between the entropy and the MMSE are attained by the BEC and BSC channels~\cite{Jiang-it08a}. For example,
\begin{align}
\cM(t) \le \cH(t) \le     h_b\left( \frac{1 - \sqrt{1- \cM(t)}}{2} \right) \label{eq:cH_to_cM}
\end{align}
where $h_b(x) \coloneqq - x \log_2(x) - (1-x) \log_2(1-x)$ is the binary entropy function. Equality on the left is attained by the BEC and equality on the right is attained by the BSC.
This type of phenomenon is also known to be somewhat typical~\cite{Land-it05}.
Another property, perhaps less known, is that the difference in entropy can be used to upper bound the difference in MMSE:
\begin{align}
    \cM(t) - \cM(s) \le  2 \ln(2) \left(  \cH(t) - \cH(s)  \right), \label{eq:cMdiff_cHdiff}
\end{align}
for $0 \le s \le t \le 1$.
To show this, one can apply Lemma~\ref{lem:H_M_q_Hu} to $\cH(t)-\cH(s)$ and then use Lemma~\ref{lem:convex_order} to verify that all terms in the resulting expansion are positive.
Keeping only the first term in the expansion gives the bound in~\eqref{eq:cMdiff_cHdiff}.
In the case of the BSC, it can be verified that
the factor $2 \ln(2)$ is tight in the limit where the crossover probability approaches 1/2 (see Figure~\ref{fig:joint_range}). Also, by applying Lemma~\ref{lem:abs_cont}, this inequality is sufficient for the absolute continuity of the entropy function to imply the absolute continuity of the MMSE function.

\subsection{GEXIT and I-MMSE Properties}
\label{sec:I-MMSE}

In this section, we present a number of useful results that characterize GEXIT functions for binary-input channels. 
GEXIT functions were introduced in~\cite{Measson-arxiv04,RU-2008,Measson-it09} and analyzed further by a variety of authors~\cite{Macris-it07a,Jiang-it08a,Kudekar-it13,Kumar-it14}.
Our treatment is based solely on the moments of the conditional expectation.

For the results in this section, it is  convenient to specify the distribution of a binary random variable $X \in \{ \pm 1\}$  in terms of it mean $\mu  = \ex{ X} \in [-1,1]$, which corresponds to  $\pr{ X = 1} = (1+\mu)/2$ and $\pr{ X = -1} = (1-\mu)/2$. The entropy of such a variable is given by $H(X) = h_b( (1+ \mu)/2)$ where we recall that $h_b(x)\coloneqq -p\log_2(p) - (1- p) \log_2(1-p)$ is the binary entropy function. The entropy is an even function of $\mu$ and it admits the following power series expansion, 
\begin{align}
h_b\left( \frac{ 1 + \mu}{ 2}    \right)  =  \sum_{k=1}^\infty \ce_k (1-\mu^{2k}) , \quad  \ce_{k} \coloneqq \frac{1}{(\ln 2)2k(2k-1)},  \label{eq:hb_series}
\end{align}
which converges uniformly for $\mu \in [-1,1]$~\cite[p.575]{Wiechman-it07}.

This expansion extends naturally to the conditional entropy of $X \in \{\pm 1\}$ given an observation $Y$ defined on the same space. In particular, replacing $\mu$ by the conditional mean $\E[X \mid Y]$ and then taking the expectation of the both sides yields
\begin{align}
    H(X \mid Y) &= \ex{h_b \left( \frac{1+\ex{X \mid Y}}{2} \right)} \\ &= \sum_{k=1}^\infty \ce_k\left(1- \| \ex{ X \mid Y} \|_{2k}^{2k} \right),  \label{eq:HXY_series1}
\end{align}
where the interchange of expectation and summation is justified by the uniform convergence of the power series. 
Notice that the conditional expectation $\ex{ X \mid Y}$ appearing in \eqref{eq:HXY_series1}  depends on both the prior mean $\mu$ as well as the channel from $X$ to $Y$. In the special case of a BMS channel, it turns out that the conditional entropy can also be expressed using a different series expansion involving $\mu$ and a sequence $\{q_k\}$ that depends only on the BMS channel. This sequence and the corresponding expansion are defined as follows.

\begin{defn}\label{defn:qk}
For a BMS channel, the moment sequence $\{q_k\}_{ k \in \mathbb{N}  }$ is defined by
\begin{align}
 q_k \coloneqq \| \E[ X_\mathrm{u} \mid Y_{\mathrm{u}}] \|_{2k}^{2k} = \E\Big[\E[X_\mathrm{u} \mid Y_\mathrm{u}]^{2k}\Big],  
     \label{eq:qk}
\end{align}
where $X_{\mathrm{u}} \in \{\pm 1\}$ is uniformly distributed and $Y_{\mathrm{u}}$ is an observation of $X_{\mathrm{u}}$ through the channel.
We use the subscript $\mathrm{u}$ to emphasize that $q_k$ is always computed using a uniformly distributed input.
\end{defn}

\begin{lem}[Two-look Formula] \label{lem:twolook}
Let $X \in \{\pm 1\}$ be a binary random variable with mean $\mu \in [-1,1]$ and prior probability  $\pr{ X =1}  = 1 -\pr{ X = -1} = (1+\mu)/2$. Let $Y$ be an observation of $X$ through a BMS channel with sequence $\{q_k\}_{k \in \mathbb{N}}$ as given in Definition~\ref{defn:qk}.  Let $U$ be another observation on the same probability space such that $U - X - Y$ forms a Markov chain. Then, the following expansions hold: 
\begin{align} 
H(X\mid Y) & =\sum_{k=1}^{\infty}\ce_k ( 1 - q_k ) \left( 1 - \mu^{2k} \right) \label{eq:two_look_formula_mu}\\
H(X\mid Y,U) & =\sum_{k=1}^{\infty}\ce_k ( 1 - q_k ) \left( 1 - \| \E[ X\mid U]\|_{2k}^{2k} \right). \label{eq:two_look_formula}
\end{align}
\end{lem}
\begin{proof}
See \hyperlink{prf:twolook}{proof} in Section~\ref{sec:proof_preliminary}. \renewcommand{\qedsymbol}{}
\end{proof}

In comparison to \eqref{eq:HXY_series1} the expansion in \eqref{eq:two_look_formula_mu} provides a decoupling between two different types of information:  the information provided by the BMS channel (encapsulated by the sequence $\{q_k\}$) and the prior information (summarized by $\mu$). 

\begin{rem}
The expansions in Lemma~\ref{lem:twolook} provide simple proofs for various properties of BMS channels. For example, by  \eqref{eq:two_look_formula_mu} the mutual information between the input and the output of the BMS channel from $X$ to $Y$ is given by
\begin{align}
    I(X;Y) = H(X) - H(X\mid Y)  = \sum_{k =1}^\infty \ce_k q_k (1- \mu^{2k}).
\end{align}
Since each term is maximized at $\mu =0$, this immediately implies the well-known fact that the capacity of the channel is attained by the uniform distribution.
Thus, the capacity satisfies $C  =\sum_{k=1}^\infty \ce_k q_k$.
\end{rem}

Before stating the technical lemmas, let us first  provide an informal overview of how Lemma~\ref{lem:twolook} will be used to bound the GEXIT function. 
Let $\{W(t) : 0 \le t \le 1\}$ be a family of BMS channels that is ordered by degradation and absolutely continuous according to Definition~\ref{defn:BS_metrics}.
Let $\{q_k(t)\}_{k \in \mathbb{N}}$ denote the sequence from Definition~\ref{defn:qk} as a function of $t$. For an input  $\vX \in \{\pm 1\}^N$, consider the output given by 
\begin{equation}
(Y_{0}(t) , \dots, Y_{i-1}(t) , Y_i(s), Y_{i+1}(t) ,\dots ,Y_{N-1}(t)  ), \label{eq:y_s_t}
\end{equation}  
where the $i$-th channel use has parameter $s$ and the other channel uses have parameter $t$. Applying  \eqref{eq:two_look_formula} with respect to the BMS channel from $X_i$ to $Y_i(s)$ and the Markov chain $Y_{\sim i}(t) - X_i - Y_i(s)$ gives
\begin{align} 
H(&X_i\mid Y_i(s), Y_{\sim i}(t)) \\ & =\sum_{k=1}^{\infty}\ce_k \left( 1 - q_k(s) \right) \left( 1 - \| \E[ X_i \mid Y_{\sim i}(t)  ]\|_{2k}^{2k} \right).  \label{eq:Hst_for_gexit}
\end{align}
Taking the $s$-derivative of both sides, interchanging the derivative and the summation, and then evaluating at $s=t$ gives the following expansion of the GEXIT function: 
\begin{align}
    G_i(t) %
    & =\sum_{k=1}^{\infty}\ce_k \left( - q_k'(t)  \right) \left( 1 - \| \E[ X_i \mid Y_{\sim i}(t) ]\|_{2k}^{2k} \right). \label{eq:GEXIT_expansion}
\end{align}
A useful property of this expansion is that the terms in the sum are all non-negative. This follows from $\| \E[ X_i \mid Y_{\sim i}(t) ]\|_{2k}^{2k} \in [0,1]$ and from the degradation ordering which ensures that each  $ q_k(t)$ is non-increasing in $t$ (see Lemma~\ref{lem:H_M_q_Hu}). This expansion plays a crucial step in Section~\ref{sec:MMSE_bounds} where it provides a link between the integral of the GEXIT function and the integral of  a related function that depends only on  the  conditional second moments. 

A further application of Lemma~\ref{lem:twolook}  appears in 
Sections~\ref{sec:two_look_bound} and \ref{sec:single_term_bound} where it is used to compare the GEXIT function $G_i(t)$ with an augmented GEXIT function $G^+_i(t)$ that depends on the original output $\vY(t)$ as well as an additional observation $U(t)$ such that $U(t) - \vX - \vY(t)$ forms a Markov chain. Applying the expansion in \eqref{eq:GEXIT_expansion} to both $G_i(t)$ and $G_i^+(t)$ and then taking the difference yields 
\begin{align}
    \!\!\! G_i&(t) - G_i^+(t)
    =\sum_{k=1}^{\infty}\ce_k \left( - q_k'(t)  \right) \\ & \; \cdot  \Big(\underbrace{ \| \E[ X_i\mid Y_{\sim i}(t) , U(t) ]\|_{2k}^{2k}  - \| \E[ X_i\mid Y_{\sim i}(t) ]\|_{2k}^{2k} }_{\ge 0} \Big). \label{eq:gexit_diff_expansion}
\end{align}
We again find that each term in the expansion is non-negative. This is due to the degradation ordering of the channel (e.g., $q_k '(t) \leq 0$ by Lemma~\ref{lem:H_M_q_Hu}) and Lemma~\ref{lem:convex_order} via the trivial Markov chain $X_i - (Y_{\sim i}(t),U(t)) - Y_{\sim i} (t)$.  Importantly, this means that keeping only the $k=1$ term provides a lower bound on the \emph{difference} of the GEXIT functions. %

\begin{lem} \label{lem:H_M_q_Hu}
Let $\{W(t) : 0 \le t \le 1\}$ be a family of BMS channels that is ordered by degradation and absolutely continuous according to Definition~\ref{defn:BS_metrics}.
Let $\{q_k(t)\}_{k \in \mathbb{N}}$ denote the sequence from Definition~\ref{defn:qk} as a function of $t$.
Then, the following properties hold: 
\begin{enumerate}[(i)]
    \item The entropy function and MMSE function from Definition~\ref{defn:BS_metrics} can be expressed as 
    \begin{align}
        \cH(t) &= %
        \sum_{k=1}^{\infty}c_{k}\left( 1 -q_{k}(t)  \right) \\ %
        \cM(t) &= 1- q_1(t)
    \end{align}

    \item Each $q_k (t)$ is non-increasing and absolutely continuous on $[0,1]$ with $q_{k}(0) =1$ and $q_{k}(1)=0$.  The derivative, $q_k '(t)$, exists almost everywhere on $[0,1]$ and satisfies $q_k '(t) \leq 0$ when it exists.

    \item The derivative of $\cH(t)$ exists almost everywhere on $[0,1]$ and is equal almost everywhere to
\begin{align} \label{eq:Ht_p}
\cH ' (t) \coloneqq \sum_{k=1}^\infty c_k (-q_k ' (t)).
\end{align}
\end{enumerate}
\end{lem}
\begin{proof}
See \hyperlink{prf:lem:H_M_q_Hu}{proof} in Section~\ref{sec:proof_preliminary}. \renewcommand{\qedsymbol}{}
\end{proof}

\begin{lem}\label{lem:gexit_expansion}
Let $\{W(t) : 0 \le t \le 1\}$ be a family of BMS channels that is ordered by degradation and absolutely continuous according to Definition~\ref{defn:BS_metrics}.
Let $\vX \in \{\pm 1\}^N$ be a binary random vector and let $\vY \in \cY^N$ be an observation of $\vX$ through the BMS channel family.
Let $U(t)$ be another observation of $\vX$, which is conditionally independent of $\vY$ given $\vX$, through a family of channels indexed by $t \in [0,1]$ and ordered by degradation.
For each $i \in [N]$, we use the parameterization in~\eqref{eq:y_s_t} to show that the GEXIT functions,
 \begin{align}
 \gexit_i(t) &\coloneqq \frac{\partial}{\partial s} H\big(\vX \mid Y_i (s), Y_{\sim i} (t) \big) \Big|_{s=t} \\
\gexit^+_i(t) &\coloneqq \frac{\partial}{\partial s} H\big(\vX \mid Y_i (s), Y_{\sim i} (t), U(t) \big) \Big|_{s=t} \, ,
 \end{align}
 exist almost everywhere and are integrable on $[0,1]$.
 These functions also satisfy (almost everywhere on $[0,1]$) the series expansions~\eqref{eq:GEXIT_expansion} and
\begin{align}
    \!\!\!\!\!\!\! G^+_i(t) 
    & =  \sum_{k=1}^{\infty}\ce_k (-q_k'(t))  \left( 1 - \| \E[ X_i \mid Y_{\sim i}(t),  U(t) ]\|_{2k}^{2k} \right). \label{eq:GEXIT_plus_expansion}
\end{align}
\end{lem}
\begin{proof}
See \hyperlink{prf:gexit_expansion}{proof} in Section~\ref{sec:proof_preliminary}. \renewcommand{\qedsymbol}{}
\end{proof}

\begin{lem}\label{lem:gexit_diff_expansion}
Under the setting of Lemma~\ref{lem:gexit_expansion}, we can lower bound~\eqref{eq:gexit_diff_expansion} with
\begin{align}
    &G_i(t) - G_i^+(t) \\
    & \; \ge c_1 (-q_1 '(t))\big(  \mmse(X_i \mid Y_{\sim i}) - \mmse(X_i \mid Y_{\sim i}, U(t) ) \big) .
\end{align}
\end{lem}
\begin{proof}
First, we note that one can rigorously  establish~\eqref{eq:gexit_diff_expansion} by subtracting~\eqref{eq:GEXIT_plus_expansion} from~\eqref{eq:GEXIT_expansion}.
The terms, $\| \E[ X_i\mid Y_{\sim i}(t) , U(t) ]\|_{2k}^{2k}  - \| \E[ X_i\mid Y_{\sim i}(t) ]\|_{2k}^{2k}$, in the resulting expansion are non-negative by the degradation ordering of the channel (e.g., $q_k '(t) \leq 0$ by Lemma~\ref{lem:H_M_q_Hu}) and Lemma~\ref{lem:convex_order} via the trivial Markov chain $X_i - (Y_{\sim i}(t),U(t)) - Y_{\sim i} (t)$.
Thus, the stated result follows from retaining only the first term.
\end{proof}

The next result is a further implication of the two-look formula that bounds one's ability to estimate a binary variable from two observations subject to an MMSE lower bound on one of the observations. 
\begin{lem}
\label{lem:BER_two_look}
Consider the setting of Lemma~\ref{lem:twolook} and let $C$ be the capacity of the BMS channel from $X$ to $Y$. Then
\begin{align}
& \BER(X \mid Y, U) \\  & \; \ge \BER(X \mid Y)   -\sqrt{ \frac{  \ln(2) (1-C)  (1 - \mmse(X \mid U))  }{ 2} }.
\end{align}
\end{lem} 
\begin{proof}
See \hyperlink{prf:BER_two_look}{proof} Section~\ref{sec:proof_preliminary}. \renewcommand{\qedsymbol}{}
\end{proof}

\begin{rem}
In comparison to~\cite{RU-2008,Measson-it09}, our GEXIT formulation is somewhat more general and requires fewer regularity assumptions.
In particular, we allow the channel family to be parameterized arbitrarily and we show that the GEXIT function exists as long as its entropy function $\cH(t)$ is absolutely continuous.
An alternative approach to the analysis of GEXIT functions, which shares some of these properties, can be found in~\cite{Kudekar-it13}.
\end{rem}

\subsection{Linear Codes on BMS Channels}

A set $\mathcal{C} \subseteq \mathbb{F}_{2}^N$ defines a binary linear code (i.e., a subspace of $ \mathbb{F}_{2}^N$) if and only if it is closed under addition, that is to say $\vu \oplus \vu' \in \cC$ for all $\vu,\vu' \in \cC$, where $\oplus$ represents element-wise modulo-2 addition.
To transmit a message over a binary channel, each codeword $\vu \in \cC$ is mapped to a channel input sequence in $\{\pm 1\}^N$ via the binary phase-shift keying (BPSK) mapping $\vu \mapsto (-1)^{\vu}$. The resulting set of BPSK-modulated codeword sequences is denoted by  $\cC_x$.

A remarkable property of linear codes on BMS channels is that many performance metrics do not depend on the transmitted codeword~\cite[p.~190]{RU-2008}.  This property greatly simplifies the analysis of coding problems because it means that one may condition on the event that the all ones input is transmitted (corresponding to the all zeros linear codeword). Note that under this event, the outputs of the BMS channel are independent random variables. 

Next, we describe a consequence of linearity and channel symmetry that is useful for our analysis. 

\begin{lem}\label{lem:bms_linear} 
Let $\vX$ be distributed uniformly on the set of BPSK-modulated code sequences $\cC_x \subseteq \{\pm 1\}^N$ associated with the binary linear code $\cC \subseteq \mathbb{F}_{2}^N$  and  %
let $\vY$ be an observation of $\vX$ through a BMS channel of the form $ \vY = \vX \odot \vZ$ where $\vZ \in \cZ^N$ is an independent vector with independent entries. 
For $i \in [N]$ and $S \subseteq [N]$,  define $f(y_S) \coloneqq \ex{ X_i \mid Y_S = y_S}$ to be the conditional expectation of the $i$-th input given the outputs  indexed by $S$. Then, for all $\vx\in \mathcal{C}_x$ and  $y_S$ in the support of $Y_S$ the following identity holds: 
\begin{align}
 f(y_S) = x_i \, f(x_S \odot y_S).
\end{align}
In particular, this implies that 
\begin{align}
    f(Y_S) = X_i\,  f(Z_S).
\end{align}
\end{lem}
\begin{proof} 
By Bayes rule, the conditional probability mass function of $\vX$ given $Y_S = y_S$ satisfies 
\begin{align}
    p_{\vX \mid Y_S}(\vx \mid y_S) \propto {p_{\vX}} (\vx) \, \prod_{i\in S} w(y_i \mid x_i), \quad \vx \in \{\pm 1\}^N
\end{align}
where $p_{\vX} (\vx)$ is the uniform distribution over the codewords, $w( y_i \mid x_i)$ is the transition probability,  and the constant of proportionality is chosen to ensure the function sums to one.  The fact that a linear code is closed under addition means that it is a subgroup of $\mathbb{F}_2^N$, and thus for any $\vu \in \cC$, a vector $\vu' \in \mathbb{F}_2^N$ satisfies $\vu' \in \cC$ if and only if $\vu \oplus \vu' \in \cC$. Using the code to input mapping $\vx=(-1)^{\vu}$, this implies that for any $\vx \in \cC_x$, a vector $\vx' \in \{\pm 1\}^N$ satisfies $\vx' \in \cC_x$ if and only if $\vx \odot \vx' \in \cC_x$.

To proceed, fix any $\vx' \in \mathcal{C}_x$ and observe that $p_{\vX} (\vx) = p_{\vX} (\vx' \odot \vx)$ for all $\vx \in \{\pm 1\}^N$.
This is because $p_{\vX}$ is uniform over the code and, since $\vx' \in \cC_x$, we see that $ \vx' \odot \vx \in \cC_x$ if and only if $\vx \in \cC_x$.
Meanwhile, the assumption of channel symmetry means that $w(y_S \mid x_S) = w(x'_S \odot y_S \mid x'_S \odot x_S)$ for all $\vx \in \{\pm 1\}^N$.
Together, these statements imply that
\begin{align}
    p_{\vX \mid Y_S}(\vx \mid y_S) = p_{\vX \mid Y_S}(\vx'\odot \vx \mid x_S'\odot y_S), \qquad x \in \{\pm 1\}^N.
\end{align}
In view of this identity,  the conditional expectation satisfies 
\begin{align}
    f(y_S) &= \sum_{\vx \in \{\pm 1\}^N}x_i\,  p_{\vX \mid Y_S}(\vx \mid y_S) \\
     & =   \sum_{\vx \in \{\pm 1\}^N}  x_i \,  p_{\vX \mid Y_S}(\vx' \odot \vx \mid  x_S' \odot y_S) \\
    & =  x_i '  \sum_{\vx\in \{\pm 1\}^N} (x_i' \, x_i) \,  p_{\vX \mid Y_S}(\vx' \odot \vx \mid  x_S' \odot y_S) \\
    & =  x_i' \, f(x'_S \odot y_S) .
\end{align}
The final statement follows from choosing $\vx' = \vx$ so that $f(y_S) = x_i f(z_S)$.
\end{proof}

Our next result provides an identity for the MMSE associated with estimating a single input.
For conceptual reasons, we find it convenient to frame the result in terms of two coupled channel outputs that are conditionally independent given the input.  
However, this approach is essentially the same as conditioning on the transmission of a particular codeword.

 \begin{lem}\label{lem:mmse_to_var} 
Let $\vX$ be distributed uniformly on the set of BPSK-modulated code sequences $\cC_x \subseteq \{\pm 1\}^N$ associated with the binary linear code $\cC \subseteq \mathbb{F}_{2}^N$  and let $\vY$ be an observation of $\vX$ through a BMS channel of the form $ \vY = \vX \odot \vZ$ where $\vZ \in \cZ^N$ is an independent vector with independent entries. 
For each $i \in [N]$ and $S \subseteq [N]$, the following identity holds
\begin{align}
 \mmse(X_i \mid Y_S) &\left(1 - \mmse(X_i \mid Y_S) \right) 
\\ \qquad & = \frac{1}{2} \big \|\E[X_i \mid Y_S]  - \E[X_i \mid Y'_S]  \big \|_2^2 \label{eq:mmse_to_var},
\end{align}
where $\vY' = \vX \odot \vZ'$ is an independent second use of the channel with the same input $\vX$. Furthermore, for every partition $(B_1, \dots, B_K)$ of $S$, we have the upper bound
\begin{align}
  \mmse(X_i \mid Y_S) & \left(1 - \mmse(X_i \mid Y_S) \right)
\\ & \le  \frac{1}{2} \sum_{k=1}^K  \big\| \E[ X_i \mid Y_S]  - \E[ X_i \mid Y^{B_k}_S]  \big\|_2^2, \label{eq:mmse_to_var_UB}
\end{align}
where $Y_S^{B_k}$ is a modified observation of $Y_S$ where the entries indexed by $B_k$ are resampled independently according to the same input $\vX$. 
\end{lem}
\begin{proof}
Define the conditional expectation  $f(y_S) \coloneqq \ex{ X_i \mid Y_S = y_S}$ and observe that
\begin{align}
   \ex{ X_i f(Y_S) \mid Y_S } =\E \big[ \ex{ X_i \mid Y_S } \ex{ X_i \mid Y_S } \mid Y_S \big] 
   = f^2 (Y_S) 
\end{align}
almost surely where the equalities follow from nested conditional expectation.  
This implies that
\begin{align}
    \var( X_i f(Y_S))
    & = \ex{ X_i^2 f^2(Y_S)} - \ex{ X_i f(Y_S)}^2\\
    & = \ex{ f^2(Y_S)} - \ex{ f^2(Y_S)}^2\\    & = \left( 1- \ex{ f^2(Y_S)} \right) \ex{f^2(Y_S)} \\
     & = \mmse( X_i \mid Y_S) \left( 1 - \mmse(X_i \mid Y_S) \right),
\end{align}
where the second step holds because $X^2_i = 1$ and the final step follows from   \eqref{eq:mmse_X_given_Y}. %
This decomposition holds generally for any random variable $X_i \in \{-1,1\}$ and any channel $X_i \to Y_S$. 

Next, we appeal to the special properties of the BMS channel and the linear code. Specifically, by  Lemma~\ref{lem:bms_linear}, it follows that $X_i f(Y_S)= f(Z_S)$. Writing $\vY = \vX \odot \vZ$ and $\vY' = \vX \odot \vZ'$ where $\vZ'$ is an independent copy of $\vZ$, we can now write 
\begin{align}
    \var( X_i f(Y_S)) & =    \var(  f(Z_S))\\
    & = \frac{1}{2} \ex{ \left( f(Z_S) - f(Z'_S) \right)^2} \\
    & = \frac{1}{2} \ex{ \left( X_i f(Z_S) - X_i f(Z'_S) \right)^2} \\
        & = \frac{1}{2} \ex{ \left(  f(Y_S) - f(Y'_S) \right)^2}, 
\end{align}
where the second line can be verified by expanding the square, the third line holds since $X_i^2 = 1$, and the last step is another application of  Lemma~\ref{lem:bms_linear}. Combining the two different expressions for $ \var( X_i f(Y_S))$ gives the desired identity. 

To prove the upper bound, recall that $f(\vZ)$ is a bounded function of independent random variables. Hence, we can apply the  Efron-Stein inequality \cite[Theorem~3.1]{Boucheron-2013} with respect to  the partition $(Z_{B_1}, \dots, Z_{B_k})$ to conclude that
\begin{align}
\var(f(\vZ)) \le \sum_{k=1}^K \frac{1}{2} \ex{ \left( f(Z_S) - f(Z_S^{B_k}) \right)^2} ,
\end{align}
where $Z_S^{B_k}$ denotes a version of $Z_S$ where the entries indexed by $B_k$ have been independently resampled from the $\vZ$~distribution. Multiplying the terms in the square by $X_i$ and then applying  Lemma~\ref{lem:bms_linear} leads to the stated bound, which is given in terms of $Y_S$ and $Y_S^{B_k}$. 
\end{proof}

Finally, we need the following result concerning the distribution of a pair of estimates based on correlated observations.

\begin{lem}\label{lem:ABC}
Let $\vX$ be distributed uniformly on the set of channel input sequences $\cC_x \subseteq \{\pm 1\}^N$ associated with the binary linear code $\cC \subseteq \mathbb{F}_{2}^N$ and let $\vY$ be an observation of $\vX$ through a BMS channel of the form $ \vY = \vX \odot \vZ$ where $\vZ \in \cZ^N$ is an independent vector with independent and identically  distributed entries. Suppose that for $i \in [N]$ there exist disjoint sets $A,B,C \in [N]\backslash \{i\}$  and a permutation matrix $\Pi$  such that  $(X_i, X_A, X_B)$ is equal in distribution to $(X_i, X_A, \Pi X_{ C})$. Then,
\begin{align}
 \big( \ex{ X_i \mid Y_{A}, Y_{B}}, & \ex{ X_i \mid Y_{A}, Y'_{B}} \big) \\ 
& \overset{d}{=} \big( \ex{ X_i \mid Y_{A}, Y_{B}} , \ex{ X_i \mid Y_{A}, Y_{C}}  \big),
\end{align} 
where $\vY'$ is an independent second use of the channel with the same input $\vX$. 
\end{lem}
\begin{proof}
Let us define the conditional expectations, 
\begin{align}
    f(u,v) & \coloneqq \ex{ X_i \mid Y_A = u, Y_{B} = v},\\
    g(u,v) & \coloneqq \ex{ X_i \mid Y_A = u, Y_C = v}.
\end{align}
The assumption that  $(X_i, X_A, X_B)$ is equal in distribution to $(X_i, X_A, \Pi X_{ C})$, combined with the assumptions on the channel, imply that $(X_i, Y_A, Y_B)$ is equal in distribution to $(X_i, Y_A, \Pi Y_{C})$, and thus 
\begin{align}
 f(u,\Pi v) = g(u,v), \label{eq:ftog}
\end{align}
for all $(u,v)$ in the support of $(Y_A,Y_C)$. 
From the channel assumptions, the channel outputs can be expressed as $\vY = \vX \odot \vZ$ and $\vY' = \vX \odot \vZ'$ where  $\vX,\vZ,\vZ'$ are independent. %
We can now write
\begin{subequations}
\label{eq:ABCproof}
\begin{align}
\big( f(Y_A, Y_B) & ,  f(Y_A, Y'_B) \big)  \nonumber \\ & =  X_i \big(  f(Z_A, Z_B) ,  f(Z_A,Z'_B) \big) \label{eq:ABCproof_a} \\
 & \overset{d}{=}  X_i \big(  f(Z_A, Z_B) ,  f(Z_A, \Pi Z_C) \big) \label{eq:ABCproof_b}  \\
  & =  X_i \big(  f(Z_A, Z_B) ,  g(Z_A, Z_C) \big) \label{eq:ABCproof_c}\\
& = \big(f(Y_A, Y_B) ,  g(Y_A, Y_C) \big), \label{eq:ABCproof_d}
\end{align}
\end{subequations}
where~\eqref{eq:ABCproof_a} relies on Lemma~\ref{lem:bms_linear},
\eqref{eq:ABCproof_b} holds because the entries of $\vZ$ and $\vZ'$ are independent and identically distributed,
\eqref{eq:ABCproof_c} is implied by~\eqref{eq:ftog},
and~\eqref{eq:ABCproof_d} again relies on Lemma~\ref{lem:bms_linear}. 
\end{proof}

\section{Proof of Main Result} 

We prove that RM codes  achieve capacity for any BMS channel in the limit of large blocklength. The first step of our proof is to embed the BMS channel of interest into a family of absolutely continuous BMS channels as described in Definition~\ref{defn:BS_metrics}. To further simplify our analysis, we will add the additional assumption that the family  of BMS channels is parameterized such that the MMSE function $\cM(t)$ defined in \eqref{eq:cM} is given by $\cM(t) = t$.

We emphasize that these assumptions are not restrictive in the sense  that, for any BMS channel, there exists a family of channels satisfying these constraints.  An explicit construction based on linear interpolation with erasure channels is described in Section~\ref{sec:construction}.  %
For the convenience of the reader we restate the channel assumptions for our main result as follows: 

\begin{assumption}\label{assumption:BMS} We have a family of BMS channels indexed by parameter $t \in [0,1]$ satisfying the following properties: 
\begin{enumerate}[A)]
\item For a random input $\vX \in \{\pm 1 \}^N$, the output associated with  $t \in [0,1]$ is given by a  BMS channel of the form 
\begin{align}
\vY(t) = \vX \odot \vZ(t)
\end{align}
where $\vZ(t) = \{Z_{i}(t) \}_{i \in [N]}$ is a random vector, independent of $\vX$, whose entries are independently and identically distributed according to a probability measure indexed by $t$. 
\item The family of BMS channels is ordered with respect to degradation where $t =0$ is the perfect channel and $t =1$ is the uninformative channel. %

\item The entropy function $\cH(t)$ defined in \eqref{eq:cH} is absolutely continuous on $[0,1]$.

\setcounter{enumi}{3}
\item %
The MMSE function $\cM(t)$ defined in \eqref{eq:cM} satisfies $\cM(t) = t$.  %
\end{enumerate}
\end{assumption}

While our main result concerns sequences of RM codes of increasing blocklength, many of the steps in our proof hold for a larger class of codes. To make these distinctions apparent, we list here the weaker properties that are sometimes used.  We note that all of these properties are satisfied by the $\mathrm{RM}(r,m)$ code with $N = 2^m$. 
In particular, we always assume that the random input vector $\vX  \in \{ \pm 1\}^N$ is distributed uniformly on the channel input sequences of a binary code.
In some cases, we also require that:
\begin{itemize}
\item the code is linear,
\item the code has a transitive symmetry group,
\item the code has a doubly transitive symmetry group.
\end{itemize}

\subsection{The Extrinsic MMSE Function}

As discussed in Section~\ref{sec:gexit_intro}, the entropy decomposition in \eqref{eq:Hdecomp2} plays an important role in the analysis of the BEC and the AWGN channels, where the partial derivatives provide natural measures of the performance of estimating a single entry of the input. However, one difficulty that arises in extending these approaches to general channels is that the GEXIT function does not seem to have an obvious estimation-theoretic interpretation. The approach taken in this paper is to study a surrogate for the GEXIT function, which we call the extrinsic MMSE function:

\begin{defn}[Extrinsic MMSE function]
Let $\vX \in \{\pm 1\}^N$ be a random vector and let  $\vY(t)$ be an observation of $\vX$ through a  BMS channel with parameter $t \in [0,1]$. The extrinsic MMSE function for input $i \in [N]$ is defined to be
\begin{align}
    M_i(t) \coloneqq \big \|X_i - \E[ X_i \mid Y_{\sim i}(t)] \|_2^2,\qquad t \in [0,1].
\end{align}

\end{defn}

The extrinsic MMSE  is similar to the EXIT function $H(X_i \mid Y_{\sim i}(t))$ in the sense that it provides a measure of the ability to estimate the $i$-th input based on the outputs from the other channels.  
As was the case for the GEXIT function, the extrinsic MMSE is identical for all $i\in [N]$ whenever the input distribution has a transitive symmetry group. In that case, we will sometimes drop the subscript $i$ and denote the extrinsic MMSE by $M(t)$.

For the purposes of proving our main capacity result with respect to the bit error rate, the code is transitive and the extrinsic MMSE has the property that $M(t)$ converges to zero (for a particular sequence of problems of increasing dimension) if and only if the bit error rate converges to zero. To prove that a code sequence, with rate converging to $R$, achieves capacity on the BMS channel family $W(t)$, it is sufficient to show that $M(t)$ converges to zero for all $t \in [0,1]$ such that the code rate $R$ is strictly less than Shannon capacity $C(t)$.

Our proof that RM codes achieve capacity on  BMS channels consists of the following steps: 
\begin{enumerate}[(1)]

    \item \textbf{Sharp threshold property:} Show that, for every sequence of RM codes with increasing blocklength, the extrinsic MMSE has a sharp threshold property with respect to $t$. Specifically, we show that
    \begin{align}
    \int_0^1 M(t) (1-  M(t)) \, dt = O\left( \frac{ \ln m}{\sqrt{m}} \right) , \label{eq:MMSE_order_bound}
    \end{align}
    which implies that $M(t)$ cannot be too different from a step function that jumps from 0 to 1.  By itself, this does not imply convergence though because the location where the function jumps is not controlled. We note that the integral exists because $M(t)$ is non-decreasing.
    
    \item  \textbf{Area theorem:} Show that, if the sequence of RM codes has limiting rate $R$, then the location of the jump in the step function must converge to the unique value  of $t$ such that the Shannon capacity $C(t)$ of the BMS channel is equal to the code rate $R$. 
\end{enumerate}
The two-step approach of first establishing a sharp threshold and then using an area theorem to localize the jump is now somewhat standard~\cite{Kudekar-it17,Reeves-it19}. The  main novelty in our approach is the reliance on the extrinsic MMSE instead of the GEXIT curve and the mechanism by which we establish convergence to a step function.

\subsection{Are Two Looks Better Than One?}

This section establishes the sharp threshold property for the extrinsic MMSE as described by \eqref{eq:MMSE_order_bound}.   Observe that, for any input distribution,   $M_i(t)$ is non-decreasing  in $t$  for each $i \in [N]$ because of the assumed channel degradation. Hence, to show that $M_i(t)$ is close to a 0/1 step function it is sufficient to show that $M_i(t)(1- M_i(t))$ is close to zero for most  (but importantly not all) values of $t$ in the unit interval. Now, we appeal to Lemma~\ref{lem:mmse_to_var} which shows that, if the input is defined by a binary linear code, then the following identity holds:
\begin{align}
    M_i(t)(1- M_i(t))  = \frac{1}{2} \| \ex{ X_i \mid Y_{\sim i}(t)} -\ex{ X_i \mid Y'_{\sim i}(t)} \|_2^2,  \label{eq:MtoVar}
\end{align}
where $\vY'(t)$ is an independent second use of the channel with the same input $\vX$.

In view of \eqref{eq:MtoVar} the entire problem of establishing a sharp threshold can be boiled down to the following question:
\begin{quote}
    Assuming that $Y_i$ is not observed, are two independent observations of the remaining code symbols likely to provide significantly different posterior estimates of $X_i$?
\end{quote}
In the setting where $X_i$ can be recovered accurately from $Y_{\sim i}$ the answer to this question is clearly negative. Conversely, in the setting where the first look is uninformative (i.e., with high probability the conditional distribution  $X_i$ given $Y_{\sim i}$ is close to the prior distribution on $X_i$) then it is unlikely that a second look will make much of a difference. The interesting setting occurs when a single look provides partial information about $X_i$, and so two looks are then better than one in a meaningful sense. Our goal is to show that, w.r.t.\ the parameter $t$, this ``interesting'' regime has measure tending to zero, that is to say most values of $t$ are ``uninteresting''. Combined with \eqref{eq:MtoVar} and monotonicity of $M_i(t)$, it follows that $M_i(t)$ converges to a 0/1 step function. 

\subsubsection{Decomposition of Variance} 
We consider a decomposition of the variance term appearing in \eqref{eq:MtoVar} with respect to a set $B \subset [N] \backslash \{i\}$, which will be specified later. For $S \subseteq [N]$, define 
\begin{align}
    \Delta^{S}_i(t) \coloneqq \frac{1}{2}  \big\| \ex{ X_i \mid Y_{\sim i}(t)} -\E[ X_i \mid   Y^S_{{\sim i}}(t)] \big\|_2^2, \label{eq:Delta_i}
\end{align}
where $\vY^S(t)$ is a modified version of $\vY(t)$ in  which the entries indexed by $S$ have been resampled according to the same input $\vX$.  The quantity $\Delta_i^S(t)$ can be interpreted as the generalized influence (see  \cite[Definition~8.22]{ODonnell-2014}) of the coordinates indexed by $S$ on the conditional mean estimator $f(y)= \ex{ X_i \mid Y_{\sim i} (t) = y}$. If the input distribution is defined by a binary linear code, then we can apply the  upper bound in Lemma~\ref{lem:mmse_to_var} to the partition given by $B$ and the singletons in $A\coloneqq [N] \setminus (B \cup \{i\})$ to obtain 
\begin{align}
    M_i(t) (1- M_i(t)) \le  \Delta_i^B(t)   + \sum_{j \in A} \Delta_i^{j}(t). \label{eq:Delta_decomp}
\end{align}
We remark that 
\eqref{eq:Delta_decomp} is general in the sense that it holds for every linear code. %

In the following, we will bound each term in \eqref{eq:Delta_decomp} by first relating it to a GEXIT function, using the results in Section~\ref{sec:I-MMSE}, and then  combining properties of the GEXIT function with some other arguments to bound the integral with respect to $t$ over the unit interval:
\begin{itemize}
     \item  The term $\Delta_i^B(t)$ is addressed in Section~\ref{sec:two_look_bound} where we establish the following. If $ N=2^m$ and the input distribution is uniform on the codewords of the $\mathrm{RM}(r,m)$ code, then, for integers $1 \le k \le m$ and $i \in [N]$, there exists a set $B \subset [N] \backslash \{i\}$ of size $2^{m} - 2^{m-k}$ such that %
    \begin{align}
    \int_0^1 \Delta_i^{B}(t)  \, dt \le   \frac{4 \ln (2) (3 k + 4)}{5 \sqrt{ m }}.
    \label{eq:two_look_bound}
\end{align}

    \item The term $\Delta_i^{j}(t)$ is addressed in Section~\ref{sec:single_term_bound}, where it is shown that if the input distribution has a doubly transitive symmetry group, then the following bound holds for all $i \ne j$,
    \begin{align}
     \int_0^1 \Delta_i^{j}(t)  \, dt \le \frac{ 4 \ln 2}{ N-1} . \label{eq:single_term_bound}
\end{align}
\end{itemize}
Combining these results leads to a family of upper bounds on the integral of \eqref{eq:Delta_decomp} that is parametrized by $k \in \{0, \dots, m\}$.  
This parameter provides a trade-off between the two terms in the bound. For large values of $k$, the bound is dominated by the difference between the rates in \eqref{eq:two_look_bound}.   Conversely, for small values of $k$,  the bound is dominated by the size of the set $A$,  %
which is given by $2^{m-k}-1$. Optimizing over the choice of the integer $k$ gives the following result: 

\begin{lem}
\label{lem:integral_decay}
Consider a family of BMS channels satisfying Assumption~\ref{assumption:BMS}. If the input distribution is uniform on the codewords of the $\mathrm{RM}(r,m)$ code, then the extrinsic MMSE satisfies 
    \begin{align}
    \int_{0}^1 M(t) (1 -M(t))  \, dt &\le  \rho(m) \coloneqq \frac{ 6 \ln(m)  + 34}{5\sqrt{m}}.
\end{align}
\end{lem}
\begin{proof}
For every integer $k \in [m]$, the bounds in  \eqref{eq:Delta_decomp}, \eqref{eq:two_look_bound}, and \eqref{eq:single_term_bound} give
    \begin{align}
    \int_0^1 M(t) (1 -M(t))  \, dt 
    & \le      4 \ln(2)  \left( \frac{2^{m-k} - 1}{2^m-1}  +  \frac{3 k + 4}{5 \sqrt{ m }} \right) \\ &\le   4 \ln(2)  \left( 2^{-k}   +  \frac{3 k + 4}{5 \sqrt{ m }}\right),
\end{align}
where the second step follows from the basic inequality  $2^{m-k} - 1 \le (2^{m} - 1) 2^{-k}$. %
Next,  we choose $k = \lceil  \frac{1}{2} \log_2 m \rceil$  and note that
\begin{align}
     2^{-k}   + \frac{3 k + 4}{5 \sqrt{m }} & \le 
\frac{1}{ \sqrt{m}}   +  \frac{3( \frac{1}{2} \log_2 (m)+1 ) +4}{5 \sqrt{m }} \\ & = \frac{ 3  \log_2(m)   + 24 }{10\sqrt{m}}.
\end{align}
The final result follows from multiplying this by $4\ln 2$ and noting that $48 \ln(2) \le 34$. 
\end{proof}

\subsubsection{Proof of Generalized Influence Bound in  (\ref{eq:two_look_bound})}
\label{sec:two_look_bound}

This section proves an upper bound on the integral of the generalized influence term $\Delta_i^B(t)$ defined in \eqref{eq:Delta_i} for a carefully chosen set $B\subset [N] \backslash \{i\}$. %
This term can be expressed as
\begin{align}
\Delta^B_i(t) \coloneqq \frac{1}{2}  \|\E[ X_i \mid Y_{A}(t), Y_{B}(t)] -\E[ X_i \mid Y_{A}(t), Y'_{B}(t)]\|_2^2, \label{eq:Delta_B_alt}
\end{align} 
where $A = [N] \backslash ( \{ i\} \cup B)$ and  $\vY'(t)$ denotes an independent second use of the BMS channel with the same input $\vX$.

Our approach to bounding this term is to view the input vector $(X_0, \dots , X_{N-1} )$, as the first $N$ entries in an extended input vector $(X_0, \dots, X_{L-1})$ of length $L > N$. With some abuse of notation we use  $X_{[N]}$ to denote the original input vector and $X_{[L]}$ to denote the extended input vector.
Associated with the extended input we define the output $Y_{[L]}(t) = (Y_0(t) \, \dots, Y_{L-1}(t))$  from the same BMS channel. If we can find an extension such that:
\begin{enumerate}[i)]
    \item the extended input $X_{[L]}$ is distributed uniformly on the codewords of a linear code; and
\item there exists a set $C \subseteq \{ N, \dots, L-1\}$ and permutation matrix $\Pi$  such that $(X_i,X_A, X_B)$ is equal in distribution to $(X_i,X_A, \Pi X_C)$,
\end{enumerate}
then we can use Lemma~\ref{lem:ABC} to conclude that 
\begin{align}
\Delta^B_i(t) %
 = \frac{1}{2} \|\E[ X_i \mid Y_{A}(t), Y_{B}(t)] -\E[ X_i \mid Y_{A}(t), Y_{C}(t)] \|_2^2 . \label{eq:Delta_B_alt2}
\end{align} 
In this expression, the second look at the entries indexed by $B$ has been replaced by an observation of the entries indexed by $C$ in the extended codeword.

To apply this, we assume that the original input is generated by a uniform distribution over the codewords of $\mathrm{RM}(r,m)$ and the extended input is generated by a uniform distribution over the codewords of $\mathrm{RM}(r,m+k)$, for some positive integer $k \le m$.
Then, the following lemma shows that the nesting property identified in Lemma~\ref{lem:rm_overlap} can be used to choose the sets $A$, $B$, and $C$ to satisfy the distributional condition defined in Lemma~\ref{lem:ABC}.
See Figure~\ref{fig:rm_in_rm_k2} for an illustration.

\begin{lem} \label{lem:extended_code}
For positive integers $(r,m,k)$ with $r,k \le m$, let $N = 2^m$ and $L = 2^{m+k}$. If $X_{[L]}$ is distributed uniformly on the codewords of the $\mathrm{RM}(r,m+k)$ code then $X_{[N]}$ is distributed uniformly on the codewords of the $\mathrm{RM}(r,m)$ code.
Furthermore, for each $i \in [N]$, there exists a partition $\{i\},A,B$ of $[N]$, a set $C\subseteq [L]\setminus [N]$, and an automorphism $\pi \colon [L]\to [L]$ of $X_{[L]}$ such that $\pi$ acts as the identity on $\{i\}\cup A$ and swaps $B$ and $C$.
This implies that there is a permutation matrix $\Pi$ such that $(X_i, X_A, X_B)$ is equal in distribution to $(X_i, X_A, \Pi X_{ C})$ and that
\begin{align} 
&\left\|\ex{ X_i \mid Y_{A}, Y_{B}} - \ex{ X_i \mid Y_{[L] \backslash \{i\}}} \right\|_2 \\ & \qquad \qquad = \left\|  \ex{ X_i \mid Y_{A}, Y_{C}}- \ex{ X_i \mid Y_{[L] \backslash \{i\}}}  \right\|_2 . \label{eq:auto_ABC_exp}
\end{align} 

\end{lem}
\begin{proof}
If $X_{[L]}$ is uniformly distributed on the codewords of $\cC = \mathrm{RM}(r,m+k)$ and $I=[N]$, then Lemma~\ref{lem:rm_overlap} shows that $X_I = X_{[N]}$ is uniformly distributed on the codewords of $\cC_I = \mathrm{RM}(r,m)$.
For $i=0$, the sets $A,B,C$ are also constructed in Lemma~\ref{lem:rm_overlap} and we will verify their properties below.
In the last step, we describe how the $i=0$ construction can be remapped to any $i\in [N]$.

To explain why the sets $A,B,C$ from Lemma~\ref{lem:rm_overlap} satisfy the stated conditions, we recall their definition from the proof of Lemma~\ref{lem:rm_overlap}.
First, we define $V=\mathbb{F}_{2}^{m}\times \{0\}^k$ and $V'=\mathbb{F}_{2}^{m-k} \times \{0\}^{k}  \times \mathbb{F}_2^{k}$ as the evaluation sets associated with the indices $I=\tau(V)$ and $I'=\tau(V')$, respectively.
Then, we define $T = I \cap I' = [2^{m-k}]$, $A = T \setminus \{0\}$, $B= I \setminus T$, and $C= I' \setminus T$.
To simplify notation, we also define $A' = \tau^{-1} (A)$, $B' = \tau^{-1} (B)$, and $C' = \tau^{-1} (C)$.

Consider the function, $\pi \colon \mathbb{F}_2^{m+k} \to \mathbb{F}_2^{m+k}$ defined by $\vv \mapsto \vv'$ with $v_i ' = v_i$ for $i\in [m-k]$, $v_i ' = v_{i+k}$ for $i\in \{m-k,\ldots,m-1\}$, and $v_i ' = v_{i-k}$ for $i\in \{m,\ldots,m+k-1\}$.
This function simply swaps bits $v_{m-k+i}$ and $v_{m+i}$ for all $i \in [k]$.
One can verify that $\pi$ is a permutation on $\mathbb{F}_2^{m+k}$ that satisfies $\pi(\vnot)=\vnot$, $\pi(\pi(\vv))=\vv$ for all $\vv\in \mathbb{F}_2^{m+k}$, $\pi(\vv)=\vv$ for all $\vv\in A'$, and $\pi(B') = C'$.
We do not discuss the precise element by element mapping from $B'$ to $C'$ because it is not important for this result (e.g., Lemma~\ref{lem:ABC} allows for an arbitrary permutation of the set $C$).

Since $\pi$ is a linear function, it defines an automorphism of $\cC$~\cite[p.~398]{Macwilliams-1977}.
For integer indices, this automorphism is given by $i \mapsto \tau(\pi(\tau^{-1}(i)))$.
Since a code automorphism preserves the distribution of the codewords, it follows that
 $(X_0,X_A,X_B)$ is equal in distribution to $(X_0,X_A,\Pi X_C)$ for some permutation matrix $\Pi$.
Moreover, $Y_{[L]}$ is a memoryless observation of $X_{[L]}$ and applying this automorphism to both $X_{[L]}$ and $Y_{[L]}$ preserves their joint distribution as well.
Thus, there is a another permutation matrix $\Pi'$ such that $(X_0,Y_A,Y_B,Y_{[L]\setminus\{0\}})$ is equal in distribution to $(X_0,Y_A,\Pi Y_C, \Pi' Y_{[L]\setminus\{0\}})$.
This implies that~\eqref{eq:auto_ABC_exp} holds for $i=0$.

For $i \in [2^m] \setminus \{0\}$, we can simply translate the sets $A'$, $B'$, and $C'$ by adding $\vi' = \tau^{-1}(i)$.
In particular, we define $\pi_{\vi'} (\vv) = \pi (\vv - \vi') + \vi'$, $A_i '  =  A' + \vi'$, $B_i '  =  B' + \vi'$, and $C_i ' =  C' + \vi'$.
Observe also that $\{\vi'\},A_i',B_i'$ forms a partition of $V$.
Next, one can verify that $\pi_{\vi'}$ is a permutation on $\mathbb{F}_2^{m+k}$ that satisfies $\pi_{\vi'} (\vi')=\vi'$, $\pi_{\vi'} (\pi_{\vi'} (\vv))=\vv$ for all $\vv\in \mathbb{F}_2^{m+k}$, $\pi_{\vi'} (\vv)=\vv$ for all $\vv\in A'$, and $\pi_{\vi'} (B_i ') = C_i '$.
Like before, since $\pi_{\vi'}$ is an affine function on $\mathbb{F}_2^{m+k}$, it defines an automorphism of $\cC$ that preserves the uniform distribution over codewords.
For $i \in [N]$, we define $A_i = \tau(A_i ')$, $B_i = \tau(B_i ')$, and $C_i = \tau(C_i ')$.
Then, the above statements imply that $\{i\},A_i,B_i$ forms a partition of $[N]$ and that $(X_i,X_{A_i},X_{B_i})$ is equal in distribution to $(X_i,X_{A_i},\Pi_i X_{C_i})$ for some permutation matrix $\Pi_i$.
Likewise, there is a permutation matrix  $\Pi_i '$ such that $(X_i,Y_{A_i},Y_{B_i},Y_{[L]\setminus\{i\}})$ is equal in distribution to $(X_i,Y_{A_i},\Pi_i Y_{C_i},\Pi_i ' Y_{[L]\setminus\{i\}})$ and this establishes~\eqref{eq:auto_ABC_exp} for $i\in [N]$.
\end{proof}

Using the nesting property, we can now  bound $\Delta_i^B(t)$ in terms of the difference between the  extrinsic MMSE functions of the original and the extended code.
Neglecting the dependence on $t$ to lighten the notation, the bound is derived by starting with \eqref{eq:Delta_B_alt} and then writing
\begin{subequations}
\label{eq:sqrInf}
\begin{align}
\sqrt{2\Delta_i^B(t)}& =  \left\| \ex{ X_i \mid Y_{A}, Y_{B}} -\ex{ X_i \mid Y_{A}, Y'_{B}} \right\|_2\label{eq:sqrInf_a}\\
& =  \left\|\ex{ X_i \mid Y_{A}, Y_{B}} -\ex{ X_i \mid Y_{A}, Y_{C}} \right\|_2 \label{eq:sqrInf_b}\\
& = \big\|\ex{ X_i \mid Y_{A}, Y_{B}}- \ex{ X_i \mid Y_{[L] \backslash \{i\}}}  \\ & \quad\; -\ex{ X_i \mid Y_{A}, Y_{C}}+ \ex{ X_i \mid Y_{[L] \backslash \{i\}}}  \big\|_2 \label{eq:sqrInf_c}\\
& \le  \left\|\ex{ X_i \mid Y_{A}, Y_{B}} - \ex{ X_i \mid Y_{[L] \backslash \{i\}}} \right\|_2  \\ & \quad\; + \left\|  \ex{ X_i \mid Y_{A}, Y_{C}}- \ex{ X_i \mid Y_{[L] \backslash \{i\}}}  \right\|_2 \label{eq:sqrInf_e}\\
& =  2 \left\|\ex{ X_i \mid Y_{[N]\backslash\{i\}}} - \ex{ X_i \mid Y_{[L] \backslash \{i\}}} \right\|_2  \label{eq:sqrInf_f}\\
& =  2 \sqrt{\big \| \E[ X_i  \mid Y_{[L]\backslash\{i\}}] \big \|_2^2 - \big \|\E[  X_i  \mid Y_{[N]\backslash\{i\}} ] \big \|_2^2}, \label{eq:sqrInf_g}\\
& =  2 \sqrt{ \mmse( X_i  \mid Y_{[N]\backslash\{i\}}) - \mmse( X_i  \mid Y_{[L]\backslash\{i\}}) }, \label{eq:sqrInf_h}
\end{align}
\end{subequations}
where  \eqref{eq:sqrInf_b} follows from combining Lemma~\ref{lem:ABC} and Lemma~\ref{lem:extended_code} to establish
\begin{align*}
 \big( \E[X_i \mid Y_{A}, Y_{B}], &\E[X_i \mid Y_{A}, Y'_{B}]\big)  
 \\ &\overset{d}{=} \big( \E[ X_i \mid Y_{A}, Y_{B}] , \E[X_i \mid Y_{A}, Y_{C}]  \big),
 \end{align*}
\eqref{eq:sqrInf_e} is given by the triangle inequality, and 
\eqref{eq:sqrInf_f} holds because $[N] = \{i\} \cup A\cup B$ and Lemma~\ref{lem:extended_code} shows that the two terms in~\eqref{eq:sqrInf_e} are equal.
The last two steps follow from~\eqref{eq:L2diff} and the fact that  $X_i - Y_{[N] \setminus \{i\}} - Y_{[L] \setminus \{i\}}$ is a Markov chain.

The next step is to use %
Lemma~\ref{lem:gexit_diff_expansion} to bound the difference in extrinsic MMSE in terms of the  difference of GEXIT functions.  Let the GEXIT functions of the original and the extended code be given by
\begin{alignat}{3}
G_i(t) \coloneqq \frac{\partial}{\partial s}  H \big( X_{[N]} \mid Y_i (s), Y_{[N]\setminus \{i\}}(t) \big) \Big|_{s=t}, &\quad & i \in [N]\\
G^\mathrm{ext}_i(t) \coloneqq \frac{\partial}{\partial s}  H \big( X_{[L]} \mid Y_i (s), Y_{[L]\setminus \{i\}}(t) \big) \Big|_{s=t}, &\quad & i \in [L].
\end{alignat}
The fact that the channel is memoryless means that, for each $i \in [N]$,
the GEXIT function for the extended code, $G_i^\mathrm{ext}(t)$, is equal to the GEXIT function for the original code augmented with the additional observations $U(t) = (Y_{N}(t) , \dots, Y_{L-1}(t))$.  In other words, $G_i^\mathrm{ext}(t)$ equals
\begin{align}
     G_i^+ (t) \coloneqq \frac{\partial}{\partial s}  H \big( X_{[N]} \mid Y_i (s), Y_{[N]\setminus \{i\}}(t) , U(t) \big) \Big|_{s=t},
\end{align}
for  $i \in [N]$.
See the first steps in the proof of  Lemma~\ref{lem:gexit_expansion} for more details.  
Since $U (t) - X_{[N]} - Y_{[N]} (t) $ forms a Markov chain, we can apply Lemma~\ref{lem:gexit_diff_expansion} to bound $G_i^+(t) - G_i(t)$ from below in terms of the difference in extrinsic MMSE. Combining this bound with the fact that  $-q'_1(t) = \cM'(t) \equiv 1$ under the assumed channel parametrization and then rearranging terms, we conclude 
that following inequality holds almost everywhere: 
 \begin{align}
\mmse( X_i  \mid Y_{[N]\backslash\{i\}}) & - \mmse( X_i  \mid Y_{[L]\backslash\{i\}}) \\
& \le 
\frac{1}{\ce_1} \left( G_i(t) - G_i^\mathrm{ext}(t) \right). \label{eq:gexit_ext_lb}
\end{align}

The remaining challenge is to argue that the  difference between the GEXIT functions is  small for most values of channel parameter $t$. Recall that, by Lemma~\ref{lem:RM_rate}, the difference in rates between  RM$(r,m)$ an $R(r,m+k)$ is at most  $(3k+4)/(5\sqrt{m})$. We will use the fact that the  difference in rate can also be expressed as the integral of the difference in GEXIT functions: 
\begin{subequations}
\label{eq:RtoGEXIT}
\begin{align}
  R(r&,m)  - R(r,m+k) = \frac{H(X_{[N]})}{N}    - \frac{ H(X_{[L]})}{L} \label{eq:RtoGEXIT_a}\\
 & = \int_0^1 \frac{d}{ d t} \left[ \frac{ H\left(X_{[N]} \mid Y_{[N]}(t)\right)}{N}    - \frac{ H\left(X_{[L]} \mid Y_{[L]}(t)\right)}{L}    \right]  \, dt \label{eq:RtoGEXIT_b}\\
  & =  \int_0^1  \frac{1}{N}  \sum_{j \in [N]} G_j(t) - \frac{1}{L}  \sum_{j \in [L]} G_j^\mathrm{ext}(t)   dt \label{eq:RtoGEXIT_c}\\
 & =  \int_0^1   G_i(t) -  G_i^\mathrm{ext}(t)   \, dt, \qquad \text{for all $i \in [N]$} \label{eq:RtoGEXIT_d},
\end{align}
\end{subequations}
where~\eqref{eq:RtoGEXIT_a} holds because $X_{[N]}$ and $X_{[L]}$ are distributed uniformly on the codewords of RM$(r,m)$ and RM$(r,m+k)$,
\eqref{eq:RtoGEXIT_b}~follows from the fundamental theorem of calculus and the assumption that $t=0$ is the perfect channel and $t =1$ is an  uninformative channel, 
\eqref{eq:RtoGEXIT_c} holds by the law of the total derivative, and 
\eqref{eq:RtoGEXIT_d} is implied by the fact that the GEXIT functions of all bits are identical (which follows from the transitive symmetry of the original and extended RM codes).

We now have all the pieces in hand to bound the integral of the influence term $\Delta_i^B(t)$. Specifically, we can write
\begin{align}
\int_0^1 & \Delta^B_i(t)  \, dt \\
&\overset{\eqref{eq:sqrInf}}{\le} 
2  \int_0^1 \left(\mmse( X_i  \mid Y_{[N]\backslash\{i\}}) - \mmse( X_i  \mid Y_{[L]\backslash\{i\}}) \right) dt\\
&  \overset{\eqref{eq:gexit_ext_lb}\;}{\le} \frac{2}{c_1} \int_0^1 \big( \gexit_i(t) - \gexit_i^\mathrm{ext}(t) \big)  \, dt \\
& \overset{\eqref{eq:RtoGEXIT}}{=} \frac{2}{c_1}  \big( R(r,m) - R(r,m+k) \big) \\
& \le \frac{2 }{c_1}  \; \frac{ 3k + 4}{ 5 \sqrt{m}},
\end{align}
where the last step follows from Lemma~\ref{lem:RM_rate}.
This concludes the proof of \eqref{eq:two_look_bound}.

\subsubsection{Proof of Generalized Influence Bound in  (\ref{eq:single_term_bound})}
\label{sec:single_term_bound}

This section proves an upper bound on the integral of the generalized influence term $\Delta_i^{j}(t)$ defined in \eqref{eq:Delta_i}. Suppressing the explicit dependence on the channel parameter $t$, this term can be expressed as
\begin{align}
\Delta_i^{j}(t) \coloneqq \frac{1}{2}\| \E[X_i \mid Y_{\sim i}] -\E[X_i \mid Y^{j}_{\sim i}] \|_2^2, 
\end{align} 
where we recall that 
\[
\vY^{j} = (Y_0, \dots, Y_{j-1}, Y'_j , Y_{j+1}, \dots, Y_{N-1})
\]
is a modified version of $\vY$ in which the  $j$-th component has been resampled according to the same input. 
We can write
\begin{subequations}
\label{eq:Delta_j_to_MMSE} 
\begin{align}
    &\sqrt{ 2 \Delta_i^j(t)} = \big  \| \ex{ X_i \mid Y_{\sim i}} - \E [ X_i \mid Y^j_{\sim i}] \big\|_2\\
    & \quad = \big  \| \big( \ex{ X_i \mid Y_{\sim i}} - \ex{ X_i \mid Y_{\sim i}, Y_j'} \big)  \\ & \quad\qquad - \big(  \E [ X_i \mid Y^j_{\sim i}]  - \ex{ X_i \mid Y_{\sim i}, Y_j'} \big) \big\|_2 \\
    & \quad \le  \big  \|  \ex{ X_i \mid Y_{\sim i}} - \ex{ X_i \mid Y_{\sim i}, Y_j'} \big\|_2  \\ & \quad\qquad + \big \|  \E [ X_i \mid Y^j_{\sim i}]  - \ex{ X_i \mid Y_{\sim i}, Y_j'}  \big\|_2 \label{eq:Delta_j_to_MMSE_a} \\
    & \quad =  2 \big  \|  \ex{ X_i \mid Y_{\sim i}} - \ex{ X_i \mid Y_{\sim i}, Y_j'} \big\|_2  \label{eq:Delta_j_to_MMSE_b} \\
   & \quad =  2  \sqrt{ \mmse( X_i \mid Y_{\sim i} ) - \mmse( X_i \mid Y_{\sim i} , Y_j') } \label{eq:Delta_j_to_MMSE_c}
\end{align}
\end{subequations}
where~\eqref{eq:Delta_j_to_MMSE_a} is the triangle inequality, \eqref{eq:Delta_j_to_MMSE_b} holds because the triples  $(X_i, Y_{\sim i} , Y'_j)$ and $(X_i, Y_{\sim i}^j, Y_j)$ have the same distribution, and \eqref{eq:Delta_j_to_MMSE_c} follows from \eqref{eq:L2diff}.

Following the same approach as in the previous section, we can use Lemma~\ref{lem:gexit_diff_expansion} to bound the difference in extrinsic MMSE  in terms of the difference of GEXIT functions. Let $G_i(t)$ be the GEXIT function of the original channel (Definition~\ref{def:gexit}) and for $j \ne i$ define 
\begin{alignat}{3}
 \gexit^{j}_i(t) &\coloneqq \frac{\partial}{ \partial s}   H\left(\vX \mid Y_i(s),   Y_{\sim i}(t), Y'_j(t) \right)  \Big|_{s = t}  
\end{alignat}
to be the GEXIT function for an augmented channel that uses the $j$-th channel twice. We can now apply Lemma~\ref{lem:gexit_diff_expansion} with  $U(t)=Y_j ' (t)$ to bound $G_i(t) - G_i^j(t)$ from below in terms of the extrinsic MMSE. Combining this bound with the fact that  $-q'_1(t) = \cM'(t) \equiv 1$ under the assumed channel parametrization and then rearranging terms, we conclude 
that following inequality holds almost everywhere: 
\begin{align}
\mmse(X_i \mid Y_{\sim i}(t))  &- \mmse(X_i \mid Y_{\sim i}(t),Y'_{j}(t))
 \\ &\le  \frac{1}{c_1}  \left(  \gexit_i(t) - \gexit^{j}_i(t)
   \right). \label{eg:MMSE_to_aug_G}
\end{align}

In view of \eqref{eq:Delta_j_to_MMSE} and \eqref{eg:MMSE_to_aug_G}, we see that the integral of the difference in GEXIT functions provides an upper bound on the integral of $\Delta_i^j(t)$. If the input distribution has a transitive symmetry group, then the integral of $\gexit_i(t)$ follows directly from the definition of the GEXIT function as discussed in Section~\ref{sec:gexit_intro}.  However, the integral of $G_i^j(t)$ does not have such a simple interpretation because the partial derivative of the  augmented channel with respect to $j$ is different than for the other channels. 

The next lemma provides a bound on the integral in question, averaged over the indices $i\ne j$.  If the input distribution has a doubly transitive symmetry, then this gives a bound that holds uniformly for all pairs of indices. Combining this result with the bounds in 
\eqref{eq:Delta_j_to_MMSE_c} and 
\eqref{eg:MMSE_to_aug_G}  %
gives the single term bound stated in \eqref{eq:single_term_bound}.

\begin{lem} \label{lem:Gij}
For every input distribution on $\{\pm 1\}^N$, 
\begin{align} \label{eq:single_term_sym}
\sum_{i,j \in [N] \, : \, i \ne j}   \int_0^1( \gexit_i(t)- \gexit_i^{j}(t) ) \, dt \le N.
\end{align}
In particular, if the input distribution has transitive symmetry, then $\gexit_{i} = \gexit_k$ for all $i,k \in [N]$.
If it has doubly transitive symmetry, then $\gexit_{i}^j = \gexit_k^\ell$ for all $i,j,k,\ell \in [N]$ with $i \ne j$ and $k \ne \ell$. Thus, we have 
\begin{align} \label{eq:single_term_dbl_sym}
  \!\!\!\! \int_0^1( \gexit_i(t)- \gexit_i^{j}(t) ) \, dt \le \frac{1}{N-1} , \;\; i,j \in [N], \; i \ne j.
\end{align}
\end{lem}

\begin{proof}
See \hyperlink{prf:Gij}{proof} Section~\ref{sec:proof_main}. \renewcommand{\qedsymbol}{}
\end{proof}

\subsection{Bounds on the Extrinsic MMSE via the Area Theorem} \label{sec:MMSE_bounds}

Since Lemma~\ref{lem:integral_decay} establishes the sharp threshold phenomenon in the sense of \eqref{eq:MMSE_order_bound}, the next step is to provide bounds on the extrinsic MMSE in terms of the rate of the code. The key tool that enables this is a relation known as the area theorem for GEXIT functions~\cite{RU-2008,Measson-it09}. Consider a family of BMS channels satisfying the assumptions in Definition~\ref{defn:BS_metrics}, and let $G(t)$ be the GEXIT function associated with a random input  $\vX$  of length $N$ whose distribution has a transitive symmetry group. Then, the generalized area theorem~\eqref{eq:Hdecomp2} implies that
\begin{align}
    \frac{1}{N} H(\vX) = \int_0^1 G(t) \, dt . \label{eq:area_theorem} 
\end{align}
This statement is an immediate consequence of the definition of the GEXIT function and the assumption of a transitive symmetry, which ensures that GEXIT function is the same for all inputs. If the distribution of $\vX$ is uniformly distributed over the input sequences of a binary code, then the LHS of \eqref{eq:area_theorem} is the rate of the code.

For the purposes of this paper, the connection between the rate and the extrinsic MMSE follows from the results in Section~\ref{sec:I-MMSE}. The details are summarized in the following result, which provides bounds on the MMSE in terms of the integral appearing in \eqref{eq:MMSE_order_bound} and the gap between the Shannon capacity and the code rate.

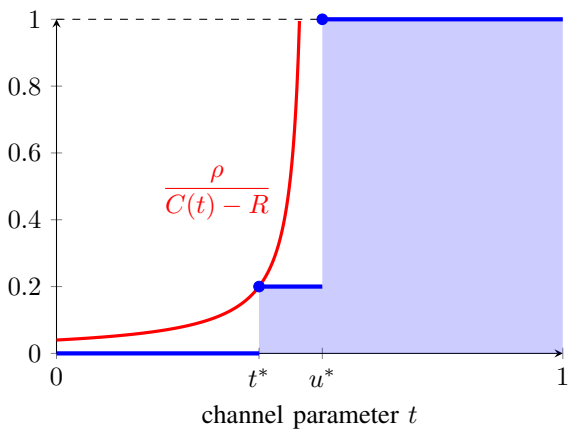
\begin{figure}[b!]
\ifextfig
\includegraphics{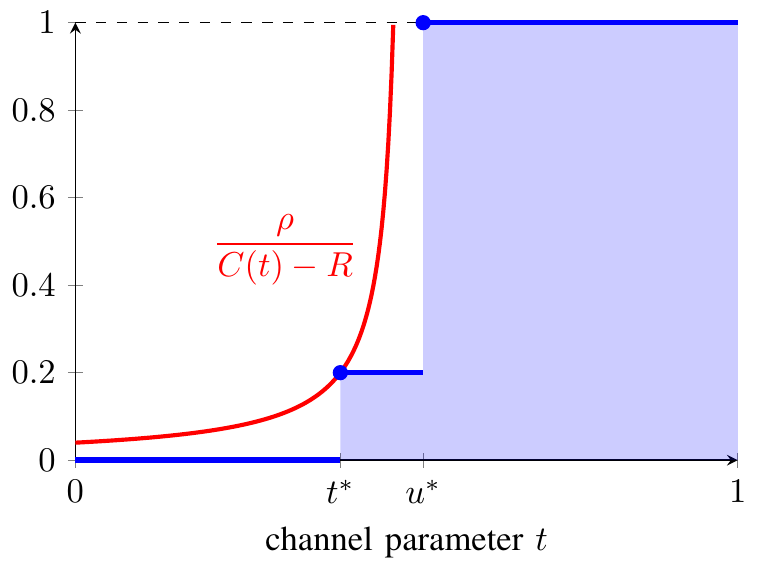}
\else
\centering
\begin{tikzpicture}

\begin{axis}[%
width=2.65in,
height=1.75in,
scale only axis,
xmin=0,
xmax=1,
xlabel={channel parameter $t$},
xtick = {0, .4, 0.5250, 1},
xticklabels ={$0$,  $t^*$, $u^*$, $1$},
ymin=0,
 axis lines =left,
 clip  = false,
]

\addplot[ dashed] coordinates { (0,1) (1,1) };

\addplot[ ultra thick, blue] coordinates { (0,0) (0.4,0) };
\addplot[ ultra thick, blue] coordinates { (.4,.2) (0.5250,.2) };
\addplot[ ultra thick, blue] coordinates {  (0.5250,1) (1,1) };
\addplot[color = blue, mark = *, only marks, mark size = 2pt] coordinates {(.4,.2) (0.5250,1)};
\addplot [draw =none, fill=blue, fill opacity=0.2] coordinates { (.4,.2) (0.5250,.2) (0.5250,1) (1,1) } \closedcycle;
\addplot [domain = 0:.5, restrict y to domain = 0:1, samples = 200, color = red, very thick] { 0.02 /(1 - x - .5)}  node[midway,above left] {$ \displaystyle\frac{\rho}{C(t) - R}$};
\end{axis}
\end{tikzpicture}
\fi

    \caption{  \label{fig:area}
    Illustration of the upper bound (red) on extrinsic MMSE given by \eqref{eq:area_UB} for a rate $R = 0.5$ code on a BEC channel with erasure rate $t$ when $\rho \coloneqq \int_0^1 M(s)(1- M(s))  ds = 0.02$.
    This bound is sharp because, for any $t^* \in [0,1-R-\rho)$, there is a non-decreasing function $\tilde{M} \colon [0,1]  \to [0,1]$ that equals the upper bound at $t=t^*$ and also satisfies the area theorem (e.g., the area in blue is equal to $R$) and the integral constraint $\int_0^1 \tilde{M}(s)(1- \tilde{M}(s))  ds = \rho$.
    The function $\tilde{M}(t)$ is shown (blue) for $t^* = 0.4$ and is given by~\eqref{eq:Mtilde} with $u^* = 0.525$.
	See Example~\ref{exa:tightness_example} for more details. 
    }
\end{figure}

\begin{lem}\label{lem:area_bounds}
Consider a family of BMS channels satisfying Assumption~\ref{assumption:BMS} and suppose that the input distribution is uniform over a code with transitive symmetry and rate $R$. There exists a unique value $t_R \in (0,1)$ such that $C(t_R) =  R$. Furthermore, the extrinsic MMSE satisfies
\begin{alignat}{3}
 \!\!\! M(t)  &\le \frac{ \kappa(t) \int_t^1 M(s) (1- M(s))\, ds }{ C(t)- R}, &\quad& t \in [0,t_R) \label{eq:area_UB}\\
   \!\!\! M(t)  &\ge 1-  \frac{ \int_{t_R}^t M(s) (1- M(s))\, ds }{ \int_{t_R}^t \psi( R- C(s)) \, ds }, && t \in (t_R,1], \label{eq:area_LB}
\end{alignat}
where $\kappa(t) \coloneqq \sup_{s \in [t,1]} \cH'(s)$, $\psi(u) \coloneqq  1- \left( 1 -  2 h_b^{-1}\left( u \right) \right)^2$, and $h_b^{-1} \colon [0,1] \to [0,1/2]$ is the inverse of the binary entropy function restricted to the domain $[0,1/2]$.  The function $\psi$ is non-negative and strictly increasing.
Thus, the denominator in~\eqref{eq:area_LB} is strictly positive for $t \in (t_R, 1]$.  
\end{lem}

\begin{proof}
See Section~\ref{sec:proof_main}. \renewcommand{\qedsymbol}{}
\end{proof}

If the input is defined by an RM code, then we can combine 
Lemma~\ref{lem:area_bounds}
with Lemma~\ref{lem:integral_decay} to obtain bounds on the extrinsic MMSE that depend only on the code rate and the blocklength. Applying these 
bounds to a sequence of RM codes with strictly increasing blocklength and code rate converging to $R \in (0,1)$, shows that the extrinsic MMSE converges to a 0/1 step function that jumps at the unique $t_R \in (0,1)$ such that $C(t_R) = R$. 

\begin{rem} 
In Appendix~\ref{sec:MMSE_sequences}, we use an alternative approach to establish the limiting behavior of the extrinsic MMSE for a sequence of RM codes.
In particular, by using the comparisons in Lemma~\ref{lem:contiguity}, one can avoid the need for explicit bounds. 
While the proofs are not necessarily shorter or simpler, we believe that the approach may be of independent interest.
\end{rem}

\begin{example} \label{exa:tightness_example} To help explain the upper bound in Lemma~\ref{lem:area_bounds}, we describe an extrinsic MMSE curve that satisfies the bound with equality (see Figure~\ref{fig:area}).
This shows that \eqref{eq:area_UB} cannot be improved without imposing some additional constraints on the extrinsic MMSE function.  Consider the family of BECs with erasure probability equal to $t$ and recall that $\cH(t) = t$, $C(t)=1-t$, and $\kappa(t) = 1$.

For all $R \in (0,1)$, $\rho \in (0,1-R)$, and $t^* \in [0,1 - R - \rho)$, one can choose $u^* = 1 -R + \rho\,  (C(t^*)-R)/(C(t^*)-R - \rho)$.
If $u^*\leq 1$, then
\begin{align} \label{eq:Mtilde}
    \tilde{M}(t) = \begin{dcases}
    0 & \text{if } t \in [0,t^*)\\
    \frac{\rho}{C(t^*)-R}  & \text{if }t \in [t^*, u^*)\\
    1  & \text{if } t \in [u^*,1]
    \end{dcases}
\end{align}
defines a non-decreasing function $\tilde{M} \colon [0,1]\to [0,1]$ with $\tilde{M}(0)=0$ and $\tilde{M}(1)= 1$ with the properties: 
\begin{itemize}
    \item For the BEC, the MMSE equals the GEXIT function and the area theorem implies
    \[ R = \int_0^1 \tilde{M}(s) \cH'(t) \, ds = (u^*-t^*) \frac{\rho}{1-t^*-R} +   1 - u^* 
    . \]
    Using the stated $u^*$, a bit of algebra shows that the last expression simplifies to $R$.
\item From the definition, we see that
\begin{align*}
\rho \coloneqq& \int_0^1 \tilde{M}(s) (1- \tilde{M}(s)) \, ds \\ =& (u^*-t^*) \left(\frac{\rho}{1-t^*-R}\right)\left(1-\frac{\rho}{1-t^*-R}\right).
\end{align*}
Using the stated $u^*$, a bit of algebra shows that the last expression simplifies to $\rho$.
    \item We have $0\leq \tilde{M}(t) \leq 1$ because $C(t^*)-R=1-t^*-R\geq \rho$ follows from $t^*<1-R-\rho$.
    \item The upper bound in \eqref{eq:area_UB} is attained at the point $t^*$, i.e., $\tilde{M}(t^*) = \rho/(C(t^*) -R)$ because $\kappa(t)=1$ for the BEC family. 
\end{itemize}
\end{example}

\subsection{RM Codes Achieve Capacity on BMS Channels} \label{sec:construction}

We are now ready to prove the main result of the paper.  To show that RM codes achieve capacity for any particular BMS channel, we need to show that the channel can be embedded into a family of BMS channels satisfying Assumption~\ref{assumption:BMS}. To this end, we may consider the following construction.

\begin{defn}[Interpolated family of BMS channels] \label{def:interp_family} For a BMS channel  with input alphabet $\cX = \{\pm1\}$, output alphabet $\cY$, and capacity $C \in (0,1)$ an interpolated family of BMS channels  satisfying Assumption~\ref{assumption:BMS} is defined by the following steps: 
\begin{itemize}
    \item  Let $\cM^* \in (0,1)$ be the MMSE of the  channel associated with a uniform input distribution. %
    
    \item For $0 \le t < \cM^*$ the output is given by the original channel with probability $t/\cM^*$ and perfect knowledge of the input otherwise.
    This can be accomplished, for example, by adding the symbols $\pm\infty$ to $\cY$ and associating them with perfect knowledge of the inputs $\pm 1$, respectively.
    
    \item For $\cM^* \le t \le 1$ the output is given by the original channel with probability $ (1-t)/(1-\cM^*)$ and is equal to the erasure symbol   otherwise.
    
\end{itemize}
The MMSE function is $\cM(t) = t$ and the entropy function is 
    \begin{align}
                \cH(t)& = \begin{dcases}
         \frac{t }{ \cM^*}(1-C) ,  & t \in [0, \cM^*)\\
         \frac{t- \cM^*}{ 1- \cM^*} C  + 1-C, & t \in [\cM^*, 1]
        \end{dcases}.%
    \end{align}
The original BMS channel corresponds to the point $t = \cM^*$. 
\end{defn}

The next result bounds the extrinsic MMSE for RM codes transmitted over a BMS channel. 

\begin{lem} \label{lem:Mt_area_thm}
Consider a BMS channel with capacity $C \in (0,1)$. The extrinsic MMSE of an $\mathrm{RM}(r,m)$ code with rate $R =R(r,m)$ satisfies
\begin{alignat}{3}
\mmse(X_i \mid Y_{\sim i}) &\le \frac{\rho(m) }{ C- R}, &\qquad& R < C\\
\mmse(X_i \mid Y_{\sim i})  &\ge 1-  \frac{ (1- C) \rho(m)}{ \psi(1-C)  \Psi(R - C) }, && R > C
\end{alignat}
for all $i \in [N]$ where $\rho(m) \coloneqq (6 \ln(m) + 34)/(5 \sqrt{m})$ and $\Psi(u) \coloneqq \int_0^u \psi(v) \, dv$ with $\psi$ given in Lemma~\ref{lem:area_bounds}.
\end{lem}
\begin{proof}
Let $\cM^*$ be the MMSE of the given BMS channel when the input is uniformly distributed on $\{\pm 1\}$.
Let $\cH(t)$ be the entropy function associated with the family of channels in Definition~\ref{def:interp_family}, and let $t_R$ be the unique value such that $1- \cH(t_R) = R$.

For the upper bound on the extrinsic MMSE, observe that if $R< C$ then $t_R > \cM^*$. Combining the bound in \eqref{eq:area_UB}, evaluated at  $t = \cM^*$, with the bound on $\int_0^1 M(s) (1- M(s)) \, ds$ in Lemma~\ref{lem:integral_decay}, we see that
\begin{align}
\mmse(X_i \mid Y_{\sim i}) &\le \frac{ \kappa(\cM^*)  \rho(m) }{  C- R} \le \frac{ C \rho(m) }{ (1-\cM^*) ( C- R)}
\end{align}
where the second step holds because  $\cH'(s) = C/(1- \cM^*)$ for $s \ge \cM^*$.
Finally, we observe that the claimed expression follows from~$C = 1-\cH(\cM^*) \geq 1- \cM(\cM^*) = 1 - \cM^*$ which is implied by~\eqref{eq:cH_to_cM}.

For the lower bound on the extrinsic MMSE, observe that, if $R > C$, then $t_R < M^*$. Combining the bound in \eqref{eq:area_LB}, evaluated at $t = \cM^*$, with  Lemma~\ref{lem:integral_decay}, we see that
\begin{align}
\mmse(X_i \mid Y_{\sim i}) &\ge 1 -  \frac{  \rho(m) }{ \int_{t_R}^{\cM^*} \psi( R -1 + s (1-C)/\cM^*  ) \, ds }.%
\end{align}
To simplify the integral, recall that $t_R$ is the unique value in $[0,1]$ such that $\cH(t_R) = 1 - R$. In particular, 
\[
R = 1 - \cH(t_R)  = 1 - \frac{t_R }{ \cM^*}(1 - C)  \;\; \iff \;\;  t_R = \frac{ \cM^*(1-R) }{ 1- C}. 
\]
Making the change of variables $u =R - 1  +  s (1- C) /\cM^*$ and noting the boundary conditions 
\begin{align*}
    s = \cM^* \quad &\implies \quad u   = R- C \\
        s = t_R \quad &\implies \quad u  = 0
\end{align*}
yields
\begin{align}
    \int_{t_R}^{\cM^*} & \psi( R -1 + s (1-C)/\cM^*  ) \, ds \\ & = \frac{ \cM^*}{ 1- C}  \int_{0}^{R-C} \psi( u ) \, du = \frac{ \cM^*}{ 1- C} \Psi(R - C). 
\end{align}

Finally, we simplify the bound to avoid dependence on $\cM^*$.
Notice that~\eqref{eq:cH_to_cM} in Section~\ref{sec:BMS_families} implies
\begin{align}
 \cM^* \le 1-C \le h_b\left( \frac{ 1- \sqrt{ 1- \cM^*}}{2} \right),
 \end{align}
 where the last inequality is equivalent to $\psi( 1- C ) \le \cM^*$. 
\end{proof}

We now state main result of the paper,  which provides non-asymptotic bounds on the BER under bit-MAP decoding for an RM code over a BMS channel. These bounds depend only on three quantities:  the capacity of the channel, the difference between the capacity and the code rate, and the blocklength. Evaluating these bounds in the limit of increasing blocklength, it follows that RM codes achieve capacity on any BMS channel. 

\begin{thm}\label{lem:BER_bound}
Consider a BMS channel with capacity $C \in (0,1)$. For every $\mathrm{RM}(r,m)$ code whose rate satisfies $R(r,m) < C$, the bit-error rate under bit-MAP decoding satisfies 
\begin{align}
\BER(X_i \mid \vY ) \le \frac{ \frac{1}{2} \rho(m)}{  C - R(r,m) }, 
\end{align}
for all $i \in [N]$ where $\rho(m) \coloneqq (6 \ln(m) + 34)/(5\sqrt{m})$. In particular, for every $R \in [0,C)$ there exists a sequence of RM codes with increasing blocklength and rate converging to $R$ such that the BER under bit-MAP decoding converges to zero.

Conversely, if $R(r,m) > C$ then 
\begin{align}
& \!\!\BER( X_i \mid \vY ) \\ &\;\ge \BER(X_i \mid Y_i ) -(1- C) \sqrt{  \frac{ \ln(2)   \rho(m)}{2 \psi(1-C)  \Psi(R(r,m) - C) } },\!%
\end{align}
for all $i \in [N]$ where $\Psi(u) \coloneqq \int_0^u \psi(v) \, dv$ with $\psi$ given in Lemma~\ref{lem:area_bounds}. In particular, for every $R \in (C,1]$ and every sequence of RM codes with increasing blocklength and rate converging to $R$, the BER under bit-MAP decoding converges to the bit-error rate associated with a single use of the channel.  
\end{thm}
\begin{proof}
The upper bound on the BER follows from combining the upper bound on the extrinsic MMSE in Lemma~\ref{lem:Mt_area_thm} with the relationship between the BER and MMSE in Lemma~\ref{lem:mmse_ber}, and then noting that $\BER(X_i \mid \vY) \le \BER(X_i \mid Y_{\sim i})$. The lower bound on the BER follows from combining the lower bound on the extrinsic MMSE in Lemma~\ref{lem:Mt_area_thm} with the relationship between the BER and MMSE in Lemma~\ref{lem:BER_two_look}.

From~\cite[Remark~24]{Kudekar-it17}, we know that for any $R\in (0,1)$, there is a sequence of RM codes with strictly increasing $m$ whose rate converges to $R$. The construction of this sequence is also discussed in Section~\ref{subsec:rm_codes} for completeness.
If, as $m\to \infty$, the code rate approaches any fixed $R<C$, then we see that the bit-error probability vanishes because $\rho(m) \to 0$.
\end{proof}

\section{Proofs}
\label{sec:proofs}

In this section, we collect proofs that have been removed from the main text due to length or importance.

\subsection{Background}
\label{sec:proof_background}

\begin{proof}[\hypertarget{prf:mmse_ber}Proof of Lemma~\ref{lem:mmse_ber}]
\label{prf:lem:mmse_ber}
Starting with the definition of the MAP decision rule, we can write
\begin{align} \label{eq:ber_mmse}
    \BER(X \mid Y) & \coloneqq \pr{ X  \ne \phi(Y)} \\
    & = \E\big[ \min\big \{ \pr{ X =1 \mid Y} , \pr{X = -1  \mid Y }\big \}  \big]\\
& = \frac{1}{2} - \frac{1}{2} \E\left[ \max \big\{\E[X|Y], - \E[X|Y] \big \} \right] \\
    &= \frac{1}{2} \left(1- \|\ex{X|Y}\|_1\right),
\end{align}
where the third step holds because $\pr{X = 1\mid Y} = \frac{1}{2}(1+ \ex{ X \mid Y})$ almost surely. For comparison, recall that the MMSE is given by
\begin{align}
\mmse(X \mid Y) & =1 - \| \E[X \mid Y] \|_2^2.    
\end{align}

The upper bound on the BER follows from the inequality $1 - u \le 1 - u^2$ for $0 \le u \le 1$,  which gives
\begin{align}
    \BER(X \mid Y) \le \frac{1}{2} \mmse(X \mid Y),
\end{align}
with equality if and only if  $\E[X \mid Y] \in \{ 0, \pm 1\}$ (i.e., the channel is equivalent to an erasure channel).

The lower bound on the BER follows from  Lyapunov's inequality: 
\begin{align*}
1 - 2 & \BER(X \mid Y)  = \| \ex{X \mid Y}\|_1 \\ &\le \| \ex{X \mid Y}\|_2 = \sqrt{ 1 - \mmse(X \mid Y)} 
\end{align*}

and thus 
\begin{align}
    \BER(X \mid Y) \ge \frac{ 1 -  \sqrt{ 1-  \mmse(X \mid Y)} }{2}, 
\end{align}
with equality if and only if $\ex{X \mid Y}$  has constant magnitude (i.e., the channel is equivalent to a BSC).
Thus, for a sequence of observations, the bit-error probability approaches 0 (respectively $\frac{1}{2}$) if and only if the MMSE approaches 0 (respectively 1).
\end{proof}

\subsection{Preliminary Results}
\label{sec:proof_preliminary}

\begin{proof}[\hypertarget{prf:twolook}Proof of Lemma~\ref{lem:twolook}]
We begin with the proof of \eqref{eq:two_look_formula_mu}.  Recall that the input  $X \in \{\pm 1\}$ has mean $\mu$ and $Y$ is an observation of $X$ through a BMS channel. We can transform the problem into one with a uniform prior using a symmetrization argument. Specifically, let 
\begin{align}
X_\mathrm{u} = V X,
\end{align}
where  $V \in \{\pm 1\}$ is a uniform binary variable that is independent of $(X,Y)$, and let $Y_\mathrm{u}$ be an observation of $X_\mathrm{u}$ through the same BMS channel such that $(X,V) - X_\mathrm{u} - Y_\mathrm{u}$  is a Markov chain. Notice that under this specification,  the symmetrized input $X_\mathrm{u}$ is uniformly distributed and the symmetrized input-output pair $(X_\mathrm{u}, Y_\mathrm{u})$ is independent of the original input $X$.  

The mutual information $I(X_\mathrm{u}; Y_\mathrm{u})$ can be decomposed according to
\begin{align}
I(X_\mathrm{u} ;Y_\mathrm{u}) & = I(V, X_\mathrm{u} ; Y_\mathrm{u})
= I(V ;  Y_\mathrm{u}) + I(X_\mathrm{u} ; Y_\mathrm{u} \mid V),  \label{eq:twolook_MI}
\end{align}
where the first step holds because of the Markov structure and the second step is the chain rule for mutual information. Using  $I(X_\mathrm{u} ;Y_\mathrm{u}) = H(X_\mathrm{u}) - H(X_\mathrm{u} \mid Y_\mathrm{u})$ and $ I(V ;  Y_\mathrm{u})  = H(V) - H(V \mid Y_\mathrm{u}) $, where $H(X_\mathrm{u}) = H(V) = 1$, and applying the series expansion of binary entropy in \eqref{eq:HXY_series1} yields 
\begin{align}
I(X_\mathrm{u} ;Y_\mathrm{u}) &%
=  \sum_{k \in \mathbb{N}} c_k \| \E[ X_\mathrm{u} \mid Y_\mathrm{u}] \|_{2k}^{2k},  \\
I(V ;Y_\mathrm{u})  %
&=\sum_{k \in \mathbb{N}} c_k \| \E[ V \mid Y_\mathrm{u}] \|_{2k}^{2k}.%
\end{align}
The second expansion can be simplified further by noting that  $V = X X_\mathrm{u}$ where $X$ is independent of $(X_\mathrm{u}, Y_\mathrm{u})$, and thus the conditional expectation decouples as the product of expectations: 
\begin{align}
\E[ V \mid Y_\mathrm{u}]  =  \E[ X X_\mathrm{u} \mid Y_\mathrm{u}]  = \mu \,  \E[X_\mathrm{u} \mid Y_\mathrm{u}] , \quad  \text{almost surely.}
\end{align}
Plugging this expression back into the expansion $I(V ;Y_\mathrm{u})$,  recalling that $q_k \coloneqq \| \E[ X_\mathrm{u} \mid Y_\mathrm{u}] \|_{2k}^{2k}$, and then rearranging the terms in \eqref{eq:twolook_MI} gives 
\begin{align}
I(X_\mathrm{u} ; Y_\mathrm{u} \mid V)  =  \sum_{k \in \mathbb{N}} c_k  q_k (1 - \mu^{2k}). 
\end{align}
In view of $I(X_\mathrm{u} ; Y_\mathrm{u} \mid V) = H(X_\mathrm{u} \mid V) - H(X_\mathrm{u} \mid V, Y_\mathrm{u})$ where $H(X_\mathrm{u} \mid V) = H(X) = h_b((1+\mu)/2)$, this expansion  of $I(X_\mathrm{u}; Y_\mathrm{u} \mid V)$ can be stated in equivalently in terms of conditional entropy according to 
\begin{align}
H(X_\mathrm{u} \mid  Y_\mathrm{u} , V)  =  \sum_{k \in \mathbb{N}} c_k (1- q_k)  (1 - \mu^{2k}).
\end{align}
Notice that the RHS is precisely the formula we are trying to prove. The LHS can be viewed as the entropy of a symmetrized binary-input channel where the input is flipped with probability one half and the status of whether it was flipped (i.e., the variable $V$) is provided at the output of the channel.
Since $Y$ is an observation of $X$ through a symmetric channel, the distribution of the channel is unaffected by this symmetrization procedure and thus $H(X_\mathrm{u} \mid Y_\mathrm{u} , V) = H(X \mid Y)$. This concludes the proof of \eqref{eq:two_look_formula_mu}.

Since~\eqref{eq:two_look_formula_mu} holds for an arbitrary prior on $X$, the proof of~\eqref{eq:two_look_formula} follows as a direct consequence of~\eqref{eq:two_look_formula_mu}.
The $U-X-Y$ Markov chain condition implies that, for any $u$ in the support of $U$, conditioning on $U=u$ only changes the prior on $X$. 
Thus, we can use \eqref{eq:two_look_formula_mu} to write
\begin{align}
    H(X \mid Y , U = u)  = \sum_{k \in \mathbb{N}} c_k (1- q_k)  (1 - \E[ X \mid U = u]^{2k}).
\end{align}
 Averaging both sides over the distribution of $U$ and interchanging the expectation with the summation (which is justified by the uniform convergence of the sum) gives the desired result.
\end{proof}

\begin{proof}[\hypertarget{prf:BER_two_look}Proof of Lemma~{\ref{lem:BER_two_look}}]

For the BMS channel from $X$ to $Y$  let $\{q_k\}_{k \in \mathbb{N}}$ be the sequence given in Definition~\ref{defn:qk}. We proceed by expanding the conditional mutual information $I(X;U \mid Y) = H(X \mid Y) - H(X \mid Y ,U) = H(X) - H(X \mid Y,U)$ two different ways.  Starting with the entropy expansion in  \eqref{eq:two_look_formula} we can write
\begin{alignat}{3}
   I(X;Y \mid U) & = \sum_{k \in \mathbb{N}} \ce_k ( 1- q_k) ( \|\E[X \mid U]\|_{2k}^{2k} - \mu^{2k}) \\
     & \le    \sum_{k \in \mathbb{N}} \ce_k ( 1- q_k) \|\E[X \mid U]\|_{2}^{2}, \\
     &  = (1-C)   (1- \mmse(X \mid U)). %
\end{alignat}
This inequality holds because $|\E[ X \mid U]| \le 1$ almost surely and thus
\[
 \|\E[X \mid U]\|_{2k}^{2k} - \mu^{2k} \le \|\E[X \mid U]\|_{2}^{2} - \mu^{2k} \le \|\E[X \mid U]\|_{2}^{2}.
\]
The last step follows from  \eqref{eq:HXY_series1}, which implies that $\sum_{k \in \mathbb{N}} \ce_k (1- q_k) = 1- C$, and \eqref{eq:mmse_X_given_Y}.

Alternatively, starting with the entropy expansion in \eqref{eq:HXY_series1} and then noting that all the terms in the expansion are non-negative (by Lemma~\ref{lem:convex_order} and the fact that $Y - (Y,U) - X$ is a Markov chain)  gives
\begin{alignat}{3}
   I(X; U \mid Y) & = \sum_{k \in \mathbb{N}} \ce_k  ( \|\E[X \mid Y, U]\|_{2k}^{2k} - \|\E[X \mid Y]\|_{2k}^{2k}) \\
     & \ge   c_1  ( \|\E[X \mid Y, U]\|_{2}^{2} - \|\E[X \mid Y]\|_{2}^{2}) \\
     &  = c_1  \|\E[X \mid Y,  U] - \E[X \mid U]\|_{2}^{2} \, ,
\end{alignat}
where the last step follows from \eqref{eq:L2diffa}. 
Combining these  upper and lower bounds on the mutual information yields
\begin{align}
     \|\E[X \mid Y,  U] &- \E[X \mid U]\|_{2}^{2} \\ &\le \frac{(1-C)    (1- \mmse(X \mid U))}{c_1}. \label{eq:MMSE_diff_MI}
\end{align}

To prove the desired inequality for the BER,  we use the identity  $\BER(X  \mid U) = \frac{1}{2} \big(1 - \| \E[ X \mid U] \|_1 \big)$, which is derived in the proof of Lemma~\ref{lem:mmse_ber}, to see that
\begin{align}
2 \big( \BER( X \mid U)   & - \,  \BER(X \mid Y, U) \big) \\
&=  \| \E[X  \mid Y , U] \|_1 - \|\E[ X  \mid Y ] \|_1 \\
 &\le  \| \E[X  \mid Y , U]- \E[X  \mid Y] \|_1 \\
 & \le \| \E[X  \mid Y , U]- \E[X  \mid Y] \|_2,
\end{align}
where the second step is the reverse triangle inequality and the third step is Lyapunov's inequality.  Combining this inequality with \eqref{eq:MMSE_diff_MI} and recalling that $c_1 = 1/(2 \ln 2)$ completes the proof. 
\end{proof}

For the next few results, the following definition and lemma will be useful.

\begin{defn}[Absolutely Continuous]
Consider  \label{def:abs_cont} a real interval $[a,b]$ and a function $f\colon [a,b] \to \mathbb{R}$.
Then, $f$ is absolutely continuous on $[a,b]$ if, for every $\epsilon >0$, there is a $\delta>0$ such that, for any sequence of disjoint intervals $\{[a_k,b_k]\}_{k\in \mathbb{N}}$ with $a\leq a_k \leq b_k \leq b$, we have
\[ \sum_{k\in \mathbb{N}} |b_k-a_k| < \delta \; \implies \; \sum_{k\in \mathbb{N}} |f(b_k)-f(a_k)| < \epsilon. \]
This definition is important because the fundamental theorem of calculus for the Lebesgue integral states that, if $f$ is absolutely continuous, then $f$ is differentiable almost everywhere on $[a,b]$ and, for all $c\in [a,b]$, the Lebesgue integral of its derivative satisfies
\[ f(c) = f(a) + \int_a^c f'(x) dx. \]
\end{defn}

\begin{lem}
Consider a function $f\colon [a,b] \to \mathbb{R}$ that is absolutely continuous on $[a,b]$ and another function $g\colon [a,b] \to \mathbb{R}$.
Then, if there is a constant $\gamma < \infty$ such that $|g(y)-g(x)| \leq \gamma |f(y)-f(x)|$ for all $x,y\in [a,b]$, then $g$ is absolutely continuous on $[a,b]$. \label{lem:abs_cont}
\end{lem}

\begin{proof}
For any $\epsilon' >0 $, we use the absolute continuity of $f$ with $\epsilon= \epsilon' / \gamma$ to obtain the desired $\delta>0$.
Thus, we find that, for any sequence of disjoint intervals $\{[a_k,b_k]\}_{k\in \mathbb{N}}$ with $a\leq a_k \leq b_k \leq b$, we find that
\[ \sum_{k\in \mathbb{N}} |b_k-a_k| < \delta \]
implies
\[ \sum_{k\in \mathbb{N}} |g(b_k)-g(a_k)| \leq \gamma \sum_{k\in \mathbb{N}} |f(b_k)-f(a_k)| < \epsilon'. \qedhere \]
\end{proof}

Now, we provide a proof for  Lemma~{\ref{lem:H_M_q_Hu}}. While the arguments for parts $(i)$ and $(ii)$ are self-contained, the proof of part $(iii)$ depends on some further results (Lemmas~\ref{lem:gexit_expansion} and~\ref{lem:H_mu}) whose proofs appear below.
We emphasize that, although the proofs of Lemmas~\ref{lem:gexit_expansion} and~\ref{lem:H_mu} depend on parts $(i)$ and $(ii)$ of Lemma~{\ref{lem:H_M_q_Hu}}, they do not depend on part $(iii)$.
Thus, the argument is not circular. 

\begin{proof}[\hypertarget{prf:lem:H_M_q_Hu}Proof of Lemma~{\ref{lem:H_M_q_Hu}} $(i)$ and $(ii)$]
By assumption, $\{W(t) : 0 \le t \le 1\}$ is a family of BMS channels that is ordered by degradation according to Definition~\ref{defn:BS_metrics}.
Since $\cH(t) = H(X_\mathrm{u} \mid Y_\mathrm{u} (t) )$ is defined with respect to an observation of a uniform input, the entropy formula given in $(i)$ follows from~\eqref{eq:two_look_formula_mu} with $\mu=0$.
Likewise, $\cM(t) = \mmse(X_\mathrm{u} \mid Y_\mathrm{u} (t) )$ is defined with respect to an observation of a uniform input and the formula given in $(i)$ follows from combining~\eqref{eq:mmse_X_given_Y} and~\eqref{eq:qk}.

For $(ii)$ and all $0\leq s < t \leq 1$, we first observe that

$q_k (s) \geq q_k (t)$ 

follows directly from the degradation ordering of the channel family and Lemma~\ref{lem:convex_order}.
Next, we observe from $(i)$ that
\begin{align}
\cH(t)-\cH(s)
&= \sum_{k\in \mathbb{N}} c_k \big( q_k (s) - q_k (t) \big) \\
&\geq  c_k \big( q_k (s) - q_k (t) \big) \quad \forall k\in \mathbb{N}, \label{eq:qk_diff_ub}
\end{align}
where the second step follows from the fact that each term in the sum is non-negative because $c_k \geq 0$ and $q_k (s) \geq q_k (t)$.
Since $\cH(t)$ is absolutely continuous on $[0,1]$ by assumption, we can combine~\eqref{eq:qk_diff_ub} and Lemma~\ref{lem:abs_cont} to show that $q_k(t)$ is also absolutely continuous on $[0,1]$.
Together, monotonicity and absolute continuity imply that $q_k ' (t) \leq 0$ when it exists.
Since $W(0)$ is a perfect channel (i.e., $\ex{X_\mathrm{u} \mid Y_\mathrm{u} (0)} = X_\mathrm{u}$) and $W(1)$ is a useless channel (i.e., $\ex{X_\mathrm{u} \mid Y_\mathrm{u} (0)} = 0$), it follows that $q_k (0) = 1$ and $q_k (1) = 0$.
\end{proof}

\begin{proof}[Proof of Lemma~{\ref{lem:H_M_q_Hu}} $(iii)$]
Statement $(iii)$ follows from Lemma~\ref{lem:H_mu} by choosing $\mu = 0$.
\end{proof}

\begin{proof}[\hypertarget{prf:gexit_expansion}Proof of Lemma~{\ref{lem:gexit_expansion}}]
Consider $H(\vX \mid Y_i(s), V(t) )$ where $V(t)$ is any side information random variable
parameterized by $t\in [0,1]$ that is ordered by degradation and conditionally independent of $Y_i$ given $X_i$.
The following chain rule plus derivative trick was introduced in~\cite{Measson-it08} for the BEC. Starting with the chain rule for entropy and then using the fact that $(V(t), X_{\sim i}) - X_i - Y_i(s)$ is a Markov chain (because the channel is memoryless) we can write
\begin{align}
 H&(\vX \mid Y_i(s), V(t) ) \\
&= H(X_i \mid Y_i(s), V(t) ) + H(X_{\sim i} \mid Y_i(s), V(t), X_i )  \\
&= H(X_i \mid Y_i(s), V(t) ) + H(X_{\sim i} \mid V(t), X_i ).
\end{align}
As the second term on the RHS does not depend on $s$, we see that the $s$-derivative can be expressed as
\begin{align*}
\frac{\partial}{\partial s} H(\vX \mid Y_i(s), V(t) )
&= \frac{\partial}{\partial s} H(X_i \mid Y_i(s), V(t) ).
\end{align*}

From statement $(ii)$ of Lemma~\ref{lem:H_M_q_Hu}, we know that, for all $k\in \mathbb{N}$, $q_k (s)$ is absolutely continuous and $-q_k ' (t)$ is non-negative almost everywhere.
Now, we can use $1- q_k (s) = \int_0^s -q_k '(u) du$ (which follows from $q_k (0) = 1$) to rewrite~\eqref{eq:Hst_for_gexit}.
For all $t\in [0,1]$, this gives
\begin{align}
    \!\!H&(X_i \mid Y_i(s), V(t) ) \\
    & =\sum_{k\in\mathbb{N}} \int_0^s \!\! \ce_k \left( - q_k'(u)  \right) \big( \underbrace{  1 - \| \E[ X_i \mid V(t) ]\|_{2k}^{2k} }_{\nu_k (t)} \big) du. \label{eq:HstV}
\end{align}
where we use $\nu_k (t) \coloneqq 1 - \| \E[ X_i \mid V(t) ]\|_{2k}^{2k}$ and neglect the index $i$ to lighten notation.
Since, for all $k\in \mathbb{N}$, the integrand is non-negative almost everywhere for $u\in[0,1]$, we can apply Tonelli's Theorem~\cite{Royden-2010} (with respect to counting measure for $k$ and Lebesgue measure for $u$) to interchange the sum and integral so that, for all $t\in[0,1]$, we see that
\[ F (u,t) \coloneqq \sum_{k\in\mathbb{N}}  \ce_k \left( - q_k'(u)  \right) \nu_k(t)
\]
exists for almost all $u \in [0,1]$. In addition, for all $s,t\in[0,1]$, it follows that $F (u,t)$ satisfies
\begin{align}
    \int_0^s F (u,t)\, du = H(X_i \mid Y_i(s), V(t) ). 
\end{align}
This proves that, for all $t\in [0,1]$, $\frac{\partial}{\partial s} H(X_i \mid Y_i(s), V(t) )$ exists for almost all $s\in[0,1]$ and is almost everywhere equal to $F (s,t)$.

Notice that if $\nu_k(t) = 1$ for all $k$  (i.e.,  $X_i$ is uniformly distributed and the $V(t)$ is independent of $X_i$) then the entropy in \eqref{eq:HstV} is equal to the function $\cH(s)$. From the assumption that $\cH(s)$ is absolutely continuous and the arguments given above it follows that there exists a set $K\subseteq [0,1]$ of measure 0 such that for all $u \in K^c$, the derivative $\cH'(u)$  exists,  is finite, and is given by %
\begin{align}
\cH'(u) = \sum_{k\in\mathbb{N}}  \ce_k \left( - q_k'(u)  \right).
\end{align}

Now, we will use the above results to argue that $F (t,t)$ equals  $\frac{\partial}{\partial s} H(X_i \mid Y_i(s), V(t) )|_{s=t}$ for almost all $t\in [0,1]$.
The issue here is that we have not ruled out the possibility that $\frac{\partial}{\partial s} H(X_i \mid Y_i(s), V(t) )$ does not exist whenever $s=t$.
To handle this detail, let us define $J_k \subset [0,1]$ to be the set of points where $q_k ' (u)$ does not exist and observe that $J_k$ has a Lebesgue measure of 0.
By the countable subadditivity of measure, it follows that $J = \cup_{k \in \mathbb{N}} J_k$ also has Lebesgue measure 0.
Hence, for all $u \in J^c$, every element of the sequence $\{-q_k ' (u)\}_{k \in \mathbb{N}}$ is well-defined and non-negative.
Let $\bar{J} = J \cup K$ and observe that $\bar{J}$ still has measure 0.
Also, if $u \in \bar{J}^c$, then the sum in $F (u,t)$ converges to a finite number when the sequence $\nu_k (t) = 1$ for all $k\in \mathbb{N}$.
But, since we always have $\nu_k (t) \in [0,1]$, the sum must also converge to a finite number for any $\nu_k (t)$ sequence.
Thus, we see that, for all $u\in \bar{J}^c$ and all $t\in [0,1]$, the sum in $F (u,t)$ is well-defined and finite.
Integrating this sum over $s$ shows that $F (s,t)$ must equal $\frac{\partial}{\partial s} H(X_i \mid Y_i(s), V(t) )$ for all $s \in \bar{J}^c$ and $t\in [0,1]$.
Thus, $\frac{\partial}{\partial s} H(X_i \mid Y_i(s), V(t) )|_{s=t}$ is almost everywhere equal to $F (t,t)$.

Finally, we consider the integrability of $F (t,t)$.
For $t\in \bar{J}^c$, define the sequence of functions $\{f_n\}_{n \in \mathbb{N}}$ according to 
\begin{align}
    f_n(t) \coloneqq \sum_{k=1}^n \ce_k (- q'_k(t)) \nu_k(t). 
\end{align}
Each $f_n$ is measurable  because $\nu_k (t)$ is measurable by monotonicity and measurability is preserved under finite sums and products. Furthermore, by the monotone convergence theorem, $f_n(t)$ converges pointwise to $F (t,t)$ for all $t \in \bar{J}^c$.  Finally, because $0 \le \nu_k(t) \le 1$ the sequence is dominated in the sense that $|f_n(t)| \le |\cH'(t)|$ holds almost everywhere (because $\bar{J}$ has measure zero). Thus, we can apply the dominated convergence theorem to conclude that the limit $F (t,t)$ is integrable.

Since $Y_{\sim i}$ is conditionally independent of $Y_i$ given $X_i$, we can choose $V(t) = Y_{\sim i} (t)$ to establish~\eqref{eq:GEXIT_expansion}.
Similarly, since $(Y_{\sim i},U(t)) - \vX - X_i - Y_i$ forms a Markov chain, we can establish~\eqref{eq:GEXIT_plus_expansion} by selecting $V(t) = (Y_{\sim i} (t),U(t))$. Finally, letting $V(t)$ be almost surely constant with $\ex{X_i \mid V(t)} = \mu \in [-1,1]$, we see that~\eqref{eq:H_mu_p} holds.

\end{proof}

\subsection{Main Results}
\label{sec:proof_main}

\begin{proof}[\hypertarget{prf:Gij}Proof of Lemma~{\ref{lem:Gij}}]
To provide some context, let us first recall the setting of the area theorem for the GEXIT function.  For any input distribution $\vX \in \{\pm 1\}^N$, the law of the total derivative gives
\begin{align}
\frac{d}{dt} H(\vX \mid \vY(t))  = \sum_{i \in [N]} G_i(t).
\end{align}
From the assumed properties of the channel family, the conditional entropy is equal to $0$ at $t=0$ and $H(\vX)$ at $t =1$, and so the integral of the above expression is equal to  $H(\vX)$. 

The desired expression in \eqref{eq:single_term_sym} differs from the setting of area theorem in two ways: 1) the $i$-th term in the summation is omitted and 2) the augmented GEXIT function $G_{i}^j(t)$ treats the $j$-th channel differently from the others. Our approach is to find a suitable definition for the augmented GEXIT function in the case $i=j$ such that the summation over all $i \in [N]$ can be expressed as the total derivative of a conditional entropy term. In particular, we will use the definition
\begin{align}
G_{i}^i (t)  \coloneqq \frac{\partial}{\partial s}  H(\vX \mid Y_i(s), Y_{\sim i}(t), Y'_i(s))\Big \vert_{s =t} ,
\end{align}
where $Y'_i(s)$ is resampled observation of the $i$ input. By the law of the total derivative and the fact that $Y_i(s)$ and $Y'_i(s)$ are identically distributed, this term can be expressed as twice the partial derivative with respect to one observation.
This gives 
\begin{align}
  G_{i}^i (t) = 2\,   \frac{\partial}{\partial s}  H(\vX \mid Y_i(s), Y_{\sim i}(t), Y'_i(t))\Big \vert_{s =t} \label{eq:Gii_alt} 
\end{align}
and the existence and expansion of this derivative follows from applying Lemma~\ref{lem:gexit_expansion} with $U(t)=Y_i ' (t)$.

Starting with~\eqref{eq:single_term_sym}, we can now add and subtract the terms with $i=j$ to obtain
\begin{align}
  \sum_{i,j \in [N] \, : \, i \ne j}    (\gexit_i(t)- \gexit_i^{j}(t) )   = &  \sum_{j \in [N]} \Bigg( \sum_{i \in [N] }  (  \gexit_i(t)- \gexit_i^{j}(t) ) \Bigg)  \\ &\;\; +   \sum_{i \in [N] }  ( \gexit^i_i(t)- \gexit_i(t) ) . \label{eq:Gij_diff_dcomp}
\end{align}
For each $j \in [N]$, one finds that the first summation over $i$ on the RHS is the total derivative of the difference in entropy terms given by 
\begin{align}
\sum_{i \in [N] }  ( \gexit_i(t) & - \gexit_i^{j}(t) )  \\ &  =   \frac{d}{ d t} \Big( H(\vX \mid \vY(t))  -  H(\vX \mid \vY(t) , Y_j'(t)) \Big) .
\end{align}

From the assumed properties of the channel family,  both of the conditional entropy terms equal 0 at $t =0$ and $H(\vX)$ at $t = 1$.
So, the integral of this term  vanishes.

The proof  has now been reduced to finding a suitable bound for the integral of the second term in \eqref{eq:Gij_diff_dcomp}, which contains only a single summation. Using the series expansions implied by Lemma~\ref{lem:gexit_expansion} for $G_i (t)$ and~\eqref{eq:Gii_alt}, we can write
\begin{align}
   G_i^i(t) - G_i(t) &= \sum_{k \in \mathbb{N} } c_k (- q'_k(t)) \Big( 1 +    \| \ex{ X_i \mid   Y_{\sim i}(t) } \|_{2k}^{2k}  \\ & \quad \qquad \;\; -  2 \| \ex{ X_i \mid Y'_i(t) , Y_{\sim i}(t) } \|_{2k}^{2k} \Big) .
\end{align}
The $-q_k '(t)$ terms are non-negative by Lemma~\ref{lem:H_M_q_Hu} and $
    \| \ex{ X_i \mid Y_{\sim i}(t) } \|_{2k} \le \| \ex{ X_i \mid Y'_i(t) ,  Y_{\sim i}(t) } \|_{2k}$,  
for $k\in \mathbb{N}$, by Jensen's inequality.
Thus, we find that
\begin{align}
      1+ \| \ex{ X_i \mid   Y_{\sim i}(t) } \|_{2k}^{2k} -  2 \| \ex{ X_i \mid Y'_i(t) , Y_{\sim i}(t) } \|_{2k}^{2k} \le 1
\end{align}
and this implies that
\begin{align}
    G_i^i(t) - G_i(t) & \le \sum_{k \in \mathbb{N} } c_k (- q'_k(t))  = \cH'(t),
\end{align}
where the sum equals the  derivative of the entropy function, $\cH'(t)$,  by Lemma~\ref{lem:H_M_q_Hu}.
Since $\cH(0) = 0$ and $\cH(1)=1$ by the assumed properties of the channel family, we have
\begin{align}
    \int_0^1 G_i^i(t) - G_i(t) \, dt \le 1. 
\end{align}
Summing this expression over $i \in [N]$ completes the proof of \eqref{eq:single_term_sym}.

If the input distribution has doubly-transitive symmetry, then $\gexit_{i}^j = \gexit_k^\ell$ for all $i,j,k,\ell \in [N]$ with $i \ne j$ and $k \ne \ell$.
This implies that, in~\eqref{eq:single_term_sym}, all terms in the sum are equal.
Thus, we can divide by $N(N-1)$ (i.e., the total number of terms) to see that each term satisfies~\eqref{eq:single_term_dbl_sym}.
\end{proof}

\begin{defn} \label{defn:H_mu}
Let $\{W(t) : 0 \le t \le 1\}$ be a family of BMS channels that is ordered by degradation according to Definition~\ref{defn:BS_metrics} and
let $\{q_k(t)\}_{k \in \mathbb{N}}$ be the sequence given in Definition~\ref{defn:qk}.
For $\mu \in [-1,1]$ and $t\in [0,1]$, let us define
\begin{align} \label{eq:H_mu}
 \cH_{\mu}(t) \coloneqq \sum_{k=1}^{\infty} c_{k} ( 1 -q_{k}(t))( 1 - \mu^{2k}).
\end{align}
If $X\in \{\pm 1\}$ is a random variable satisfying $\ex{X}=\mu \in [-1,1]$ and $Y(t)$ is an observation of $X$ through $W(t)$, then~\eqref{eq:two_look_formula_mu} implies that $\cH_{\mu} (t) = H(X \mid Y(t))$.
We also note that $\cH_0 (t) = \cH (t)$ by Lemma~\ref{lem:H_M_q_Hu}.  
\end{defn}

\begin{lem} \label{lem:H_mu}
For $\mu \in [-1,1]$ and $t\in [0,1]$, the function $\cH_{\mu} (t)$ from Definition~\ref{defn:H_mu} is non-decreasing in $t$, absolutely continuous in $t$, and non-increasing in $\mu^2$. Its derivative with respect to $t$, denoted by $\cH_\mu ' (t)$, exists almost everywhere and is almost everywhere equal to
\begin{align} \label{eq:H_mu_p}
\cH_{\mu} ' (t) = \sum_{k\in \mathbb{N}} c_k (- q_k ' (t)) (1-\mu^{2k}).
\end{align}
In addition,  $\cH_\mu ' (t)$ is non-negative and non-increasing in $\mu^2$ for almost all $t\in [0,1]$.
\end{lem}

\begin{proof}
The function $\cH_\mu (t)$ exists and is bounded because the $k$-th term in the sum is non-negative and upper bounded by $c_k$ (which is summable).
The monotonicity of $\cH_\mu (t)$ in $t$ and $\mu$ follows directly from~\eqref{eq:H_mu} given the monotonicity of $q_k (t)$ in $t$ and $\mu^{2k}$ in $\mu^2$.
The proof of Lemma~\ref{lem:gexit_expansion} establishes the absolute continuity of $\cH_\mu (t)$ and the power series expansion for its derivative.
Given the expansion, we know that $\cH_{\mu} ' (t) \geq 0$ almost everywhere because, for $k\in \mathbb{N}$, we have $-q_k '(t) \geq 0$ almost everywhere by statement $(ii)$ in Lemma~\ref{lem:H_M_q_Hu}.
We emphasize that this proof depends on parts $(i)$ and $(ii)$ of Lemma 19 (which are used in Lemma 20), but does not depend on part $(iii)$ of Lemma 19. Thus, using Lemma 40 to prove part $(iii)$ of Lemma 19 is not circular.

Likewise, this expansion shows that $\cH_{\mu} ' (t)$ is non-increasing in $\mu^2$ for almost all $t\in [0,1]$.
\end{proof}

\begin{rem}
As described above, $\cH_{\mu}(t)$ represents the conditional entropy of a random variable $X \in \{\pm 1\}$ with mean $\mu$ observed through the BMS channel $W(t)$. For a different interpretation, consider the setting where $X$ is uniformly distributed and $U$ is an observation through a BSC with crossover probability $p= (1-\mu)/2$. In this case $\ex{X \mid U} \in \{ \pm \mu\}$ almost surely, and since $\cH_{\mu}(t)$ is an even function of $\mu$, it follows that $H_{\mu}(t) = H (X \mid Y(t), U)$. 
\end{rem}

\begin{proof}[Proof of Lemma~{\ref{lem:area_bounds}}]
The existence and uniqueness of $t_R$ follow because $C(\cdot)$ is continuous and strictly increasing. Combining the area theorem~\eqref{eq:area_theorem} with the integral representation $C(t) = 1- \cH(t) = \int_t^1 \cH'(s) \,ds$ leads to the following decomposition:
\begin{align} \label{eq:Ct_minus_R_decomp}
\!\!\!\!\! C(t) - R &=  - \int_0^t G(s) \, ds + \int_t^1 \left(\cH'(s)  - G(s)  \right) \, ds.
\end{align}
Notice that this difference is strictly positive on $[0,t_R)$ and strictly negative on $(t_R, 1]$. 

Combining the expansions in~\eqref{eq:Ht_p} and~\eqref{eq:GEXIT_expansion}, we see that $\cH'(s) - G (s)$ satisfies
\begin{align}
\cH'(s) - G (s)
&= \sum_{k\in \mathbb{N}} c_k (- q_k '(s)) \| \E[ X_i \mid Y_{\sim i}(s) ]\|_{2k}^{2k} \label{eq:Ht_p_minus_G} \\
&\leq \sum_{k\in \mathbb{N}} c_k (- q_k '(s)) \underbrace{\| \E[ X_i \mid Y_{\sim i}(s) ]\|_{2}^{2}}_{1-M(s)} \\
&= (1-M(s)) \cH ' (s) 
, \label{eq:Ht_p_minus_G_ub}
\end{align}
where the inequality follows from $q_k ' (s) \leq 0$ almost everywhere and  $\| \E[ X_i \mid Y_{\sim i}(s) ]\|_{2k}^{2k} \leq \| \E[ X_i \mid Y_{\sim i}(s) ]\|_{2}^{2}$.
To prove the upper bound on $M(t)$, we combine~\eqref{eq:Ht_p_minus_G_ub} with the non-negativity of $G(s)$ (which follows from~\eqref{eq:GEXIT_expansion}) to obtain
\begin{align}
 C(t) - R &\le   \int_t^1 \left(1  - M(s)  \right) \cH'(s)  \, ds.
\end{align}
Multiplying both sides by $M(t)$ and recalling the $M(\cdot)$ is non-decreasing allows us to write
\begin{align}
M(t) \left( C(t) - R  \right) &\le   \int_t^1 M(s) \left(1  - M(s)  \right) \cH'(s)  \, ds\\
 &\le  \kappa(t)  \int_t^1 M(s) \left(1  - M(s)  \right)  \, ds.
\end{align}
If $t < t_R$ then $C(t) > R$ and we can divide both sides by $C(t) - R$ to obtain \eqref{eq:area_UB}.

Using the expansion \eqref{eq:GEXIT_expansion}, we observe that
\begin{align} \label{eq:G_lb}
G (s)
&= \sum_{k\in \mathbb{N}} c_k (- q_k '(s)) \left( 1- \| \E[ X_i \mid Y_{\sim i}(s) ]\|_{2k}^{2k} \right) \\
&\leq \sum_{k\in \mathbb{N}} c_k (- q_k '(s)) 
\big( 1- \underbrace{\| \E[ X_i \mid Y_{\sim i}(s) ]\|_{2}^{2k}}_{(1-M(s))^{k}} \big) \\
&= \cH_{\sqrt{1-M(s)}}' (s), \label{eq:G_ub}
\end{align}
where the inequality follows from $\| \E[ X_i \mid Y_{\sim i}(s) ]\|_{2k}^{2k} \geq \| \E[ X_i \mid Y_{\sim i}(s) ]\|_{2}^{2k}$ and the final step follows from~\eqref{eq:H_mu_p}.

Now, we focus on the lower bound in \eqref{eq:area_LB}.
We start by multiplying both sides of~\eqref{eq:Ct_minus_R_decomp} by negative one, applying~\eqref{eq:G_ub} to upper bound on $G(s)$, and then using the lower bound $\cH'(s) - G(s) \geq 0$ (which follows from~\eqref{eq:Ht_p_minus_G}).
This gives
\begin{align} \label{eq:R_Ct_bound_Hmu}
R - C(t) &\le  \int_0^t \cH'_{\sqrt{1-M(s)}}(s)  \, ds.%
\end{align}
Since $M(s)$ is non-decreasing in $s$ and $\cH'_\mu(s)$ is non-increasing in $\mu^2$ (see  Lemma~\ref{lem:H_mu})   we have
\begin{align}
    \cH'_{\sqrt{1- M(s)}}(s) \le  \cH'_{\sqrt{1- M(t)}}(s) , \qquad 0 \le s \le t \le 1.
\end{align}
Integrating both sides gives
\begin{align}
   \int_0^t \cH'_{\sqrt{1- M(s)}}(s) \, ds &\le  \int_0^t \cH'_{\sqrt{1- M(t)}}(s) \, ds \\
   &\le  \int_0^1 \cH'_{\sqrt{1- M(t)}}(s) \, ds \\
   & = h_b\left( \frac{ 1- \sqrt{ 1- M(t)}}{2} \right), \label{eq:H_u_hb_ub}
\end{align}
where the first inequality follows from $M(s) \leq M(t)$ for $s\in[0,t]$, the second inequality holds because $\cH'_\mu(\cdot)$ is non-negative, and the equality follows from $\cH_\mu(1)=h_b( (1- \mu)/2)$.
Since the mapping $z \mapsto h_b\left( \frac{ 1- \sqrt{ 1- z}}{2} \right)$ is strictly increasing on $[0,1]$ with inverse given by $\psi(\cdot)$,
we can combine this with~\eqref{eq:R_Ct_bound_Hmu} to see that
\begin{align}
    M(t) \ge \psi\left( R - C(t) \right), \qquad t \in [t_R, 1].
\end{align}
We can also strengthen this lower bound by incorporating knowledge about the area under the $M(s)(1-M(s))$ curve.
To do this, we write
\begin{align}
\int_{t_R}^t  \! M(s) (1 - M(s)) \, ds & \ge (1- M(t)) \int_{t_R}^t \! M(s)  \, ds\\
& \ge (1- M(t)) \int_{t_R}^t \! \psi(R - C(s))   \, ds.
\end{align}
Since $\psi(\cdot)$ is non-negative and strictly increasing, the integral is strictly positive and so we can rearrange terms to obtain the bound given in \eqref{eq:area_LB}.
\end{proof}

\appendix

\section{Additional Material}

\subsection{BMS Channels with General Output Alphabets}
\label{sec:bms_general}

For the purpose of our proof, it is convenient to focus on BMS channels satisfying the conditions in Definition~\ref{def:bms}, i.e., the output alphabet is equal to the extended reals and the transition probability satisfies $w(y \mid +1) = w(-y \mid -1)$. In this section, we provide a more general definition of BMS channels with respect to an arbitrary output alphabet $\cY$ and show any channel satisfying this definition can be mapped to one satisfying the conditions of Definition~\ref{def:bms}.

 Let $W_{Y\mid X}$ be a binary channel with input alphabet $\cX = \{\pm 1\}$, output alphabet $\cY$, and let $w (y \mid x)$ denote the conditional density of $Y$ with respect to a fixed dominating measure. It well-known that a minimal sufficient statistic for estimating $X$ from $Y$ is provided by log-likelihood ratio $\ell \colon \cY \to \bar{ \mathbb{R}}$, which is defined by 
\begin{align}
\ell(y) \coloneqq \log \frac{w(y \mid +1)}{w(y \mid -1)}.
\end{align}
Note that in cases where the output uniquely defines the input (e.g., the perfect channel), the log-likelihood ratio can take the values $\pm \infty$ in the extended real numbers.

\begin{defn}[Channel Symmetry] A binary channel $W_{Y|X}$ with input alphabet $ \cX = \{\pm 1\}$ and log-likelihood ratio $\ell$ is called symmetric if the conditional distribution of $\ell(Y)$ given the input is $+1$ is equal to the conditional distribution of $-\ell(Y)$ given the input is $-1$.
\end{defn}

For a symmetric channel, the relevant properties of the channel are completely  summarized by the distribution of the log-likelihood ratio when the input is $+1$. This distribution is often referred to as the $L$-density of the channel \cite{RU-2008}. As a consequence,  the specific details of the channel and the output space $\cY$ can be neglected and one may assume, without loss of generality, that the output alphabet is a subset of the extended reals. For example, if a random variable $X \in \{\pm 1\}$ is transmitted through a symmetric binary channel $W_{Y \mid X}$ that produces an output $Y$,  then the sufficient statistic  $\ell(Y)$ can be expressed as the product of the input $X$ and an \emph{independent} noise term $Z$ according to:
\begin{align}
\ell(Y) = X Z 
\end{align}
where $Z \coloneqq X \, \ell(Y)$ is drawn according to the conditional distribution of $\ell(Y)$ when the input is $+1$.

\subsection{Degradation Ordering of Channels}
\label{sec:degradation}

This section reviews some facts about channel degradation. The basic idea is that a channel $W_{Z \mid X}$ is degraded with respect to a channel $W_{Y \mid X}$ if the output of $W_{Z \mid X}$ can be simulated by post-processing the output of $W_{Y \mid X}$. 

\begin{defn}[Channel Degradation {\cite[p.~204]{RU-2008}}] Consider channels  $W_{Y \mid X}$ and $W_{Z \mid X}$  defined on the same input alphabet $\cX$. The channel $W_{Z \mid X}$ is said to be (stochastically) degraded with respect to $W_{Y\mid X}$ if there exists a third channel $W_{Z \mid Y}$  with input alphabet $\cY$ and output alphabet $\cZ$ such that $W_{Z \mid X}$ is equal to the composition of $W_{Y\mid X}$ and $W_{Z \mid Y}$. For example, if $w_{Y \mid X}(y \mid x)$ is a probability density function this means that 
\begin{align}
    w_{Z \mid X} (z \mid  x) = \int_{\cY} w_{Z \mid Y}(z  \mid y) \, w_{Y \mid X} (y \mid x) \, d y,
\end{align}
for all $x \in \cX$ and $z \in \mathcal{Z}$.
Likeiwse, if $w(y \mid x)$ is a probability mass function then the same expression holds with the integral replaced by a summation. 
\end{defn}

In some cases, the relationship between random variables is described  without specifying the channel explicitly. If  $Y$ and $Z$ represent two observations of a third random variable $X$, we say that  $Y$ is stochastically degraded w.r.t.\ $Z$ if the channel $W_{Z|X}$ is degraded w.r.t.\ the channel $W_{Y|X}$. 

The above definition is equivalent~\cite[p.~205]{RU-2008} to the statement that, for any distribution $p_X$ on the input alphabet $\cX$,  there exists a joint distribution on random variables $(X,Y,Z) \in \cX \times \cY \times \cZ$ such that:
\begin{itemize}
    \item $X$ has distribution $p_X$,
    \item $Y$ is an observation of $X$ through channel $W_{Y \mid X}$
    \item $Z$ is an observation of $X$ through channel $W_{Z \mid X}$; and
    \item $X - Y - Z$ forms a Markov chain.
\end{itemize}

The following is closely related to previous characterizations of channel degradation~\cite[p.~206]{RU-2008}.

\begin{lem}[Convex Order] \label{lem:convex_order}
Let $\cX$ be a vector space over $\mathbb{R}$ and let $X \in \cX$ be a random variable that is transmitted through two channels $W_{Y \mid X}$ and $W_{Z \mid X}$ whose outputs are $Y$ and $Z$, respectively.  If $W_{Z \mid X}$ is degraded with respect to $W_{Y \mid X}$, then for all convex functions $\phi: \cX \to \mathbb{R}$, we have
\begin{align*}
\ex{\phi\left(\ex{X\mid Y}\right)} \ge \ex{\phi\left(\ex{X\mid Z} \right)},
\end{align*}
provided that the expectations exist. 
In particular, if $X$ is real-valued then %
\begin{align*}
\ex{\ex{X \mid Y}^{2k}}
\geq \ex{\ex{X\mid Z}^{2k}}, \qquad k \in \mathbb{N}.
\end{align*}
\end{lem}
\begin{proof}
We note that expectations are defined using the vector space structure on $\cX$.
Observe that the expectations in the inequality depend only on the marginal distributions of the pairs $(X,Y)$ and $(X,Z)$ and thus we are free to consider any  joint distribution on $(X,Y,Z)$ with the same pairwise marginals. From the definition of channel degradation, there exists a joint distribution such that $X - Y - Z$ forms a Markov chain. Under the distribution, the conditional expectation satisfies $\ex{ X \mid Y, Z} = \ex{ X \mid Y}$ almost surely and so the first result follows from writing
\begin{align*}
\ex{\phi\left(\ex{X\mid Y}\right)} 
& = \ex{\phi\left(\ex{X\mid Y,Z}\right)} \\
&= \ex{ \ex{\phi\left(\ex{X\mid Y, Z}\right) \mid Z}} \\
&\geq \ex{\phi\left(\ex{\ex{X\mid Y, Z}\mid Z }\right)} \\
&= \ex{\phi\left(\ex{X\mid Z} \right)},
\end{align*}
where the third step follows from Jensen's inequality and the convexity of $\phi$.  The second result holds because $\phi(x) = x^{2k}$ is convex on $\mathbb{R}$ for all positive integers $k$. 
\end{proof}

\subsection{Comparison with Earlier Proof for the BEC}
\label{sec:bec_comparison}

This section discusses the relationship between the approach used in this paper, which is applicable to any BMS channel, and the approach used in earlier work which applies only to the BEC~\cite{Kudekar-it17}.
Recall that the proof in this paper depends crucially the nesting property of RM codes described Section~\ref{sec:new_obs}. In comparison,  the approach in \cite{Kudekar-it17} combines special properties of the BEC with results from the theory of boolean functions~\cite{Kahn-focs88,Bourgain-ijm92} to prove that any sequence of codes with a doubly transitive symmetry group achieves capacity. 

To make the comparison, we first simplify the approach used in this paper for the special case of the BEC.
For the BEC, let $t$ denote the erasure rate and recall that the GEXIT function simplifies to the EXIT function in this case.
Thus, we have
\[ G_i (t) = H(X_i \mid Y_{\sim i}(t)). \]
In addition, for any received sequence, the channel input $X_i$ is either recoverable or unknown.
It follows that the $\ex{X_i | Y_{\sim i} (t) }^2 \in \{0,1\}$ and the extrinsic MMSE also satisfies
\begin{align*}
    M_i (t) &= 1 - \ex{\ex{X_i | Y_{\sim i} (t) }^2} \\
    &= \Pr \big( \ex{X_i | Y_{\sim i} (t) } = 0 \big) = H(X_i \mid Y_{\sim i}(t)).
\end{align*}
Now, we assume that the code has transitive symmetry so that we can restrict our attention to $M(t) \coloneqq  M_0 (t)$ and use Lemma~\ref{lem:mmse_to_var} to upper bound the variance of the estimate.

Next, we will evaluate $\Delta_0^j \coloneqq \Delta_0^{\{j\}}$ by starting from its definition in~\eqref{eq:Delta_i}.
Suppressing $t$, we can rewrite this as $\Delta_0^j  = \frac{1}{2} \ex{D_j (\vY,Y_j ' )}$, where $Y_j ' $ is an independent observation of $X_j$ through the same channel and $D_j (\vy,y')$ defined to be
\begin{align*}
      \ex{ \left(\ex{ X_0 \mid Y_{\sim 0}} -\ex{ X_0 \mid  Y_{{\sim 0,j}}, Y_j '} \right)^2 \middle| \vY = \vy,  Y_j ' = y' }.
\end{align*}
Now, we observe that $D_j (\vy,y') \in \{0,1\}$ and it equals 0 unless $y_j \neq y'$.
If $y_j \neq y'$, then this quantity is related to the influence (from the theory of boolean functions) and we see that
\begin{align*}
    &\ex{D_j (\vY,Y_j ' ) \, | \, Y_j  \neq Y_j' } \\ & \qquad = \Pr(\ex{X_0|Y_{\sim 0}} \neq \ex{X_0|Y_{\sim 0,j},Y_j',Y_j'\neq Y_j}) =  I_j  ,
\end{align*}
where $I_j $ is influence of the $j$-th received value on the EXIT function as defined in~\cite{Kudekar-it17}.
Since we have $\Pr (Y_j \neq Y_j ') = 2t(1-t)$, it follows that
\begin{equation} \label{eq:delta_inf}
\Delta_0^j (t) = \frac{2 t (1-t)}{2} I_j (t).
\end{equation}

From~\cite[Remark~18]{Kudekar-it17}, we also know that
\[ I_j (t) = \left\{\frac{d}{d s_j} H(X_0 | Y_{\sim 0} (s_0,\ldots,s_{N-1})) \right\}_{(s_0,\ldots,s_{N-1})=(t,\ldots,t)}\!\!\!\!. \]
Notice that $I_0 (t) = 0$ because $Y_{\sim 0}$ does not depend on $Y_0$.
Assuming doubly transitive symmetry, we see that $I_j (t) = I_1 (t)$ for all $j \in [N] \setminus \{0\}$.
Thus, the total derivative formula implies that
\[ \frac{d}{dt} H(X_0|Y_{\sim 0} (t)) = (N-1) I_1 (t). \]
Following the approach in this paper, we can use~\eqref{eq:delta_inf} to see that 
\begin{align*}
  \int_0^1 \Delta_0^j (t) \, dt &\leq \frac{1}{4} \int_0^1 I_j (t) \, dt \\ &= \frac{1}{4(N-1)} \int_0^1 \left(\frac{d}{dt} H(X_0|Y_{\sim 0} (t)) \right) \, dt \\ &= \frac{1}{4(N-1)},
\end{align*}
where the integral equals 1 if the minimum distance of the code is at least 2.
We can also apply this bound to a subset $A \subseteq [N]$ by summing over all $j\in A$. From this, we see that the total contribution will vanish as long as $|A|/N$ vanishes for the chosen sequence of codes.

In contrast, the proof in~\cite{Kudekar-it17} is based on results from the theory of boolean functions~\cite{Kahn-focs88,Bourgain-ijm92} that imply $I_1 (t) \geq C \frac{\ln N}{N} H(X_0|Y_{\sim 0} (t)) \big(1-H(X_0|Y_{\sim 0} (t))\big)$ for some constant $C>0$.
Thus, the proof in~\cite{Kudekar-it17} shows that, for any $\delta >0$, the quantity $H(X_0|Y_{\sim 0} (t))$ must transition from $\delta$ to $1-\delta$ over an interval whose width is roughly $1/(C \ln N)$.

In this paper, the remaining terms in~\eqref{eq:Delta_decomp} are grouped together.
To analyze $\cC = \mathrm{RM}(r,m)$ with $N=2^m$, we choose $k\geq 1$ and define $A = [2^{m-k}]$.
By Lemma~\ref{lem:rm_punct}, we see that $X_A$ is a uniform random codeword from $\mathrm{RM}(r,m-k)$.
Then, we define $B = [N] \setminus A$ and recall, from Section~\ref{sec:two_look_bound}, that $\Delta^B_0(t)$ equals
\begin{align}
\frac{1}{2}\ex{ \left(\ex{ X_0 \mid Y_{A}(t), Y_{B}(t)} -\ex{ X_0 \mid Y_{A}(t), Y'_{B}(t)} \right)^2}, 
\end{align} 
where $\vY'(t)$ denotes an independent second observation of $\vX$ through a BEC with the same erasure probability.
Since we are working on the BEC, both inner conditional expectations can only take values in the set $\{-1,0,1\}$ with 0 indicating erasure and $\pm 1$ indicating successful recovery.
Thus, we can simplify $\Delta^B_0(t)$ by expanding the square and taking expectations to get
\begin{align*}
\Delta^B_0(t) &= \ex{ \ex{ X_0 \mid Y_{A}(t), Y_{B}(t)}^2} \\ & \qquad - \expt \big[ \ex{ X_0 \mid Y_{A}(t), Y_{B}(t)} \ex{ X_0 \mid Y_{A}(t), Y_{B} ' (t)} \big] \\
&\leq \ex{ \ex{ X_0 \mid Y_{A}(t), Y_{B}(t)}^2} \\ & \qquad - \expt \big[ \ex{ X_0 \mid Y_{A}(t), Y_{B}(t), Y_{B} ' (t)} \big] \\
&= H(X_0 | Y_A (t), Y_B (t) ) - H(X_0 | Y_A (t), Y_B (t), Y_B ' (t) ) \\
& \leq H(X_0 | Y_A (t), Y_B (t) ) - H(X_0 ' | U_{\sim 0} (t) ).
\end{align*}
The first inequality holds because it may be possible to recover $X_0$ by jointly processing $Y_A, Y_B, Y_B '$ even when it cannot be recovered separately from either $Y_A,Y_B$ or $Y_A,Y_B '$.
The second inequality follows from assuming that $\vU(t)$ is the observation of a uniform random codeword $\vX'$ from $\mathrm{RM}(r,m+k)$ and that $(X_0,Y_A,Y_B,Y_B ')$ is equal in distribution to $(X_0 ', U_A, U_B, U_C)$ (e.g., see Lemma~\ref{lem:rm_overlap} and Section~\ref{sec:two_look_bound}).

Finally, we can put things together.
First, we can integrate the upper bound on $ \Delta^B_0(t)$ to see that
\begin{align*}
\int_0^1 & \Delta^B_0(t) \, dt \\
&\leq \int_0^1 H(X_0 | Y_A (t), Y_B (t) ) \, dt - \int_0^1 H(X_0 ' | U_{\sim 0} (t) ) \, dt \\
&= R(r,m) - R(r,m+k) \\
&\leq \frac{3k+4}{5\sqrt{m}},
\end{align*}
where the last step follows from Lemma~\ref{lem:RM_rate}.
Then, we can integrate~\eqref{eq:Delta_decomp} to see that
\begin{align*}
    \int_0^1 \! M_i(t) (1- M_i(t)) \, dt
    & \le  \int_0^1 \! \Delta_i^B(t) \, dt   + \sum_{j \notin  B} \int_0^1 \! \Delta_i^{j}(t) \, dt \\
    & \le \frac{3k+4}{5\sqrt{m}} + \frac{2^{m-k}}{4 (2^m - 1)}.
\end{align*}
This upper bound vanishes if we consider a code sequence where $m\to \infty$ with $k$ chosen according to $k = \lfloor \log_2 m \rfloor$.
Thus, the EXIT function has a sharp threshold and the EXIT area theorem (e.g., see~\cite[Proposition~11]{Kudekar-it17}) implies that $M(t) = H(X_0|Y_{\sim 0} (t))$ will jump at $1-R$ in the limit.

\subsection{Localization of Jump in Extrinsic MMSE via Sequences} \label{sec:MMSE_sequences}

In Section~\ref{sec:MMSE_bounds}, we provide non-asymptotic bounds on the extrinsic MMSE associated with a family of BMS channels and an RM$(r,m)$ code. Applying these bounds to a sequence of RM codes with strictly increasing blocklength and code rate converging to $R \in (0,1)$, shows that the extrinsic MMSE converges to a 0/1 step function that jumps at the unique point $t_R$ such that $C(t_R) = R$. For that result, this section provides an alternative proof which may be of independent interest.

In particular, we make use of Lemma~\ref{lem:contiguity} below which shows that convergence of the extrinsic MMSE to 0 or 1 is equivalent to convergence of the GEXIT to its lower and upper bounds, respectively.

Let $\{ \cC^{(n)}\}_{n \in \mathbb{N}}$ be a sequence of transitive codes with strictly increasing blocklength and rate converging to $R \in (0, 1)$.  For a BMS family satisfying Assumption~\ref{assumption:BMS}, let $\{ (G^{(n)}, M^{(n)})\}_{n \in \mathbb{N}}$ be the corresponding sequence of GEXIT functions and extrinsic MMSE functions.
The bounds given here and in Section~\ref{sec:MMSE_bounds} depend primarily on the quantity \[a_n = \int_0^1 M^{(n)}(s) (1-M^{(n)}(s))\, ds.\]
We will see that a code sequence achieves capacity on the family of BMS channels if $a_n \to 0$.

The approach taken in this section is a proof by contradiction.
Suppose that $a_n \to 0$ but the sequence of extrinsic MMSE functions, $M^{(n)}(t)$, does not converge to a 0/1 step function that jumps at $t=t_R$.
Then, one of two things must happen. 
Either there is a $t' < t_R$, an $\epsilon \in (0,1)$, and a subsequence $M^{(n_k)}(t)$ such that $M^{(n_k)} (t') \geq \epsilon $ for all $k\in \mathbb{N}$.
Or, there is a $t' > t_R$, an $\epsilon \in (0,1)$, and a subsequence $M^{(n_k)}(t)$ such that $M^{(n_k)} (t') \leq 1-\epsilon $ for all $k\in \mathbb{N}$.

The following lemma implies that both possibilities lead to contradictions.
To see this, we recall that the area theorem implies
\[ \int_{0}^1 G^{(n)}(t)\, dt \to R. \]
This also implies that the limit is the same for any subsequence $G^{(n_k)}(t)$.
Now, for the $t' < t_R$ case, we see~\eqref{eq:lemma42_1}  implies that the limit inferior of the sequence of GEXIT integrals is at least $C(t')>R$ which gives a contradiction.
By comparing~\eqref{eq:GEXIT_expansion} and~\eqref{eq:Ht_p}, it is easy to see that $G^{(n)} (t) \leq \cH'(t)$.
Thus, for the $t' > t_R$ case, we see the sequence of GEXIT integrals is upper bounded by $C(t') = \int_{t'}^1 \cH' (t) \, dt < R$ which gives a contradiction.

The lemma is obtained by combining an upper bound on $a_n$ (e.g., see Lemma~\ref{lem:integral_decay}) with the comparison between the GEXIT function $G^{(n)}(t)$ and the extrinsic MMSE $M^{(n)}(t)$ established in Lemma~\ref{lem:contiguity}.
Thus, the sequence of extrinsic MMSE functions, $M^{(n)}(t)$, converges to a 0/1 step function that jumps at $t=t_R$.
Finally, applying Lemma~\ref{lem:contiguity} again shows that the sequence of GEXIT functions $G^{(n)} (t)$ converges almost everywhere to a function that jumps from 0 to $\cH'(t)$ at $t=t_R$.

\begin{lem}\label{lem:Mn_convergence}
Under the assumptions stated above, if $a_n \to 0$, then, for every $t' \in (0,1)$, we have
\begin{alignat}{3}
\liminf_{n \to \infty} M^{(n)}(t') > 0  \; &\implies \;     \int_{t'}^1 G^{(n)}(t) \, dt \to C(t') \;\;\; \label{eq:lemma42_1}\\
\limsup_{n \to \infty} M^{(n)}(t') < 1  \; &\implies \;     \int_{0}^{t'} G^{(n)}(t) \, dt \to 0. \label{eq:lemma42_2}
\end{alignat}
\end{lem}
\begin{proof} If  $\liminf_{n \to \infty} M^{(n)}(t') > 0 $ then there exists an $\eps \in (0,1)$ and an integer $N$  such that $M^{(n)}(t') \ge \eps$ for all  $n \ge N$. For $t \in (t',1]$ and $n \ge N$, we can write
\begin{align}
    &\int_0^1 M^{(n)}(s) (1 - M^{(n)}(s)) \, ds \\ &\;\ge     \int_{t'}^t \! M^{(n)}(s) (1 - M^{(n)}(s)) \, ds  \ge  \eps (t- t') (1 - M^{n}(t)),
\end{align}
where the second inequality follows from $\eps \le M^{(n)}(s) \le M^{(n)}(t)$ for $s \in [t',t]$. By assumption, the LHS (i.e., $a_n$) converges to 0 and this proves that $M^{(n)}(t) \to 1$ for all $t \in (t',1]$. By Lemma~\ref{lem:contiguity}, it follows that
\begin{align}
 \int_{t'}^{1} \big( \cH'(t) - G^{(n)}(t) \big) \, dt \to 0,
\end{align}
which is equivalent to the stated result in view of the fact that $C(t') = \int_{t'}^1\cH'(t) \, dt$.

For the second statement, the argument is essentially the same.
If  $\limsup_{n \to \infty} M^{(n)}(t') < 1 $ then there exists an $\eps \in (0,1)$ and an integer $N$  such that $M^{(n)}(t') \le 1-\eps$ for all  $n \ge N$. For $t \in [0,t')$ and $n \ge N$, we can write
\begin{align}
    &\int_0^1 \! M^{(n)}(s) (1 - M^{(n)}(s)) \, ds \\ & \quad\ge     \int_{t}^{t'} \! M^{(n)}(s) (1 - M^{(n)}(s)) \, ds  \ge  M^{(n)}(t') (t'- t) \eps,
\end{align}
where the second inequality follows from $1-M^{(n)}(s) \ge 1 - M^{(n)}(t') \ge \epsilon$ for $s \in [t,t']$. By assumption, the LHS (i.e., $a_n$) converges to 0 and this proves that $M^{(n)}(t) \to 0$ for all $t \in [0,t')$.
Similarly, the second result follows from applying Lemma~\ref{lem:contiguity}.
\end{proof}

\begin{lem}
\label{lem:contiguity}
Using the setup from Lemma~\ref{lem:gexit_expansion}, assume that $\cM (t)$ is strictly increasing and consider a sequence of problems where the BMS channel family is fixed but the code is changing (e.g., $\vX$ depends on $n$).
Let $\{ (G^{(n)}, M^{(n)})\}_{n \in \mathbb{N}}$ be the corresponding sequence of GEXIT and extrinsic MMSE functions for the same symbol (say $X_0$).
Then, for any $t' \in (0,1)$, we have
\begin{align}
\int_0^{t'} \!\! G^{(n)} (s) \, ds \to 0 \; \Leftrightarrow \; \forall s \in [0,t'), \; M^{(n)} (s) \to 0, \;\;\; \label{eq:gexit_contig_0}
\end{align}
\begin{align}
\int_{t'}^{1} \! \big( \cH'(s) - G^{(n)} (s) \big) \, ds \to 0  \Leftrightarrow \; \forall s\in (t',1], \, M^{(n)} (s) \to 1  . \label{eq:gexit_contig_1}
\end{align}
\end{lem}

\begin{proof}
Without loss of generality, we assume that $G^{(n)} (t)$ and $M^{(n)} (t)$ are the GEXIT and extrinsic MMSE functions of $X_0$ for the $n$-th problem in the sequence.
We will need two bounds on the GEXIT function to proceed.
The first is derived in~\eqref{eq:Ht_p_minus_G_ub} and
rewriting it in the notation of this lemma gives
\begin{align} \label{eq:gexit_mmse1}
\cH'(t) - G^{(n)} (t) \leq \left(1 - M^{(n)} (t)\right) \mathcal{H}'(t).
\end{align}
The second will be derived shortly and can be stated as
\begin{align} \label{eq:gexit_mmse2}
\cH'(t) - G^{(n)} (t)  \geq \frac{(1-M^{(n)} (t))\cM'(t)}{2\ln 2}.
\end{align}
To see this, we can subtract~\eqref{eq:GEXIT_expansion} from~\eqref{eq:H_mu_p}
and lower bound by the first term in the resulting sum because all terms are non-negative.
Then,~\eqref{eq:gexit_mmse2} holds because $- q_k ' (t) = \cM'(t)$ and $\| \E[ X_i \mid Y_{\sim i} (t) ]\|_{2}^{2} = 1-M^{(n)}(t)$.

Since $1-M^{(n)} (s)$ is non-increasing, it follows that $1-M^{(n)}(t') \leq 1-M^{(n)} (s) \leq 1-M^{(n)} (t)$ for $s\in [t,t']$ and $0 \leq t \leq t' \leq 1$.
Integrating~\eqref{eq:gexit_mmse1} and~\eqref{eq:gexit_mmse2} over the interval $[a,b]$ shows that
\begin{align*} \label{eq:gexit_mmse1_3}
\big( 1- M^{(n)} (b) \big) &\frac{\cM(b) - \cM(a)}{2\ln 2} \leq \int_{a}^{b} \big( \cH'(s) - G^{(n)} (s) \big) \, ds
\\ & \leq \big( 1- M^{(n)} (a) \big) \big( \cH(b) - \cH(a) \big). 
\end{align*}

{
\emph{Proof of $\Longleftarrow$ in \eqref{eq:gexit_contig_1}:} Starting with the fact that $G^{(n)}(t) \le \cH'(t)$ almost everywhere, we can write 
\begin{align}
0 &\le \int_{t'}^1 (\cH'(s) - G^{(n)}(s) )\,ds\\
&= \int_{t'}^t (\cH'(s) - G^{(n)}(s) )\,ds  + \int_{t}^1 (\cH'(s) - G^{(n)}(s) )\,ds \\
&\le  (1- M^{(n)}(t')) \underbrace{ (\cH(t)- \cH(t'))}_{< \eps}  \\ & \qquad \qquad \qquad \qquad \qquad + \underbrace{(1- M^{(n)}(t)) }_\text{$< \eps$ for all $n>N$}  (1- \cH(t)), 
\end{align}
where the last step follows from two applications of  \eqref{eq:gexit_mmse1_3}.
By the continuity of $\cH(\cdot)$, for any $\eps >0$, there exists $t \in (t', 1]$ such that $\cH(t) - \cH(t') < \eps$.
Since $M^{(n)}(t) \to 1$, there exists $N \in \mathbb{N}$ such that $1-M^{(n)}(t) < \eps$ for all $n>N$.
Thus, the RHS converges to 0 because, for any $\eps >0$, there is an $N\in \mathbb{N}$ such that the RHS is less than $2\epsilon$ for all $n>N$.
}

\emph{Proof of $\Longrightarrow$ in \eqref{eq:gexit_contig_1}:}
Consider the left-hand inequality of~\eqref{eq:gexit_mmse1_3}.
Since the integrand is non-negative, the integral from $a=t'$ to $b=t'+\delta$, with $\delta \in (0,1-t']$, is upper bounded by the integral from $a=t'$ to $b=1$.
Thus, for all $\delta \in (0,1-t']$, we see that
\begin{align*}
& M^{(n)} (t'+\delta) \\ & \qquad \geq 1 - \frac{2\ln 2}{\cM(t'+\delta) - \cM(t')} \int_{t'}^{1} \big( \cH'(s) - G^{(n)} (s) \big) \, ds.
\end{align*}
Since the integral on the RHS converges to 0 and $\cM(t+\delta) - \cM(t) > 0$, it follows that $M^{(n)} (t'+\delta) \to 1$ for all $\delta \in (0,1-t']$.

\emph{Proof of $\Longrightarrow$ in \eqref{eq:gexit_contig_0}:}
Using the expansion in~\eqref{eq:GEXIT_expansion}, it follows from non-negativity of each term and $\E[ X_i \mid Y_{\sim i} (s) ]^{2k} \leq \E[ X_i \mid Y_{\sim i} (s) ]^{2}$ that $G^{(n)} (s) \geq M^{(n)} (s) \cH' (s)$ almost everywhere.  Thus, for any $\delta \in (0,t']$, we can write
\begin{align*}
\int_{0}^{t'} G^{(n)} (s) \, ds
&\geq \int_{t'-\delta}^{t'} G^{(n)} (s) \, ds \\
& \geq
\int_{t'-\delta}^{t'} M^{(n)} (s) \mathcal{H}'(s) \, ds \\ &\geq
\big( \cH(t') - \cH(t'-\delta) \big) M^{(n)} (t'-\delta), 
\end{align*}
where the first inequality follows from $G^{(n)}(s) \geq 0$ almost everywhere, the second inequality is due to the lower bound mentioned above, and the third inequality follows from integrating after applying  $M^{(n)} (s) \geq M^{(n)} (t'-\delta)$ for $s \in [t'-\delta,t']$ and integration.
Since $\cM(t)$ is strictly increasing,~\eqref{eq:cMdiff_cHdiff} implies that $\cH(t') - \cH(t'-\delta) \geq \big( \cM(t') - \cM(t'-\delta) \big) / (2\ln 2) > 0$ for $\delta \in (0,t']$.
Thus, if $\int_{0}^{t'} G^{(n)} (s) \, ds \to 0$, then $M^{(n)} (t'-\delta) \to 0$ for all $\delta \in (0,t']$.

\emph{Proof of $\Longleftarrow$ in \eqref{eq:gexit_contig_0}:}
Since $M^{(n)} (t)$ is non-decreasing and Lemma~\ref{lem:H_mu} establishes that $\cH_{\mu} ' (t)$ is non-increasing in $\mu$, we can upper bound the integral of $G^{(n)} (t)$ over the interval $[0,t]$ with
\begin{align*}
\int_0^{t} G^{(n)} (s) \, ds
&\leq \int_0^{t} \mathcal{H}_{\sqrt{1-M^{(n)} (s)}} ' (s) \, ds \\
&\leq \int_0^{t} \mathcal{H}_{\sqrt{1-M^{(n)} (t)}} ' (s) \, ds \\
&\leq \int_0^{1} \mathcal{H}_{\sqrt{1-M^{(n)} (t)}} ' (s) \, ds \\
&= h_b \left(\frac{1-\sqrt{1-M^{(n)} (t)}}{2}\right),
\end{align*}
where the first inequality is given by~\eqref{eq:G_ub}, the second inequality holds because $\cH_{\sqrt{1-m}}'  (t)$ is non-decreasing in $m$ for almost all $t\in[0,1]$, the third inequality follows from the fact that $\cH_{\mu} ' (t) \geq 0$ almost everywhere, and the final equality is given by~\eqref{eq:H_u_hb_ub}.
To complete the proof, for any $t\in [0,t')$, we write
\begin{align}
\int_{0}^{t'}  G^{(n)}&(s)\,ds
= \int_{0}^{t} G^{(n)}(s) \,ds  + \int_{t}^{t'} G^{(n)}(s) \,ds \\
&\le  \underbrace{h_b \left(\frac{1-\sqrt{1-M^{(n)} (t)}}{2}\right)}_\text{$< \eps$ for all $n>N$} + \underbrace{ (\cH(t')- \cH(t))}_{< \eps}.
\end{align}
By the continuity of $\cH(\cdot)$, for any $\eps >0$, there exists $t \in [0,t')$ such that $\cH(t') - \cH(t) < \eps$.
Since $M^{(n)}(t) \to 0$, continuity of the $h_b$-term in $M^{(n)} (t)$ implies that there is an $N \in \mathbb{N}$ such that it is less than $\eps$ for all $n>N$.
Thus, the RHS converges to 0 because, for any $\eps >0$, there is an $N\in \mathbb{N}$ such that the RHS is less than $2\epsilon$ for all $n>N$.
\end{proof}

\section*{Acknowledgements}
 The authors would like to thank Shrinivas Kudekar for insightful comments on a draft of this manuscript.
 They are also indebted to the anonymous reviewers whose comments greatly improved the quality of the presentation.

\ifbib
\bibliographystyle{ieeetr}
\bibliography{WCLabrv,WCLbib,WCLnewbib,bib_gr}
\else
\bibliography{main.bbl}
\fi

\ifbio
\begin{IEEEbiographynophoto}{Henry D. Pfister} (S'99-M'03-SM'09) received his Ph.D. in Electrical Engineering in 2003 from the University of California, San Diego and is currently a professor in the Electrical and Computer Engineering Department of Duke University with a secondary appointment in Mathematics.  Prior to that, he was an associate professor at Texas A\&M University (2006-2014), a post-doctoral fellow at the \'{E}cole Polytechnique F\'{e}d\'{e}rale de Lausanne (2005-2006), and a senior engineer at Qualcomm Corporate R\&D in San Diego (2003-2004).  His current research interests include information theory, error-correcting codes, quantum computing, and machine learning.

He received the NSF Career Award in 2008 and a Texas A\&M ECE Department Outstanding Professor Award in 2010.  He is a coauthor of the 2007 IEEE COMSOC best paper in Signal Processing and Coding for Data Storage, a coauthor of a 2016 Symposium on the Theory of Computing (STOC) best paper, and a recipient of the 2021 Information Theory Society Paper Award.  He has served the IEEE Information Theory Society as a member of the Board of Governors (2019-2022), an Associate Editor for the IEEE Transactions on Information Theory (2013-2016), and a Distinguished Lecturer (2015-2016).  He was the General Chair of the 2016 North American School of Information Theory and a Technical Program Committee Co-Chair of the 2021 International Symposium on Information Theory.
\end{IEEEbiographynophoto}
\fi

\end{document}